\documentclass[11pt,a4paper]{article}
\pdfoutput=1
\usepackage{a4wide}
\usepackage{amsmath}
\usepackage{cite}
\usepackage{color}
\usepackage{amssymb}
\usepackage{amsfonts}
\usepackage{booktabs}
\usepackage{multirow}
\usepackage{pdflscape}
\usepackage{gensymb}
\usepackage{graphicx}
\usepackage{bigstrut}
\usepackage{pifont}
\usepackage{enumerate}
\usepackage[small,bf]{caption}
\def\1{\mathbf{1}}
\def\3{\mathbf{3}}
\def\2{\mathbf{2}}

\def\D{\Delta}

\def\th{\theta}

\def\o{\omega}
\def\O{\Omega}

\def\gtap{\ \raisebox{-.4ex}{\rlap{$\sim$}} \raisebox{.4ex}{$>$}\ }
\def\ltap{\ \raisebox{-.4ex}{\rlap{$\sim$}} \raisebox{.4ex}{$<$}\ }
\def\di{{\rm d}}

\numberwithin{equation}{section}

\DeclareMathOperator{\diag}{diag}

\DeclareMathOperator{\im}{Im}
\DeclareMathOperator{\re}{Re}
\DeclareMathOperator{\arccot}{arccot}

\newcommand{\bo}[1]{\boldsymbol{#1}}

\newcommand{\be}{\begin{equation}}
\newcommand{\ee}{\end{equation}}

\newcommand{\mefff}{\mbox{$ \langle\!\, m \,\!\rangle $}}
\newcommand{\meff}{\mbox{$\left|  \langle \!\,  m  \,\!  \rangle \right| $}}
\newcommand{\betabeta}{\mbox{$(\beta \beta)_{0 \nu}  $}}

\newcommand{\btb}{\bigstrut}

\definecolor{darkgreen}{RGB}{34,181,34}
\definecolor{darkred}{RGB}{225,0,0}
\newcommand{\cmark}{\color{darkgreen} \text{\ding{51}}}
\newcommand{\xmark}{\color{darkred} \text{\ding{55}}}

\begin{document}
\begin{titlepage}

\vspace*{-15mm}
\begin{flushright}
SISSA 20/2017/FISI \\
IPMU17-0057
\end{flushright}
\vspace*{0.7cm}

\begin{center}
{\bf\Large Neutrino Mixing and Leptonic CP Violation} \\[2mm]
{\bf\Large from $\bo{S_4}$ Flavour and Generalised CP Symmetries} \\[8mm]
J.~T.~Penedo$^{~a,}$\footnote{E-mail: \texttt{jpenedo@sissa.it}}, 
S.~T.~Petcov$^{~a,b,}$\footnote{Also at:
Institute of Nuclear Research and Nuclear Energy,
Bulgarian Academy of Sciences, 1784 Sofia, Bulgaria.} 
and
A.~V.~Titov$^{~a,}$\footnote{E-mail: \texttt{atitov@sissa.it}} \\
\vspace{8mm}
$^{a}$\,{\it SISSA/INFN, Via Bonomea 265, 34136 Trieste, Italy} \\
\vspace{2mm}
$^{b}$\,{\it Kavli IPMU (WPI), University of Tokyo, 5-1-5 Kashiwanoha, 277-8583 Kashiwa, Japan}
\end{center}
\vspace{8mm}

\begin{abstract}
\noindent
We consider a class of models of neutrino mixing 
with $S_4$ lepton flavour symmetry combined with 
a generalised CP symmetry, which are broken to 
residual  $Z_2$ and $Z_2 \times H^\nu_{\rm CP}$ symmetries in 
the charged lepton and neutrino sectors, respectively,
$H^\nu_{\rm CP}$ being a remnant CP symmetry of the neutrino  
Majorana mass term.
In this set-up the neutrino mixing angles and CP violation (CPV) phases 
of the neutrino mixing matrix 
depend on three real parameters~---~two angles and a phase. 
We classify all phenomenologically viable mixing patterns 
and derive predictions for the Dirac and Majorana CPV phases. 
Further, we use the results obtained on the neutrino mixing angles and 
leptonic CPV phases to derive predictions for the effective 
Majorana mass in neutrinoless double beta decay.  
\end{abstract}

\end{titlepage}

\setcounter{footnote}{0}

\section{Introduction}
\label{sec:intro}

  Understanding the origin of the pattern of neutrino mixing 
that emerged from the neutrino oscillation data in the recent years 
(see, e.g., \cite{Olive:2016xmw})
is one of the most challenging problems in neutrino physics.
It is part of the more general fundamental problem
in particle physics of understanding the origins of
flavour in the quark and lepton sectors, i.e., of the patterns
of quark masses and mixing, and of the charged lepton 
and neutrino masses and of neutrino mixing.

  The idea of extending the Standard Model (SM) with 
a non-Abelian discrete flavour symmetry has been widely 
exploited in attempts to make progress towards the 
understanding the origin(s) of flavour 
(for reviews on the discrete symmetry 
approach to the flavour problem see, e.g.,  
\cite{Altarelli:2010gt,Ishimori:2010au,King:2014nza}).
In this approach it is assumed that at a certain high-energy 
scale the theory possesses 
a flavour symmetry, which is broken at lower energies 
to  residual symmetries of the charged lepton and neutrino sectors, 
yielding certain predictions for the values of, and/or 
correlations between, the low-energy neutrino 
mixing parameters. 
In the reference 3-neutrino mixing scheme 
we are going to consider in what follows 
(see, e.g., \cite{Olive:2016xmw}),
i) the values of certain pairs of, or of all three,  
neutrino mixing angles 
are predicted to be correlated, and/or
ii) there is a correlation between the value 
of the Dirac CP violation (CPV) phase $\delta$ 
in the neutrino mixing matrix and the values 
of the three neutrino mixing angles~%
\footnote{Throughout the present study we 
use the standard parametrisation of the 
Pontecorvo, Maki, Nakagawa and Sakata (PMNS) 
neutrino mixing matrix (see, e.g., \cite{Olive:2016xmw}). 
}, 
$\th_{12}$, $\th_{13}$ and $\th_{23}$,
which includes also symmetry dependent
fixed parameter(s) (see, e.g., 
\cite{Petcov:2014laa,Girardi:2015vha,Girardi:2015rwa,Marzocca:2013cr,Tanimoto:2015nfa,Ballett:2013wya,Ge:2011qn,Antusch:2011ic} and references quoted therein). 
These correlations are usually referred to as 
``neutrino mixing sum rules''.
As we have already indicated, the sum rules for the Dirac 
phase $\delta$, in particular, depend on the
underlying symmetry form of the PMNS matrix
\cite{Petcov:2014laa,Girardi:2015vha,Girardi:2015rwa,Marzocca:2013cr,Tanimoto:2015nfa} 
(see also, e.g., \cite{Ballett:2013wya,Ge:2011qn,Antusch:2011ic}),
which  in turn is determined by the assumed lepton
favour symmetry that typically has to be broken, 
and by the residual unbroken symmetries in the charged lepton and neutrino sectors        
(see, e.g., \cite{Altarelli:2010gt,Ishimori:2010au,King:2014nza,Girardi:2015rwa,Tanimoto:2015nfa}). 
They can be tested experimentally 
(see, e.g., \cite{Girardi:2015vha,Ballett:2013wya,Girardi:2014faa,Hanlon:2013ska}).
These tests can provide unique information about the 
possible existence of a new fundamental symmetry 
in the lepton sector, which determines the pattern 
of neutrino mixing \cite{Petcov:2014laa}.  
Sufficiently precise experimental data on 
the neutrino mixing angles 
and on the Dirac CPV  phase can also be used 
to distinguish between 
different possible underlying 
flavour symmetries leading to viable 
patters of neutrino mixing.

 While in the discrete flavour symmetry approach 
at least some of the neutrino mixing angles 
and/or the Dirac phase are determined (directly or 
indirectly via a sum rule)
by the flavour symmetry, the Majorana CPV phases 
$\alpha_{21}$ and $\alpha_{31}$ \cite{Bilenky:1980cx}  
remain unconstrained. 
The values of the Majorana CPV phases are 
instead constrained to lie in 
certain narrow intervals, or are predicted,  
in theories which in addition to a flavour symmetry 
possess at a certain high-energy scale  
a generalised CP (GCP) symmetry \cite{Branco:1986gr}.
The GCP symmetry should be implemented in a theory 
based on a discrete flavour symmetry in a way 
that  is consistent with the flavour symmetry 
\cite{Feruglio:2012cw,Holthausen:2012dk}.
At low energies the GCP symmetry is broken, in general, 
to residual CP symmetries of the charged lepton 
and neutrino sectors. 

 In the scenarios involving a GCP symmetry, which were 
most widely explored so far 
(see, e.g., \cite{Feruglio:2012cw,Ding:2013hpa,Ding:2013bpa,Li:2015jxa,DiIura:2015kfa,Ballett:2015wia}),  
a non-Abelian flavour symmetry $G_f$ consistently combined with a 
GCP symmetry $H_{\rm CP}$ is broken to residual Abelian symmetries 
$G_e = Z_n$, $n > 2$, or $Z_m \times Z_k$, $m,\,k \geq 2$,  
and $G_\nu = Z_2 \times H^\nu_{\rm CP}$ of the charged lepton and 
neutrino mass terms, respectively~%
\footnote{We note that in refs.~\cite{Ding:2013bpa,Li:2015jxa} 
the residual symmetry $G_e$ of the charged lepton mass term 
is augmented with a remnant CP symmetry $H^e_{\rm CP}$ as well.
}. 
The factor $H^\nu_{\rm CP}$ in $G_\nu$ stands 
for a remnant GCP symmetry of the neutrino mass term. 
In such a set-up, $G_e$ fixes completely the 
form of the unitary matrix $U_e$ 
which diagonalises the product $M_e M_e^\dagger$ 
and enters into the expression of the PMNS matrix, 
$M_e$ being the charged lepton mass matrix
(in the charged lepton mass term written in the left-right convention). 
At the same time, $G_\nu$ fixes the unitary 
matrix $U_\nu$, diagonalising the neutrino Majorana mass 
matrix $M_\nu$ up to a single free 
real parameter~---~%
a rotation angle $\th^\nu$. 
Given the fact that the 
PMNS neutrino mixing matrix $U_{\rm PMNS}$ is given 
by the product 
\be
U_{\rm PMNS} = U_e^\dagger\,U_\nu\,,
\ee
%
all three neutrino mixing angles
are expressed in terms of this rotation angle.
In this class of models one obtains 
specific correlations between the values of the 
three neutrino mixing angles, while 
the leptonic CPV phases are typically 
predicted to be exactly 
$0$ or $\pi$, or else $\pi/2$ or $3\pi/2$. 
For example, in the set-up considered in \cite{Feruglio:2012cw} 
(see also \cite{Ding:2013hpa}),  
based on $G_f \rtimes H_{\rm CP} = S_4 \rtimes H_{\rm CP}$ 
broken to   
$G_e = Z_3^T$ and $G_{\nu} = Z_2^S \times H^\nu_{\rm CP}$ with 
$H^\nu_{\rm CP} = \{U,SU\}$~%
\footnote{$S$, $T$ and $U$ are the generators of $S_4$ 
in the basis 
for its 3-dimensional representation 
we employ in this work (see subsection~\ref{subsec:formsPMNS}).}, 
the authors find:
\begin{align}
\sin^2\th_{13} = \frac{2}{3}\sin^2\th^\nu\,,\quad 
&\sin^2\th_{12} = \frac{1}{2 + \cos2\th^\nu} = \frac{1}{3\left(1 - \sin^2\th_{13}\right)}\,,\quad
\sin^2\th_{23} = \frac{1}{2}\,,\label{Fer2012}\\[2mm]
&|\sin\delta| = 1\,,\quad
\sin\alpha_{21} = \sin\alpha_{31} = 0\,.
\end{align}
%
It follows, in particular, from the results on the neutrino 
oscillation parameters~---~best fit values, $2\sigma$ and $3\sigma$ 
allowed ranges~---~%
obtained in the latest global fit of neutrino oscillation 
data \cite{Capozzi:2017ipn} and summarised in Table~\ref{tab:parameters}, 
to be used in our further analysis~%
\footnote{The results on the neutrino oscillation parameters 
obtained in the global fit performed in \cite{Esteban:2016qun} 
differ somewhat from, but are compatible at
$1\sigma$ confidence level (C.L.) with,  
those found in \cite{Capozzi:2017ipn} and
given in Table~\ref{tab:parameters}.}, 
that the predictions quoted in eq.~\eqref{Fer2012} 
for $\sin^2\theta_{12} $ and $\sin^2\theta_{23}$ lie 
outside of their respective currently allowed $2\sigma$ ranges~%
\footnote{We have used the best fit value of 
$\sin^2\theta_{13}$ to obtain the prediction of 
$\sin^2\theta_{12}$ leading to the quoted conclusion.
Using the  $2\sigma$ allowed range for $\sin^2\theta_{13}$ 
leads to a minimal value of $\sin^2\theta_{12} = 0.340$,  
which is above the maximal allowed value of 
$\sin^2\theta_{12}$ at $2\sigma$ C.L., but inside its $3\sigma$ range.}.
\begin{table}
\centering
\renewcommand{\arraystretch}{1.2}
\begin{tabular}{lccc} 
\toprule
Parameter & Best fit value & $2\sigma$ range & $3\sigma$ range \\ 
\midrule
$\sin^2\theta_{12}/10^{-1}$ & $2.97$ & $2.65 - 3.34$ & $2.50 - 3.54$ \\
$\sin^2\theta_{13}/10^{-2}$~(NO) & $2.15$ & $1.99 - 2.31$ & $1.90 - 2.40$\\
$\sin^2\theta_{13}/10^{-2}$~(IO) & $2.16$ & $1.98 - 2.33$ & $1.90 - 2.42$\\
$\sin^2\theta_{23}/10^{-1}$~(NO) & $4.25$ & $3.95 - 4.70$ & $3.81 - 6.15$\\
$\sin^2\theta_{23}/10^{-1}$~(IO) & $5.89$ & $3.99 - 4.83 \oplus 5.33 - 6.21$ & $3.84 - 6.36$\\
$\delta/\pi$~(NO) & $1.38$ & $1.00 - 1.90$ & $0 - 0.17 \oplus 0.76 - 2$\\
$\delta/\pi$~(IO) & $1.31$ & $0.92 - 1.88$ & $0 - 0.15 \oplus 0.69 - 2$\\
\midrule
$\Delta m_{21}^{2}/10^{-5}$~eV$^2$ & $7.37$ & $7.07 - 7.73$ & $6.93 - 7.96$\\
$\Delta m_{31}^{2}/10^{-3}$~eV$^2$~(NO) & $2.56$ & $2.49 - 2.64$ & $2.45 - 2.69$\\
$\Delta m_{23}^{2}/10^{-3}$~eV$^2$~(IO) & $2.54$ & $2.47 - 2.62$ & $2.42 - 2.66$\\
\bottomrule
\end{tabular}
\caption{The best fit values, $2\sigma$ and 3$\sigma$ ranges of the 
neutrino oscillation parameters obtained in the global 
analysis of the neutrino oscillation data 
performed in~\cite{Capozzi:2017ipn}. 
}
\label{tab:parameters}
\end{table}
%

 Another example of one-parametric models is the extensive study 
performed in \cite{Yao:2016zev}, in which the authors have considered 
two different residual symmetry patterns. The first pattern is the one described above, 
and the second pattern has 
$G_e = Z_2 \times H^e_{\rm CP}$ and $G_\nu = Z_2 \times Z_2 \times H^\nu_{\rm CP}$ 
as residual symmetries in the charged lepton and neutrino sectors, respectively.
The authors have performed an exhaustive scan over discrete groups of order less than 
2000, which admit faithful 3-dimensional irreducible representations, and classified 
phenomenologically viable mixing patterns.

 Theoretical models based on the approach to 
neutrino mixing that combines discrete 
symmetries and GCP invariance, 
in which the neutrino mixing angles and the leptonic 
CPV phases are functions of two or three parameters have also 
been considered in the literature 
(see, e.g., \cite{Girardi:2013sza,Turner:2015uta,Girardi:2016zwz,Lu:2016jit}).
In these models 
the residual symmetry $G_e$ of the charged lepton mass term 
is typically assumed to be a $Z_2$ symmetry or to be fully broken.
In spite of the larger number of parameters in terms of which 
the neutrino mixing angles and the leptonic CPV phases are 
expressed, the values of the CPV phases are still predicted 
to be correlated with the values of the    
three neutrino mixing angles.
A set-up with $G_e = Z_2 \times H^e_{\rm CP}$ and 
$G_\nu = Z_2 \times H^\nu_{\rm CP}$ has been 
considered in \cite{Lu:2016jit}. 
The resulting PMNS matrix in such a scheme depends on two 
free real parameters~---~two angles $\th^\nu$ and $\th^e$.
The authors have obtained  several 
phenomenologically viable neutrino mixing patterns 
from $G_f = S_4$ combined with $H_{\rm CP}$, 
broken to all possible residual symmetries of the type indicated above.
Models allowing for three free parameters 
have been investigated  in \cite{Girardi:2013sza,Turner:2015uta,Girardi:2016zwz}.
In, e.g., \cite{Turner:2015uta},
the author has considered $G_f = A_5$ combined with $H_{\rm CP}$, 
which are broken to $G_e = Z_2$ and $G_\nu = Z_2 \times H^\nu_{\rm CP}$.
In this case, the matrix $U_e$ depends on an angle $\th^e$ and 
a phase $\delta^e$, 
while the matrix $U_\nu$ depends on an angle $\th^\nu$.
In these two scenarios the leptonic CPV phases
possess  non-trivial values. 

  The specific correlations between the values of the three 
neutrino mixing angles, which characterise the one-parameter 
models based on $G_e = Z_n$, $n > 2$, or $Z_m \times Z_k$, $m,\,k \geq 2$,  
and $G_\nu = Z_2 \times H^\nu_{\rm CP}$,
do not hold in the two- and three-parameter models.
In addition, the Dirac CPV phase in the two- and three-parameter 
models is predicted to have non-trivial values which are 
correlated with the values of the three neutrino mixing angles 
and differ from  $0$, $\pi$, $\pi/2$ and $3\pi/2$, 
although the deviations from, e.g., $3\pi/2$  can be relatively small.
The indicated differences between the predictions 
of the models based on 
$G_e = Z_n$, $n > 2$, or $Z_m \times Z_k$, $m,\,k \geq 2$,  
and on $G_e = Z_2$ symmetries make it possible 
to distinguish between them experimentally 
by improving the precision on each of the three measured neutrino 
mixing angles $\theta_{12}$, $\theta_{23}$ and  $\theta_{13}$, 
and by performing a sufficiently precise measurement of the 
Dirac phase $\delta$.

 In the present article, we investigate the possible 
neutrino mixing patterns generated by a    
 $G_f = S_4$ symmetry combined with an $H_{\rm CP}$ symmetry 
when these symmetries are broken down 
to $G_e = Z_2$ and $G_\nu = Z_2 \times H^\nu_{\rm CP}$. 
In Section~\ref{sec:framework}, we describe a general framework 
for deriving the form of the PMNS matrix, dictated by the chosen 
residual symmetries. Then, in Section~\ref{sec:S4}, 
we apply this framework to  
$G_f = S_4$ combined with $H_{\rm CP}$ and obtain all 
phenomenologically viable mixing patterns. 
Next, in Section~\ref{sec:bb0nu}, using the obtained predictions
for the neutrino mixing angles and the Dirac and Majorana CPV phases, 
we derive predictions for the neutrinoless double beta decay 
effective Majorana mass. 
Section~\ref{sec:conclusions} contains the conclusions of the present 
study.

\section{The Framework}
\label{sec:framework}

 We start with a non-Abelian flavour symmetry group $G_f$, 
which admits a faithful irreducible 3-dimensional representation $\rho$. 
The three generations of left-handed (LH) leptons are assigned to 
this representation. 
Apart from that, the high-energy theory respects also 
the GCP symmetry $H_{\rm CP}$, which is implemented consistently 
along  with the flavour symmetry. 
At some flavour symmetry breaking scale 
$G_f \rtimes H_{\rm CP}$ gets broken down to residual symmetries 
$G_e$ and $G_\nu$ 
of the charged lepton and neutrino mass terms, respectively. 
The residual flavour symmetries are Abelian subgroups of $G_f$.
The symmetries $G_e$ and $G_\nu$ significantly constrain the form of 
the neutrino mixing matrix $U_{\rm PMNS}$, as we demonstrate below.

\subsection{The PMNS Matrix from $\bo{G_e = Z_2}$ and 
$\bo{G_\nu = Z_2 \times H^\nu_{\rm CP}}$}
\label{subsec:PMNS}
 
 We choose $G_e$ to be a $Z_2$ symmetry. We will denote it 
as $Z_2^{g_e} \equiv \{1,g_e\}$, $g_e^2 = 1$ being an element of $G_f$ 
of order two, generating the $Z_2^{g_e}$ subgroup. 
The invariance of the charged lepton mass term under $G_e$ implies 
\be
\rho(g_e)^\dagger M_e M_e^\dagger\, \rho(g_e) = M_e M_e^\dagger\,.
\label{eq:Ge}
\ee
Below we show how this invariance constrains the form 
of the unitary matrix $U_e$, diagonalising $M_e M_e^\dagger$:
\be
U_e^\dagger M_e M_e^\dagger\, U_e = \diag(m_e^2, m_\mu^2, m_\tau^2)\,.
\label{eq:mediag}
\ee

 Lets $\O_e$ be a diagonalising unitary matrix of $\rho(g_e)$, such that 
\be
\O_e^\dagger\, \rho(g_e)\, \O_e = \rho(g_e)^\di \equiv \diag(1,-1,-1)\,.
\label{eq:Oe}
\ee
%
This result is obtained as follows. 
The diagonal entries of $\rho(g_e)^{\rm d}$
are constrained to be $\pm 1$,
since this matrix must still furnish
a representation of $Z_2$
and hence its square is the identity.
We have assumed that
the trace of $\rho(g_e)$ is $-1$,
for the relevant elements $g_e$,
as it is the case for the
3-dimensional representation of $S_4$
we will consider later on~%
\footnote{
For the other 3-dimensional irreducible
representation of $S_4$
the trace can be either $-1$ or
$+1$, depending on $g_e$. 
Choosing $+1$ would simply imply a change of
sign of $\rho(g_e)^{\rm d}$,
which however does not lead to new constraints.
The conclusions we reach in what follows
are then independent of the choice
of 3-dimensional representation.
}.
Note that we can take the order of the eigenvalues of $\rho(g_e)$ 
as given in  eq.~\eqref{eq:Oe} 
without loss of generality, as will become clear later. 

Expressing $\rho(g_e)$ from eq.~\eqref{eq:Oe} and 
substituting it in eq.~\eqref{eq:Ge}, we obtain 
\be
\rho(g_e)^\di\, \O_e^\dagger\, M_e M_e^\dagger\, \O_e\, \rho(g_e)^\di = \O_e^\dagger\, M_e M_e^\dagger\, \O_e\,.
\ee
%
This equation implies that $\O_e^\dagger\, M_e M_e^\dagger\, \O_e$ has the 
block-diagonal form
\be
\begin{pmatrix}
\times & 0 & 0 \\
0 & \times & \times \\
0 & \times & \times
\end{pmatrix}\,,
\label{eq:blockdiag23}
\ee
%
and, since this matrix is hermitian, 
it can be diagonalised by a unitary matrix with 
a $U(2)$ transformation acting on the 
2-3 block. In the general case, 
the $U(2)$ transformation can be parametrised 
as follows:
\be
\begin{pmatrix}
\cos\th^e & -\sin\th^e\, e^{-i\delta^e} \\
\sin\th^e\, e^{i\delta^e} & \cos\th^e 
\end{pmatrix}\,
\begin{pmatrix}
e^{i\beta_1^e} & 0 \\
0 & e^{i\beta_2^e}
\end{pmatrix}\,.
\ee
%
The diagonal phase matrix is, however, unphysical, 
since it can be eliminated by rephasing of the charged lepton fields, 
and we will not keep it in the future.
Thus, we arrive to the conclusion that the matrix $U_e$ 
diagonalising $M_e M_e^\dagger$ reads
\be
U_e = \O_e\, U_{23}(\th^e,\delta^e)^\dagger\, P_e^T\,,
\label{eq:Ue}
\ee
%
with
\be
U_{23}(\th^e,\delta^e) = 
\begin{pmatrix}
1 & 0 & 0 \\
0 &\cos\th^e & \sin\th^e\, e^{-i\delta^e} \\
0 & - \sin\th^e\, e^{i\delta^e} & \cos\th^e
\end{pmatrix}\,,
\label{eq:U23}
\ee
%
and $P_e$ being one of six permutation matrices, 
which need to be taken into account, 
since in the approach under consideration the order 
of the charged lepton masses 
is unknown. The six permutation matrices read:
\begin{align}
& P_{123} = 
\begin{pmatrix}
1 & 0 & 0 \\
0 & 1 & 0 \\
0 & 0 & 1
\end{pmatrix}\,, \quad
P_{132} = 
\begin{pmatrix}
1 & 0 & 0 \\
0 & 0 & 1 \\
0 & 1 & 0
\end{pmatrix}\,, \quad
P_{213} = 
\begin{pmatrix}
0 & 1 & 0 \\
1 & 0 & 0 \\
0 & 0 & 1
\end{pmatrix}\,, \\
& P_{231} = 
\begin{pmatrix}
0 & 1 & 0 \\
0 & 0 & 1 \\
1 & 0 & 0
\end{pmatrix}\,, \quad
P_{312} = 
\begin{pmatrix}
0 & 0 & 1 \\
1 & 0 & 0 \\
0 & 1 & 0
\end{pmatrix}\,, \quad
P_{321} = 
\begin{pmatrix}
0 & 0 & 1 \\
0 & 1 & 0 \\
1 & 0 & 0
\end{pmatrix}\,.
\end{align}
%
Note that the order of indices in $P_{ijk}$ stands 
for the order of rows, i.e., 
when applied from the left to a matrix, it gives  
the desired order, $i$-$j$-$k$, of 
the matrix rows. The same is also true for columns, 
when $P_{ijk}$ is applied from 
the right, except for $P_{231}$ which leads 
to the $3$-$1$-$2$ order of columns 
and $P_{312}$ yielding the $2$-$3$-$1$ order.

 In the neutrino sector we have a $G_\nu = Z_2 \times H^\nu_{\rm CP}$ 
residual symmetry. We will denote the $Z_2$ symmetry of 
the neutrino mass matrix as 
$Z_2^{g_\nu} \equiv \{1,g_\nu\}$, with $g_\nu^2 = 1$ being an element of $G_f$, 
generating the $Z_2^{g_\nu}$ subgroup. 
$H^\nu_{\rm CP} = \{X_\nu\}$ is the set of remnant GCP unitary
transformations $X_\nu$
forming a residual CP symmetry of the neutrino mass matrix.
$H^\nu_{\rm CP}$ is contained
in $H_{\rm CP} = \{X\}$ which is the GCP symmetry 
of the high-energy theory consistently
defined along with the flavour symmetry $G_f$~%
\footnote{It is worth to comment here on the notation $H^\nu_{\rm CP}$ we use. 
When we write in what follows $H^\nu_{\rm CP} = \{X_{\nu1}, X_{\nu2}\}$, we mean a set of GCP transformations ($X_{\nu1}$ and $X_{\nu2}$) compatible with the residual 
flavour $Z_2^{g_\nu}$ symmetry (see eq.~\eqref{eq:consistency}). 
However, when writing $G_\nu = Z_2^{g_\nu} \times H^\nu_{\rm CP}$, 
$H^\nu_{\rm CP}$ is intended to be a group generated by $X_{\nu1}$. 
Namely, following Appendix~B in \cite{Feruglio:2012cw}, 
$H^\nu_{\rm CP}$ is isomorphic to $\{\mathcal{I},\mathcal{X}_{\nu1}\}$, 
where $\mathcal{I}$ is the unit matrix and 
$$\mathcal{X}_{\nu1} = 
\begin{pmatrix}
0 & X_{\nu1} \\
X_{\nu1}^* & 0
\end{pmatrix}\,,
$$
both of them acting on $(\varphi,\varphi^*)^T$.
Then, $Z_2^{g_\nu}$ is isomorphic to $\{\mathcal{I},\mathcal{G}_\nu\}$, where
$$\mathcal{G}_{\nu} = 
\begin{pmatrix}
\rho(g_\nu) & 0 \\
0 & \rho^*(g_\nu)
\end{pmatrix}
$$
acts again on $(\varphi,\varphi^*)^T$. 
Finally, it is not difficult to convince oneself that the full residual symmetry group 
$G_\nu$ is given by a direct product $Z_2^{g_\nu} \times H^\nu_{\rm CP}$, and the 
second GCP transformation $X_{\nu2} = \rho(g_\nu)\, X_{\nu1}$ is contained in it.
The same logic applies to the notation $H_{\rm CP}$, and, as has been shown in 
Appendix~B of \cite{Feruglio:2012cw}, the full symmetry group is a semi-direct product 
$G_f \rtimes H_{\rm CP}$. 
Note that these notations are widely used in the literature.}.
The invariance under $G_\nu$ of the neutrino mass matrix implies that 
the following two equations hold: 
\begin{align}
& \rho(g_\nu)^T M_\nu\, \rho(g_\nu) = M_\nu\,,
\label{eq:Gnu} \\
& X_\nu^T M_\nu\, X_\nu = M_\nu^*\,.
\label{eq:GCP}
\end{align}
%
In addition, the consistency condition between 
$Z_2^{g_\nu}$ and $H^\nu_{\rm CP}$ 
has to be respected:
\be
X_\nu\, \rho^*(g_\nu)\, X_\nu^{-1} = \rho(g_\nu)\,.
\label{eq:consistency}
\ee
%
To derive the form of the unitary matrix $U_\nu$ 
diagonalising the neutrino Majorana mass matrix $M_\nu$ as
\be
U_\nu^T M_\nu\, U_\nu = \diag(m_1,m_2,m_3)\,,
\label{eq:mnudiag}
\ee
%
$m_j > 0$ being the neutrino masses, 
we will follow the method presented in \cite{Lu:2016jit}.

 Lets $\O_{\nu1}$ be a diagonalising unitary matrix 
of $\rho(g_\nu)$, such that
\be
\O_{\nu1}^\dagger\, \rho(g_\nu)\, \O_{\nu1} = \rho(g_\nu)^\di \equiv \diag(1,-1,-1)\,.
\label{eq:Onu1}
\ee
%
Expressing $\rho(g_\nu)$ from this equation and substituting it in 
the consistency condition, eq.~\eqref{eq:consistency}, we find
\be
\rho(g_\nu)^\di\, \O_{\nu1}^\dagger\, X_\nu\, \O_{\nu1}^*\, \rho(g_\nu)^\di = \O_{\nu1}^\dagger\, X_\nu\, \O_{\nu1}^*\,,
\label{eq:Onu1X}
\ee
%
meaning that $\O_{\nu1}^\dagger\, X_\nu\, \O_{\nu1}^*$
is a block-diagonal matrix, having the form of eq.~\eqref{eq:blockdiag23}.
Moreover, this matrix is symmetric, since the GCP transformations $X_\nu$ 
have to be symmetric in order for all the three neutrino 
masses to be different \cite{Feruglio:2012cw,Ding:2013hpa}, 
as is required by the data. 
In Appendix~\ref{app:symmetryofX} we provide a proof of this.
Being a complex (unitary) symmetric matrix, it is diagonalised 
by a unitary matrix $\O_{\nu2}$ via the transformation:
\be
\O^\dagger_{\nu2}\, (\O_{\nu1}^\dagger\, X_\nu\, \O_{\nu1}^*)\, \O_{\nu2}^* = 
(\O_{\nu1}^\dagger\, X_\nu\, \O_{\nu1}^*)^\di\,.
\label{Onu2}
\ee
%
The matrix $(\O_{\nu1}^\dagger\, X_\nu\, \O_{\nu1}^*)^\di$ 
is, in general, a diagonal phase matrix. 
However, we can choose 
$(\O_{\nu1}^\dagger\, X_\nu\, \O_{\nu1}^*)^\di = {\rm diag}(1,1,1)$
as the phases of $(\O_{\nu1}^\dagger\, X_\nu\, \O_{\nu1}^*)^\di$ 
can be moved to the matrix $\O_{\nu 2}$.
With this choice we obtain the Takagi factorisation of 
the $X_\nu$ (valid for unitary symmetric matrices):
\be
X_\nu = \O_\nu\, \O_\nu^T\,,
\label{eq:XOnu}
\ee
%
with $\O_\nu = \O_{\nu1}\, \O_{\nu2}$.

Since, as we have noticed earlier, 
$\O_{\nu1}^\dagger\, X_\nu\, \O_{\nu1}^*$ has the form of 
eq.~\eqref{eq:blockdiag23}, the matrix $\O_{\nu2}$
can be chosen without loss of generality to
have the form of eq. \eqref{eq:blockdiag23}
with a unitary $2\times 2$ matrix in the 2-3 block.
This implies that the matrix $\O_\nu = \O_{\nu1}\, \O_{\nu2}$ 
also diagonalises  $\rho(g_\nu)$. Indeed, 
\be
\O_{\nu}^\dagger\, \rho(g_\nu)\, \O_{\nu} = 
 \O^\dagger_{\nu2}\, \rho(g_\nu)^\di\, \O_{\nu2} 
= \rho(g_\nu)^\di\,,
\label{eq:Onu}
\ee
%
where we have used eq. (\ref{eq:Onu1}).

 We substitute next $X_\nu$ from eq.~\eqref{eq:XOnu} in the GCP invariance 
condition of the neutrino mass matrix, eq.~\eqref{eq:GCP}, and find that 
the matrix $\O_\nu^T\, M_\nu\, \O_\nu$ is real.
Furthermore, this is a symmetric matrix, since the neutrino Majorana 
mass matrix $M_\nu$ is symmetric. A real symmetric matrix can 
be diagonalised by a real orthogonal transformation. 
Employing eqs.~\eqref{eq:Onu} and \eqref{eq:Gnu},  we have
\be
\rho(g_\nu)^\di\, \O_\nu^T\, M_\nu\, \O_\nu\, \rho(g_\nu)^\di = 
\O_\nu^T\, M_\nu\, \O_\nu\,,
\ee
%
implying that $\O_\nu^T\, M_\nu\, \O_\nu$ is a block-diagonal matrix 
as in eq.~\eqref{eq:blockdiag23}.
Thus, the required orthogonal transformation is a rotation 
in the 2-3 plane on an angle $\th^\nu$: 
\be
R_{23}(\th^\nu) = 
\begin{pmatrix}
1 & 0 & 0 \\
0 &\cos\th^\nu & \sin\th^\nu \\
0 & - \sin\th^\nu & \cos\th^\nu
\end{pmatrix}\,.
\label{eq:R23}
\ee
%
Finally, the matrix $U_\nu$ diagonalising $M_\nu$ reads 
\be
U_\nu = \O_\nu\, R_{23}(\th^\nu)\, P_\nu\, Q_\nu\,,
\label{eq:Unu}
\ee
%
where $P_\nu$ is one of the six permutation matrices, 
which accounts for different order of $m_j$, and 
the matrix $Q_\nu$ renders them positive. 
Without loss of generality $Q_\nu$ can be parametrised as follows: 
\be
Q_\nu = \diag(1, i^{k_1}, i^{k_2})\,, \quad
{\rm with} \quad k_{1,2} = 0, 1\,.
\label{eq:Qnu}
\ee

 Assembling together the results for $U_e$ and $U_\nu$, 
eqs.~\eqref{eq:Ue} and \eqref{eq:Unu}, we obtain 
for the form of the PMNS matrix:
\be
U_{\rm PMNS} =  
P_e\, U_{23}(\th^e,\delta^e)\, \O_e^\dagger\, 
\O_\nu\, R_{23}(\th^\nu)\, P_\nu\, Q_\nu\,.
\label{eq:UPMNS}
\ee
%

Thus, in the approach we are following 
the PMNS matrix depends on three free real 
parameters~%
\footnote{It should be noted that the matrix $\O_{\nu 2}$ in eq.~\eqref{Onu2} with
$(\O_{\nu1}^\dagger\, X_\nu\, \O_{\nu1}^*)^\di = {\rm diag}(1,1,1)$, and thus
the matrix $\O_\nu =\O_{\nu 1}\, \O_{\nu 2}$ in eq. (\ref{eq:XOnu}),
is determined up to a multiplication by an orthogonal matrix $O$ on the right. 
The matrix $\O_{\nu 2}\, O$ must be unitary since it 
diagonalises a complex symmetric matrix, 
which implies that $O$ must be unitary in addition of being orthogonal, 
and therefore must be a real matrix.
Equation~\eqref{eq:Onu} restricts further this real orthogonal matrix $O$ 
to have the form of a real rotation in the 2-3 plane, which can be
``absorbed'' in the $R_{23}(\th^\nu)$ matrix in eq.~\eqref{eq:UPMNS}.}~%
---~the two angles $\th^e$ and $\th^\nu$ 
and the phase $\delta^e$. One of the elements of the PMNS matrix 
is fixed to be a constant by the employed residual symmetries.
We note finally that, since 
$R_{23}(\th^\nu + \pi) = R_{23}(\th^\nu)\, \diag(1,-1,-1)$, 
where the diagonal matrix can be absorbed into $Q_\nu$, and 
$U_{23}(\th^e + \pi,\delta^e) = \diag(1,-1,-1)\, U_{23}(\th^e,\delta^e)$, 
where the diagonal matrix contributes to the unphysical charged lepton phases, 
it is sufficient to consider $\th^e$ and $\th^\nu$ 
in the interval $[0,\pi)$.

\subsection{Conjugate Residual Symmetries}
\label{subsec:conjsym}

 In this subsection we briefly recall why the residual symmetries 
$G_e'$ and $G_\nu'$ conjugate to $G_e$ and $G_\nu$, respectively, under the 
same element of the flavour symmetry group $G_f$ lead to the same PMNS matrix 
(see, e.g., \cite{Feruglio:2012cw,Ding:2013bpa}).
Two pairs of residual symmetries 
$\{Z_2^{g_e}, Z_2^{g_\nu}\}$ and $\{Z_2^{g_e'}, Z_2^{g_\nu'}\}$ 
are conjugate to each other under $h \in G_f$ if
\be
h\, g_e \,h^{-1} = g_e' 
\quad 
{\rm and}
\quad
h\, g_\nu \,h^{-1} = g_\nu'\,.
\label{eq:similarity}
\ee
At the representation level this means
\be
\rho(h)\, \rho(g_e) \,\rho(h)^\dagger = \rho(g_e' )
\quad 
{\rm and}
\quad
\rho(h)\, \rho(g_\nu) \,\rho(h)^\dagger = \rho(g_\nu')\,.
\ee
Substituting $\rho(g_e)$ and $\rho(g_\nu)$ from these equalities 
to eqs.~\eqref{eq:Ge} and \eqref{eq:Gnu}, respectively, we obtain
\be
\rho(g_e')^\dagger M_e' M_e'^\dagger\, \rho(g_e') = M_e' M_e'^\dagger
\quad
{\rm and}
\quad
\rho(g_\nu')^T M_\nu'\, \rho(g_\nu') = M_\nu'\,,
\ee
where the primed mass matrices are related to the original ones as
\be
M_e' M_e'^\dagger = \rho(h)\, M_e M_e^\dagger\, \rho(h)^\dagger
\quad
{\rm and}
\quad
M_\nu' = \rho(h)^*\, M_\nu\, \rho(h)^\dagger\,.
\ee
As can be understood from eq.~\eqref{eq:GCP} (or eq.~\eqref{eq:consistency}), 
the matrix $M_\nu'$ will respect a remnant CP symmetry $H^{\nu'}_{\rm CP} = \{X_\nu'\}$, 
which is related to $H^\nu_{\rm CP} = \{X_\nu\}$ as follows:
\be
X_\nu' = \rho(h)\, X_\nu\, \rho(h)^T\,.
\ee
Obviously, the unitary transformations $U_e'$ and $U_\nu'$ 
diagonalising the primed mass matrices are given by 
\be
U_e' = \rho(h)\,U_e
\quad
{\rm and}
\quad
U_\nu' = \rho(h)\,U_\nu\,,
\ee
thus yielding 
\be
U_{\rm PMNS}' = U_e'^\dagger\, U_\nu' = U_e^\dagger\, U_\nu = U_{\rm PMNS}\,.
\ee

\subsection{Phenomenologically Non-Viable Cases}
\label{subsec:nonviable}

 Here we demonstrate that at least two types of residual 
symmetries $\{G_e, G_\nu\} = \{Z_2^{g_e}, Z_2^{g_\nu} \times H^\nu_{\rm CP}\}$, 
characterised by certain $g_e$ and $g_\nu$, 
cannot lead to phenomenologically viable form of the PMNS matrix. \\

 $\bullet$ \textbf{Type I:~~$\bo{g_e = g_\nu}$.} 
In this case, 
we can choose $\O_e = \O_\nu\, P$, 
with $P_{123}$ or $P_{132}$. 
Then, eq.~\eqref{eq:UPMNS} yields
\be
U_{\rm PMNS} =  
P_e\, U_{23}(\th^e,\delta^e)\, P\, 
R_{23}(\th^\nu)\, P_\nu\, Q_\nu\,.
\label{eq:UPMNSge=gnu}
\ee
This means that up to permutations of the rows and columns $U_{\rm PMNS}$ has 
the form of eq.~\eqref{eq:blockdiag23}, i.e., contains four zero entries,  
which are ruled out by neutrino oscillation data \cite{Capozzi:2017ipn,Esteban:2016qun}. \\

 $\bullet$ \textbf{Type II:~~$\bo{g_e,\,g_\nu \in Z_2 \times Z_2 \subset G_f}$.} 
Now we consider two different order two elements $g_e \neq g_\nu$, which 
belong to the same $Z_2 \times Z_2 = \{1,\, g_e,\, g_\nu,\, g_e\,g_\nu\}$ subgroup of $G_f$. 
In this case, since $g_e$ and $g_\nu$ commute, 
there exists a unitary matrix simultaneously diagonalising both 
$\rho(g_e)$ and $\rho(g_\nu)$. Note, however, that the order 
of eigenvalues in the resulting diagonal matrices will be different. 
Namely, lets $\O_{\nu1}$ be a diagonalising matrix of 
$\rho(g_\nu)$ and $\rho(g_e)$, and 
lets $\O_{\nu1}$ diagonalise $\rho(g_\nu)$ as in eq.~\eqref{eq:Onu1}. 
Then, $\O_{\nu1}^\dagger\, \rho(g_e)\, \O_{\nu1}$ can yield either 
$\diag(-1,1,-1)$ or $\diag(-1,-1,1)$, but not $\diag(1,-1,-1)$.
Hence, $\O_e$ diagonalising $\rho(g_e)$ as 
in eq.~\eqref{eq:Oe}, must read 
\begin{align}
\O_e = \O_{\nu1}\,P\,, \quad {\rm with} \quad 
& P = P_{213}~{\rm or}~P_{312} \quad {\rm if} \quad 
\O_{\nu1}^\dagger\, \rho(g_e)\, \O_{\nu1} = \diag(-1,1,-1)\,, \\
{\rm and} \quad & P = P_{231}~{\rm or}~P_{321} \quad {\rm if} \quad 
\O_{\nu1}^\dagger\, \rho(g_e)\, \O_{\nu1} = \diag(-1,-1,1)\,.
\end{align}
Taking into account that $\O_\nu = \O_{\nu1}\, \O_{\nu2}$, 
with $\O_{\nu2}$ of the block-diagonal form given in eq.~\eqref{eq:blockdiag23}, 
we obtain
\be
U_{\rm PMNS} =  
P_e\, U_{23}(\th^e,\delta^e)\, P^T\, \O_{\nu2}\, 
R_{23}(\th^\nu)\, P_\nu\, Q_\nu\,,
\label{eq:UPMNSgegnuK4}
\ee
where $P^T\, \O_{\nu2}$, depending on $P$, can take one of the following forms:
\be
\begin{pmatrix}
0 & \times & \times \\
\times & 0 & 0 \\
0 & \times & \times
\end{pmatrix}
\quad {\rm or} \quad
\begin{pmatrix}
0 & \times & \times \\
0 & \times & \times \\
\times & 0 & 0
\end{pmatrix}\,.
\ee
As a consequence, $U_{\rm PMNS}$ up to permutations of the rows and columns
has the form 
\be
\begin{pmatrix}
0 & \times & \times \\
\times & \times & \times \\
\times & \times & \times
\end{pmatrix}\,,
\ee
containing one zero element, which is ruled out by the data.

\section{Mixing Patterns from $\bo{G_f \rtimes H_{\rm CP} = S_4 \rtimes H_{\rm CP}}$ 
Broken to \\
$\bo{G_e = Z_2}$ and $\bo{G_\nu = Z_2 \times H^\nu_{\rm CP}}$}
\label{sec:S4}

\subsection{Group $S_4$ and Residual Symmetries}
\label{subsec:S4}

 $S_4$ is the symmetric group of permutations of four objects. 
This group is isomorphic to the group of rotational symmetries of the cube. 
$S_4$ can be defined in terms of three generators $S$, $T$ and $U$, satisfying \cite{Hagedorn:2010th}
\be
S^2 = T^3 = U^2 = (ST)^3 = (SU)^2 = (TU)^2 = (STU)^4 = 1\,.
\label{eq:S4gener}
\ee
From 24 elements of the group there are nine elements of order two, 
which belong to two of five conjugacy classes of $S_4$ (see, e.g., \cite{Ding:2013hpa}):
\begin{align}
& 3\,\mathcal{C}_2: \{S\,,~TST^2\,,~T^2ST\}\,, 
\label{eq:C2}\\
& 6\,\mathcal{C}_2': \{U\,,~TU\,,~SU\,,~T^2U\,,~STSU\,,~ST^2SU\}\,.
\label{eq:C2prime}
\end{align}
Each of these nine elements generates a corresponding $Z_2$ subgroup of $S_4$. 
Each subgroup can be the residual symmetry of $M_e M_e^\dagger$, and,
combined with compatible CP transformations, yield 
the residual symmetry of $M_\nu$. 
Hence, we have 81 possible pairs of only residual flavour symmetries 
(taking into account remnant CP symmetries increases the number of possibilities). 
Many of them, however, being conjugate to each other, will lead to  
the same form of the PMNS matrix, as explained in subsection~\ref{subsec:conjsym}.
Thus, we first identify the pairs of elements $\{g_e,g_\nu\}$, which are not related 
by the similarity transformation given in eq.~\eqref{eq:similarity}.
We find nine distinct cases for which $\{g_e,g_\nu\}$ can be chosen as
\begin{align}
& \{S,S\}\,,\quad\{U,U\}\,,\quad\{T^2ST,S\}\,,\quad\{S,U\}\,,\quad\{U,S\}\,,\quad\{SU,U\}\,, 
\label{eq:nonviable} \\
& \{S,TU\}\,,\quad\{TU,S\}\,,\quad\{TU,U\}\,. 
\label{eq:viable}
\end{align}
The pair $\{S,S\}$ is obviously conjugate to $\{TST^2,TST^2\}$ and $\{T^2ST,T^2ST\}$, while 
$\{U,U\}$ is conjugate to $\{g_e,g_\nu\}$ with $g_e = g_\nu$ being one of the remaining five 
elements from conjugacy class $6\,\mathcal{C}_2'$ given in eq.~\eqref{eq:C2prime}. 
The pairs $\{T^2ST,S\}$, $\{S,U\}$, $\{U,S\}$ and $\{SU,U\}$ are conjugate to five pairs each, 
and $\{S,TU\}$ and $\{TU,S\}$ to eleven pairs each.
Finally, $\{TU,U\}$ is conjugate to 23 pairs. 
As it should be, the total number of pairs yields 81. 
The complete lists of pairs of elements which are conjugate to each of these nine pairs are given in Appendix~\ref{app:conjugatepairs}.

 The cases in eq.~\eqref{eq:nonviable} do not lead to phenomenologically 
viable results. The first two of them belong to the cases of 
Type I (see subsection~\ref{subsec:nonviable}).
The remaining four belong to Type II, since $S_4$ contains 
$Z_2^S \times Z_2^{TST^2} = \{1,\,S,\,TST^2\,,T^2ST\}$ and 
$Z_2^S \times Z_2^{U} = \{1,\,S,\,U\,,SU\}$ subgroups 
(see, e.g., \cite{Li:2014eia}).
Thus, we are left with three cases in eq.~\eqref{eq:viable}.

 We have chosen $g_\nu$ in such a way that it is $S$, $U$ or $TU$ for 
all the cases in eq.~\eqref{eq:viable}. Now we need to identify the remnant CP 
transformations $X_\nu$ compatible with each of these three elements.
It is known that the GCP symmetry $H_{\rm CP} = \{X\}$ compatible with $G_f = S_4$ is 
of the same form of $G_f$ itself \cite{Holthausen:2012dk}, i.e., 
\be
X = \rho(g), \quad g \in S_4\,.
\ee
Thus, to find $X_\nu$ compatible with $g_\nu$ of interest, 
we need to select those $X = \rho(g)$, which 
i) satisfy the consistency condition in eq.~\eqref{eq:consistency} and 
ii) are symmetric in order to avoid partially degenerate neutrino mass spectrum, 
as was noted earlier.
The result reads~%
\footnote{For notation simplicity we will not write the representation 
symbol $\rho$, keeping in mind that $X_\nu = g$ meas $X_\nu = \rho(g)$ with $g \in G_f$.}: 
\begin{align}
& X_\nu = 1\,,~(S)\,,~U\,,~(SU)\,,~TST^2U\,,~(T^2STU) \quad {\rm for} \quad g_\nu = S\,; \\
& X_\nu = 1\,,~(U)\,,~S\,,~(SU) \quad {\rm for} \quad g_\nu = U\,; \\
& X_\nu = U\,,~(T)\,,~STS\,,~(T^2STU) \quad {\rm for} \quad g_\nu = TU\,.
\end{align}
A GCP transformation in parentheses appears automatically to be a remnant CP symmetry of $M_\nu$, 
if $X_\nu$ which precedes this in the list is a remnant CP symmetry. 
This is a consequence of eqs.~\eqref{eq:Gnu} and \eqref{eq:GCP}, which imply that 
if $X_\nu$ is a residual CP symmetry of $M_\nu$, then $\rho(g_\nu) X_\nu$ is a residual CP symmetry as well.
Therefore, we have three sets of remnant CP transformations compatible with $Z_2^{S}$, namely, 
$H^\nu_{\rm CP} = \{1,S\}$, $\{U,SU\}$ and $\{TST^2U,T^2STU\}$, 
two sets compatible with $Z_2^{U}$, which are $H^\nu_{\rm CP} = \{1,U\}$ and $\{S,SU\}$, 
and two sets consistent with $Z_2^{TU}$, which read $H^\nu_{\rm CP} = \{U,T\}$ and $\{STS,T^2STU\}$.
Taking them into account, we end up with seven possible pairs of residual symmetries 
$\{G_e, G_\nu\} = \{Z_2^{g_e}, Z_2^{g_\nu} \times H^\nu_{\rm CP}\}$, with $\{g_e,g_\nu\}$ as in 
eq.~\eqref{eq:viable}.  
In what follows, we will consider them case by case and classify all phenomenologically 
viable mixing patterns they lead to. 

 Before starting, however, let us recall the current knowledge on the 
 absolute values of the PMNS matrix elements, which we will use in
 what follows. The $3\sigma$ ranges of the absolute values of
 the PMNS matrix elements read \cite{Capozzi2016}
\be
|U_{\rm PMNS}|_{3\sigma} = 
\begin{pmatrix}
0.796 \rightarrow 0.855 & 0.497 \rightarrow 0.587 & 0.140 \rightarrow 0.153 \\
0.245 \rightarrow 0.513 & 0.543 \rightarrow 0.709 & 0.614 \rightarrow 0.768 \\
0.244 \rightarrow 0.510 & 0.456  \rightarrow 0.642 & 0.624 \rightarrow 0.776
\end{pmatrix}
\label{eq:PMNS3sigmaNO}
\ee
for the neutrino mass spectrum with normal ordering (NO), and
\be
|U_{\rm PMNS}|_{3\sigma} = 
\begin{pmatrix}
0.796 \rightarrow 0.855 & 0.497 \rightarrow 0.587 & 0.140 \rightarrow 0.153 \\
0.223 \rightarrow 0.503 & 0.452 \rightarrow 0.703 & 0.614 \rightarrow 0.783 \\
0.257 \rightarrow 0.526 & 0.464 \rightarrow 0.712 & 0.605 \rightarrow 0.775
\end{pmatrix}
\label{eq:PMNS3sigmaIO}
\ee
for the neutrino mass spectrum with inverted ordering (IO).
The ranges in eqs.~\eqref{eq:PMNS3sigmaNO} and \eqref{eq:PMNS3sigmaIO} 
differ a little from the results obtained in \cite{Esteban:2016qun}.

\subsection{Explicit Forms of the PMNS Matrix}
\label{subsec:formsPMNS}

 First, we present an explicit example of constructing the PMNS matrix 
in the case of $g_e = S$, $g_\nu =TU$ and 
$H^\nu_{\rm CP} = \{U,T\}$, which is the first case out of 
the seven potentially viable cases indicated above. 
We will work in the basis for $S_4$ from \cite{King:2009mk}, in which 
the matrices for the generators $S$, $T$ and $U$ in the 3-dimensional representation 
read
\be
S = \frac{1}{3}
\begin{pmatrix}
-1 & 2 & 2 \\
2 & -1 & 2 \\
2 & 2 & -1
\end{pmatrix}\,,
\quad
T = 
\begin{pmatrix}
1 & 0 & 0 \\
0 & \o^2 & 0 \\
0 & 0 & \o
\end{pmatrix}
\quad
{\rm and}
\quad
U = -
\begin{pmatrix}
1 & 0 & 0 \\
0 & 0 & 1 \\
0 & 1 & 0
\end{pmatrix}\,,
\label{eq:generators}
\ee
where $\omega = e^{2\pi i/3}$.
For simplicity we use the same notation ($S$, $T$ and $U$) for the generators 
and their 3-dimensional representation matrices.
We will follow the procedure described in subsection~\ref{subsec:PMNS}.
The matrix $\O_e$ which diagonalises $\rho(g_e) = S$ (see eq.~\eqref{eq:Oe}) is given by 
\be
\O_e = \frac{1}{\sqrt{6}}
\begin{pmatrix}
\sqrt{2} & -\sqrt{3} & -1 \\
\sqrt{2} & 0 & 2 \\
\sqrt{2} & \sqrt{3} & -1
\end{pmatrix}\,.
\ee
The matrix $\O_\nu$, such that
$\O_\nu\, \O_\nu^T = U$ (see eq.~\eqref{eq:XOnu}), reads 
\be
\O_\nu = \frac{1}{\sqrt{2}}
\begin{pmatrix}
0 & 0 &\sqrt{2}i \\
e^{\frac{2\pi i}{3}} & - e^{\frac{i \pi}{6}} & 0 \\
e^{\frac{i \pi}{3}} & e^{-\frac{i \pi}{6}} & 0
\end{pmatrix}\,.
\ee
Using the master formula in eq.~\eqref{eq:UPMNS}, we obtain that 
up to permutations of the rows and columns $U_{\rm PMNS}$ has the 
form 
\be
\begin{pmatrix}
\frac{i}{\sqrt{2}} & \times & \times \\
\times & \times & \times \\
\times & \times & \times
\end{pmatrix}\,,
\ee
%
where ``$\times$'' entries are functions of the free parameters 
$\th^\nu$, $\th^e$ and $\delta^e$.
Taking into account the current data, eqs.~\eqref{eq:PMNS3sigmaNO} and 
\eqref{eq:PMNS3sigmaIO}, the fixed element with the absolute value of 
$1/\sqrt{2} \approx 0.707$ can be $(U_{\rm PMNS})_{\mu2}$, $(U_{\rm PMNS})_{\mu3}$, 
$(U_{\rm PMNS})_{\tau2}$ or $(U_{\rm PMNS})_{\tau3}$.
Note that $|(U_{\rm PMNS})_{\tau2}| = 0.707$ is
outside the $3\sigma$ range 
in the case of the NO neutrino mass spectrum, while 
$|(U_{\rm PMNS})_{\mu2}| = 0.707$ is 
at the border of the  $3\sigma$ allowed ranges for both the
NO and IO spectra.

 Let us consider as an example the first possibility, i.e., 
$P_e = P_\nu = P_{213}$, leading to $|(U_{\rm PMNS})_{\mu2}| = 1/\sqrt{2}$.
In this case the mixing angles of the standard parametrisation of the PMNS matrix are related to the free parameters $\th^\nu$, $\th^e$ and $\delta^e$ as follows:
\begin{align}
\sin^2\th_{13} & = |(U_{\rm PMNS})_{e3}|^2 = 
\frac{1}{24} \bigg[\cos2\theta^\nu \left(\sin2\theta^e \left(3 \sin\delta^e + 
4\sqrt{3}\cos\delta^e\right) + 4\cos2\theta^e - 1\right) \nonumber\\
& +\sqrt{2}\sin2\theta^\nu \left(\sin2\theta^e 
\left(\sqrt{3} \cos\delta^e - 6 \sin\delta^e\right) + \cos2\theta^e + 2\right) 
 - 3 \sin\delta^e \sin2\theta^e + 9\bigg]\,,
\label{eq:ssth13}\\
\sin^2\th_{23} & = \frac{|(U_{\rm PMNS})_{\mu3}|^2}{1-|(U_{\rm PMNS})_{e3}|^2} = 
\frac{3 - 2\sqrt{2} \sin2\theta^\nu + \cos2\theta^\nu}{12\cos^2\th_{13}}\,, 
\label{eq:ssth23}\\
\sin^2\th_{12} & = \frac{|(U_{\rm PMNS})_{e2}|^2}{1-|(U_{\rm PMNS})_{e3}|^2} = 
\frac{1 + \sin\delta^e \sin2\theta^e}{4\cos^2\th_{13}}\,. 
\label{eq:ssth12}
\end{align}
Moreover, from $|(U_{\rm PMNS})_{\mu2}| = 1/\sqrt{2}$ we obtain a sum rule for $\cos\delta$:
\be
\cos\delta = \frac{2\cos^2\theta_{12} \cos^2\theta_{23} + 
2\sin ^2\theta_{12} \sin^2\theta_{23} \sin^2 \theta_{13} - 1}
{\sin2\th_{12} \sin2\th_{23} \sin\th_{13}}\,. 
\label{eq:sumrule}
\ee

 Let us comment now on the following issue. 
Once one of the elements of the PMNS matrix is fixed to be a constant, 
we still have four possible configurations, namely, 
a permutation of two remaining columns, a permutation of two remaining rows 
and both of them.
For instance, in the case considered above, except for $P_e = P_\nu = P_{213}$, 
we can have a fixed $(U_{\rm PMNS})_{\mu2}$ with 
$(P_e, P_\nu) = (P_{213}, P_{231})$, $(P_{312}, P_{213})$ and $(P_{312}, P_{231})$.
These combinations of the permutation matrices will not lead, however, 
to different mixing patterns by virtue of the following relations:
\begin{align}
& R_{23}\left(\th^\nu\right) P_{231} = R_{23}\left(\th^\nu + \pi/2\right) P_{213}\, 
\diag\left(-1, 1, 1\right)\,, \label{eq:unph1} \\
& P_{312}\, U_{23}\left(\th^e, \delta^e\right) = 
\diag\left(e^{i\delta^e}, 1, -e^{-i\delta^e}\right) P_{213}\, 
U_{23}\left(\th^e+\pi/2, \delta^e\right)\,. \label{eq:unph2}
\end{align}
Indeed, e.g., in the case of $(P_e,P_\nu) = (P_{312}, P_{231})$, 
defining $\hat\th^{\nu} = \th^\nu + \pi/2$, $\hat\th^e = \th^e + \pi/2$ 
and absorbing the matrix $\diag\left(-1, 1, 1\right)$ in the matrix $Q_\nu$,
we obtain the same PMNS matrix as in the case of $(P_e,P_\nu) = (P_{213}, P_{213})$:
\be
U_{\rm PMNS} =  
P_{213}\, U_{23}(\hat\th^e,\delta^e)\, \O_e^\dagger\, 
\O_\nu\, R_{23}(\hat\th^\nu)\, P_{213}\, Q_\nu\,.
\ee
The phases in the matrix $\diag\left(e^{i\delta^e}, 1, -e^{-i\delta^e}\right)$ are unphysical, 
and we have disregarded them.

\begin{table}[t!]
\hspace{-0.5cm}
\begin{tabular}{|c|c|c|c|c|}
\hline
$g_e$ & $\O_e$ & $g_\nu$ & $H^\nu_{\rm CP}$ & $\O_\nu$ \btb\\
\hline
\multirow{6}[36]{*}{$S$} & 
\multirow{6}[36]{*}{$\dfrac{1}{\sqrt{6}}
\begin{pmatrix}
\sqrt{2} & -\sqrt{3} & -1 \\
\sqrt{2} & 0 & 2 \\
\sqrt{2} & \sqrt{3} & -1
\end{pmatrix}$} & 
\multirow{6}[36]{*}{$TU$} &
\multirow{3}[18]{*}{$\{U,T\}$} & \\
& & & & 
$\dfrac{1}{\sqrt{2}}
\begin{pmatrix}
0 & 0 &\sqrt{2}i \\
e^{\frac{2\pi i}{3}} & - e^{\frac{i \pi}{6}} & 0 \\
e^{\frac{i \pi}{3}} & e^{-\frac{i \pi}{6}} & 0
\end{pmatrix}$\\ 
& & & & \\
\cline{4-5}
& & & 
\multirow{3}[18]{*}{$\{STS, T^2STU\}$} & \\
& & & &
$\dfrac{1}{\sqrt{6}}
\begin{pmatrix}
0 & 2i & \sqrt{2} \\
\sqrt{3}e^{\frac{i\pi}{6}} & e^{\frac{i\pi}{6}} & -\sqrt{2}e^{-\frac{i\pi}{3}} \\
\sqrt{3}e^{-\frac{i\pi}{6}} & -e^{-\frac{i\pi}{6}} & -\sqrt{2}e^{\frac{i\pi}{3}}
\end{pmatrix}$\\
& & & & \\
\hline
\multirow{15}[75]{*}{$TU$} & 
\multirow{15}[75]{*}{$\dfrac{1}{\sqrt{2}}
\begin{pmatrix}
0 & 0 & \sqrt{2} \\
e^\frac{i\pi}{3} & e^{-\frac{2\pi i}{3}} & 0 \\
1 & 1 & 0
\end{pmatrix}$} & 
\multirow{9}[47]{*}{$S$} & 
\multirow{3}[18]{*}{$\{1, S\}$} & \\
& & & & 
$\dfrac{1}{\sqrt{6}}
\begin{pmatrix}
\sqrt{2} & -\sqrt{3} & -1 \\
\sqrt{2} & 0 & 2 \\
\sqrt{2} & \sqrt{3} & -1
\end{pmatrix}$ \\ 
& & & & \\
\cline{4-5}
& & & 
\multirow{3}[18]{*}{$\{U, SU\}$} & \\
& & & & 
$\dfrac{i}{\sqrt{6}}
\begin{pmatrix}
\sqrt{2} & -2 & 0 \\
\sqrt{2} & 1 & -\sqrt{3}i \\
\sqrt{2} & 1 & \sqrt{3}i
\end{pmatrix}$\\
& & & & \\
\cline{4-5}
& & &
\multirow{3}[18]{*}{$\{TST^2U, T^2STU\}$} & \\
& & & &
$\dfrac{1}{\sqrt{3}}
\begin{pmatrix}
1 & i & 1 \\
1 & e^{-\frac{i\pi}{6}} & -e^{-\frac{i\pi}{3}} \\
1 & -e^{\frac{i\pi}{6}} & -e^{\frac{i\pi}{3}}
\end{pmatrix}$\\
& & & & \\
\cline{3-5}
 & & 
\multirow{6}[30]{*}{$U$} & 
\multirow{3}[18]{*}{$\{1,U\}$} & \\
& & & &
$\dfrac{1}{\sqrt{2}}
\begin{pmatrix}
0 & 0 &\sqrt{2} \\
-1 & 1 & 0 \\
1 & 1 & 0
\end{pmatrix}$ \\
& & & & \\
\cline{4-5}
& & & 
\multirow{3}[18]{*}{$\{S, SU\}$} & \\
& & & & 
$-\dfrac{i}{\sqrt{6}} 
\begin{pmatrix}
0 & \sqrt{2}i & -2 \\
\sqrt{3} & \sqrt{2}i & 1 \\
-\sqrt{3} & \sqrt{2}i & 1
\end{pmatrix}$\\
& & & & \\
\hline
\end{tabular}
\caption{The matrices $\O_e$ and $\O_\nu$ dictated by the residual 
symmetries $G_e = Z_2^{g_e}$ and $G_\nu = Z_2^{g_\nu} \times H^\nu_{\rm CP}$ 
for all seven phenomenologically viable pairs of $G_e$ and $G_\nu$.
For each pair $H^\nu_{\rm CP} = \{X_{\nu1},X_{\nu2}\}$
of remnant GCP transformations,
the given matrix $\Omega_\nu$ provides the Takagi factorisation of
the first element, i.e., $X_{\nu1} = \O_\nu\, \O_\nu^T$~\protect\footnotemark{}.
}
\label{tab:OeOnu}
\end{table}
%
 We list in Table~\ref{tab:OeOnu} 
the matrices $\O_e$ and $\O_\nu$
for all seven phenomenologically viable 
pairs of residual symmetries 
$\{G_e, G_\nu\} = \{Z_2^{g_e}, Z_2^{g_\nu} \times H^\nu_{\rm CP}\}$. 
It turns out, however, that four of these seven pairs, namely, 
$\{G_e, G_\nu\} = \{Z_2^{S}, Z_2^{TU} \times H^\nu_{\rm CP}\}$ with 
$H^\nu_{\rm CP} = \{U,T\}$ and $\{STS,T^2STU\}$, and
$\{G_e, G_\nu\} = \{Z_2^{TU}, Z_2^{S} \times H^\nu_{\rm CP}\}$ with 
$H^\nu_{\rm CP} = \{U,SU\}$ and $\{TST^2U,T^2STU\}$, 
lead to the same predictions for the mixing parameters.
We demonstrate this in Appendix~\ref{app:equivalence}.

\subsection{Extracting Mixing Parameters and Statistical Analysis}
\label{subsec:statanalysis}

 In this subsection we perform a statistical analysis of the predictions 
for the neutrino mixing angles and CPV phases for each of the four 
distinctive sets of the residual flavour and CP symmetries, which are 
$\{G_e, G_\nu\} = \{Z_2^{TU}, Z_2^{S} \times H^\nu_{\rm CP}\}$ with 
$H^\nu_{\rm CP} = \{1,S\}$ and $\{U,SU\}$, and 
$\{G_e, G_\nu\} = \{Z_2^{TU}, Z_2^{U} \times H^\nu_{\rm CP}\}$ with 
$H^\nu_{\rm CP} = \{1,U\}$ and $\{S,SU\}$. 
This allows us to derive predictions for the 
three neutrino mixing angles and the three leptonic 
CPV phases, which, in many of the cases 
analysed in the present study is impossible to obtain 
purely analytically. 

  Once a pair of residual symmetries and the permutation matrices 
$P_e$ and $P_\nu$ are specified, we have the expressions for 
$\sin^2\th_{ij}$ in terms of $\th^\nu$, $\th^e$ and $\delta^e$ of the type of 
eqs.~\eqref{eq:ssth13}\,--\,\eqref{eq:ssth12}. 
Moreover, employing a sum rule for $\cos\delta$ analogous to 
that in eq.~\eqref{eq:sumrule} and computing the rephasing invariant
\be
J_{\rm CP} = \im\left\{(U_{\rm PMNS})^*_{e1}\, (U_{\rm PMNS})^*_{\mu3}\, 
(U_{\rm PMNS})_{e3}\, (U_{\rm PMNS})_{\mu1}\right\}\,,
\ee
%
which determines the magnitude of CPV 
effects in neutrino oscillations \cite{Krastev:1988yu} and 
which in the standard parametrisation of the PMNS matrix 
is proportional to $\sin\delta$,
\be
J_{\rm CP} = \frac{1}{8} \sin2\th_{12} \sin2\th_{23} \sin2\th_{13} \cos\th_{13} \sin\delta\,,
\ee
%
we know the value of $\delta$ for any  $\th^\nu$, $\th^e$ and $\delta^e$.
\footnotetext{$X_{\nu2}$ is instead factorised as 
$X_{\nu2} = \tilde\O_\nu\, \tilde\O_\nu^T$,
with $\tilde\O_\nu = \O_\nu \diag(1, i, i)$, 
as follows from 
$X_{\nu2} = \rho(g_\nu)\, X_{\nu1} = 
\O_\nu\, \O_\nu^\dagger\, \rho(g_\nu)\, \O_\nu\, \O_\nu^T = 
\O_\nu\, \rho(g_\nu)^\di\, \O_\nu^T$, 
with $\rho(g_\nu)^\di$ defined in eq.~\eqref{eq:Onu1}.}%
Similarly, making use of the two charged lepton rephasing invariants~%
\footnote{In their general form, 
when one keeps explicit the unphysical phases $\xi_j$ in the Majorana condition 
$C\,\overline{\nu_j}^T = \xi_j\,\nu_j$, $j=1,2,3$, 
the rephasing invariants related to the Majorana phases involve $\xi_j$ and 
are invariant under phase transformations of both the charged lepton and neutrino fields 
(see, for example, eqs.~(22)\,--\,(28) in \cite{Bilenky:2001rz}). We have set $\xi_j = 1$.},
associated with the Majorana phases
\cite{Nieves:1987pp,AguilarSaavedra:2000vr,Bilenky:2001rz,Nieves:2001fc},
\be
I_1 = \im\left\{(U_{\rm PMNS})^*_{e1}\, (U_{\rm PMNS})_{e2}\right\} 
\quad {\rm and} \quad 
I_2 = \im\left\{(U_{\rm PMNS})^*_{e1}\, (U_{\rm PMNS})_{e3}\right\}\,,
\label{eq:inv}
\ee
%
and the corresponding real parts
\be
R_1 = \re\left\{(U_{\rm PMNS})^*_{e1}\, (U_{\rm PMNS})_{e2}\right\} 
\quad {\rm and} \quad 
R_2 = \re\left\{(U_{\rm PMNS})^*_{e1}\, (U_{\rm PMNS})_{e3}\right\}\,,
\ee
%
which in the standard parametrisation of the PMNS matrix read:
\begin{align}
& I_1 = \sin\th_{12} \cos\th_{12} \cos^2\th_{13} \sin\left(\alpha_{21}/2\right)\,, 
\quad
I_2 = \cos\th_{12} \sin\th_{13} \cos\th_{13} \sin\left(\alpha_{31}/2 - \delta\right)\,,\\
& R_1 = \sin\th_{12} \cos\th_{12} \cos^2\th_{13} \cos\left(\alpha_{21}/2\right)\,, 
\quad
R_2 = \cos\th_{12} \sin\th_{13} \cos\th_{13} \cos\left(\alpha_{31}/2 - \delta\right)\,,
\label{eq:invstd}
\end{align}
%
we also obtain the values of $\alpha_{21}$ and $\alpha_{31}$ for any 
$\th^\nu$, $\th^e$ and $\delta^e$.

  Further, we scan randomly over $\th^\nu \in [0,\pi)$, $\th^e \in [0,\pi)$ 
and $\delta^e \in [0,2\pi)$ and calculate the values of 
$\sin^2\th_{ij}$ and the CPV phases.
We require $\sin^2\th_{ij}$ to lie in the corresponding $3\sigma$ ranges 
given in Table~\ref{tab:parameters}. 
The obtained values of $\sin^2\th_{ij}$ and $\delta$ can be characterised by 
a certain value of the $\chi^2$ function constructed as follows:
\be
\chi^2 \left(\vec{x}\right) = \sum_{i=1}^4 \chi^2_i \left(x_i\right)\,,
\label{eq:chisq}
\ee
%
where $\vec{x} = \{x_i\} = (\sin^2\th_{12}, \sin^2\th_{13}, \sin^2\th_{23}, \delta)$ and 
$\chi^2_i$ are one-dimensional projections for NO and IO taken 
from \cite{Capozzi:2017ipn}~%
\footnote{We note that according to the latest global oscillation data, 
there is an overall preference for NO over IO 
of $\D\chi^2_{\rm IO-NO} \approx 3.6$. 
Nevertheless, we take a conservative approach and treat both orderings 
on an equal footing. A discussion on this issue can be found in \cite{Capozzi:2017ipn}.}. 
Thus, we have a list of points 
$(\sin^2\th_{12}, \sin^2\th_{13}, \sin^2\th_{23}, \delta,$ $\alpha_{21}, \alpha_{31}, \chi^2)$. 
To see the restrictions on the mixing parameters imposed by flavour and CP symmetries 
we consider all 15 different pairs $(a,b)$ of the mixing parameters. For each pair we divide 
the plane $(a,b)$ into bins and find a minimum of the $\chi^2$ function in each bin. 
We present results in terms of heat maps with colour representing a minimal value 
of $\chi^2$ in each bin. 
The results obtained in each case are discussed in the following subsection.

\subsection{Results and Discussion}
\label{subsec:results}

 In this subsection we systematically go through all different potentially viable cases 
and summarise their particular features. All these cases can be divided in four groups 
corresponding to a particular pair of residual symmetries $\{G_e, G_\nu\}$. 

In each case we concentrate on results for the ordering for which 
a better compatibility with the global data is attained.
Note that results for NO and IO differ only 
i) due to the fact that the $3\sigma$ ranges of $\sin^2\th_{13}$ and $\sin^2\th_{23}$ 
depend slightly on the ordering
and
ii) in the respective $\chi^2$ landscapes. 
Moreover, we present numerical results for the Majorana phases 
obtained for $k_1 = k_2 = 0$, where $k_1$ and $k_2$ are defined in eq.~\eqref{eq:Qnu}. 
However, one should keep in mind that 
all four $(k_1,k_2)$ pairs, where $k_i = 0,1$, are allowed.  
Whenever $k_{1(2)} = 1$, the predicted range for $\alpha_{21(31)}$ shifts by $\pi$. 
 The values of the $k_i$ are important for the predictions of 
the neutrinoless double beta decay 
effective Majorana mass (see, e.g., 
\cite{Bilenky:2001rz,Bilenky:1987ty,bb0nuth}), 
which we obtain in Section \ref{sec:bb0nu}.
\bigskip\bigskip

\noindent\textbf{Group A: $\bo{\{G_e, G_\nu\} = \{Z_2^{TU}, Z_2^{S} \times H^\nu_{\rm CP}\}}$ with 
$\bo{H^\nu_{\rm CP} = \{1,S\}}$.} 
Using the corresponding matrices $\O_e$ and $\O_\nu$ from Table~\ref{tab:OeOnu} 
and the master formula for the PMNS matrix in eq.~\eqref{eq:UPMNS}, we find 
the following form of the PMNS matrix 
(up to permutations of rows and columns and the phases in the matrix $Q_\nu$):
\be
U_{\rm PMNS}^{\rm A} = \frac{1}{2\sqrt3} \begin{pmatrix}
\sqrt6\, e^{-\frac{i\pi}{6}} & \sqrt3\,e^{i\th^\nu} & \sqrt3\, e^{-i\th^\nu} \\[0.1cm]
\sqrt2\, c^e e^{\frac{i\pi}{3}} + 2\, s^e e^{-i\delta^e} & 
a_1 \left(\th^\nu, \th^e, \delta^e\right) & a_2 \left(\th^\nu, \th^e, \delta^e\right) \\[0.1cm]
2\, c^e - \sqrt2\, s^e e^{\frac{i\pi}{3}} e^{i\delta^e} & 
a_3 \left(\th^\nu, \th^e, \delta^e\right) & a_4 \left(\th^\nu, \th^e, \delta^e\right)
\end{pmatrix}\,,
\label{eq:UPMNSA}
\ee
%
with $c^e \equiv \cos\th^e$, $s^e \equiv \sin\th^e$, 
$c^\nu \equiv \cos\th^\nu$, $s^\nu \equiv \sin\th^\nu$ and
\begin{align}
a_1 \left(\th^\nu, \th^e, \delta^e\right) &= \left[\sqrt3 c^\nu + \left(2 - i \sqrt3\right) s^\nu\right]c^e + 
\sqrt2 \left(s^\nu - \sqrt3\, c^\nu\right) s^e e^{-i\delta^e}\,, \\
a_2 \left(\th^\nu, \th^e, \delta^e\right) &= \left[\sqrt3 s^\nu - \left(2 - i \sqrt3\right) c^\nu\right]c^e - 
\sqrt2 \left(c^\nu + \sqrt3\, s^\nu\right) s^e e^{-i\delta^e}\,, \\
a_3 \left(\th^\nu, \th^e, \delta^e\right) &= \sqrt2 \left(s^\nu - \sqrt3\, c^\nu\right)c^e 
-  \left[\sqrt3 c^\nu + \left(2 - i \sqrt3\right) s^\nu\right] s^e e^{i\delta^e} \,, \\
a_4 \left(\th^\nu, \th^e, \delta^e\right) &= - \sqrt2 \left(c^\nu + \sqrt3\, s^\nu\right)c^e - 
\left[\sqrt3\, s^\nu - \left(2 - i \sqrt3\right) c^\nu\right] s^e e^{i\delta^e} \,.
\end{align}
%
From eq.~\eqref{eq:UPMNSA}, we see that 
the absolute values of the elements of the first row are fixed.
Namely, the modulus of the first element is equal to $1/\sqrt{2}$, 
while the moduli of the second and third elements equal $1/2$. 
Taking into account the current knowledge of the mixing parameters, 
eqs.~\eqref{eq:PMNS3sigmaNO} and 
\eqref{eq:PMNS3sigmaIO}, this implies that 
there are only two potentially viable cases: 
i) with $|(U_{\rm PMNS})_{\mu1}| = |(U_{\rm PMNS})_{\mu2}| = 1/2$ and 
$|(U_{\rm PMNS})_{\mu3}| = 1/\sqrt{2}$, and 
ii) with $|(U_{\rm PMNS})_{\tau1}| = |(U_{\rm PMNS})_{\tau2}| = 1/2$ and 
$|(U_{\rm PMNS})_{\tau3}| = 1/\sqrt{2}$.
\bigskip

$\bullet$ \textbf{Case A1: 
$\bo{|(U_{\rm PMNS})_{\mu1}| = |(U_{\rm PMNS})_{\mu2}| = 1/2}$, 
$\bo{|(U_{\rm PMNS})_{\mu3}| = 1/\sqrt{2}}$ 
($\bo{P_e = P_{213}}$, $\bo{P_\nu = P_{321}}$).} 
In this case we obtain 
\begin{align}
\sin^2\th_{23} & = \frac{1}{2\left(1 - \sin^2\th_{13}\right)} 
\label{eq:ss23A1}\\
& = \frac{1}{2} \left(1+\sin^2\th_{13}\right) + \mathcal{O}\left(\sin^4\th_{13}\right)\,.
\end{align}
This means that only a narrow interval $\sin^2\th_{23} \in [0.510,0.512]$ 
is allowed using the $3\sigma$ region for $\sin^2\th_{13}$. 
From the equality $|(U_{\rm PMNS})_{\mu1}| = 1/2$, 
which we find to hold in this case, it follows that 
$\cos\delta$ satisfies the following sum rule:
\be
\cos\delta = \frac{1 - 4 \sin ^2\theta_{12} \cos^2\theta_{23} 
- 4 \cos^2\theta_{12} \sin^2\theta_{23} \sin^2\th_{13}}
{2\sin2\th_{12} \sin2\th_{23} \sin\th_{13}}\,, 
\label{eq:cosdeltaA1}
\ee
%
where the mixing angles in addition are correlated among themselves.
We find that $\sin^2\th_{13}$ is constrained to lie in the interval
$(0.0213, 0.0240(2)]$ for NO (IO) and, hence, 
$\sin^2\th_{23}$ in $[0.5109, 0.5123(4)]$. 
This range of values of $\sin^2\th_{23}$ is not compatible with its current $2\sigma$ range.
Moreover, $\sin^2\th_{12}$ is found to be between approximately $0.345$ and $0.354$, 
which is outside its current $2\sigma$ range as well. 
What concerns the CPV phases, the predicted values of $\delta$ are distributed 
around $0$, namely, $\delta \in [-0.11\pi,0.11\pi]$, of 
$\alpha_{21}$ around $\pi$, $\alpha_{21} \in (0.93\pi, 1.07\pi)$, while 
the values of $\alpha_{31}$ fill the whole range, i.e., $\alpha_{31} \in [0,2\pi)$. 
These numbers, presented for the NO spectrum, remain practically unchanged  
for the IO spectrum.
However, the global minimum $\chi^2_{\rm min}$ of the $\chi^2$ function, defined in eq.~\eqref{eq:chisq}, yields approximately 22~(19) for NO (IO), which implies that this case is disfavoured by the global data at more than $4\sigma$.
\bigskip

$\bullet$ \textbf{Case A2: 
$\bo{|(U_{\rm PMNS})_{\tau1}| = |(U_{\rm PMNS})_{\tau2}| = 1/2}$,
$\bo{|(U_{\rm PMNS})_{\tau3}| = 1/\sqrt{2}}$ 
$\bo{(P_e = P_\nu}$ $\bo{= P_{321})}$.} 
 This case shares the predicted ranges for $\sin^2\th_{12}$, $\sin^2\th_{13}$, 
$ \alpha_{21}$ and $\alpha_{31}$ with case A1, but differs in the predictions 
for $\sin^2\th_{23}$ and $\delta$.
Again, there is a correlation between $\sin^2\th_{13}$ and $\sin^2\th_{23}$:
\begin{align}
\sin^2\th_{23} & = \frac{1 - 2\sin^2\th_{13}}{2\left(1 - \sin^2\th_{13}\right)} 
\label{eq:ss23A2}\\
& = \frac{1}{2} \left(1 - \sin^2\th_{13}\right) + \mathcal{O}\left(\sin^4\th_{13}\right)\,,
\end{align}
which, in particular, implies that $\sin^2\th_{23} \in [0.4877(6), 0.4891]$, 
which is not compatible with its present $2\sigma$ range. 
We also find that $|(U_{\rm PMNS})_{\tau1}| = 1/2$. This equality leads to 
the following sum rule:
\be
\cos\delta = \frac{4 \sin ^2\theta_{12} \sin^2\theta_{23} 
+ 4 \cos^2\theta_{12} \cos^2\theta_{23} \sin^2\th_{13} - 1}
{2\sin2\th_{12} \sin2\th_{23} \sin\th_{13}}\,.
\label{eq:cosdeltaA2}
\ee
It is worth noting that we should always keep in mind 
the correlations between the mixing angles in expressions of this type.
The values of $\delta$ in this case lie around $\pi$, in the interval $[0.89\pi, 1.11\pi]$.
As in the previous case, the global minimum of $\chi^2$ is somewhat large, 
$\chi^2_{\rm min} \approx 18.5~(15)$ for NO (IO), 
meaning that this case is also disfavoured. 
\bigskip
\bigskip

\noindent\textbf{Group B: $\bo{\{G_e, G_\nu\} = \{Z_2^{TU}, Z_2^{S} \times H^\nu_{\rm CP}\}}$ with 
$\bo{H^\nu_{\rm CP} = \{U,SU\}}$.} 
For this choice of the residual symmetries, 
the PMNS matrix reads 
(up to permutations of rows and columns and the phases in the matrix $Q_\nu$):
\be
U_{\rm PMNS}^{\rm B} = \frac{1}{2 \sqrt3} \begin{pmatrix}
\sqrt6\, e^{\frac{i\pi}{3}} & \sqrt3 \left(c^\nu + s^\nu\right) e^{\frac{i\pi}{3}} & 
\sqrt3 \left(s^\nu - c^\nu\right) e^{\frac{i\pi}{3}} \\[0.1cm]
- \sqrt2\, c^e e^{-\frac{i\pi}{6}}  + 2\, i\, s^e e^{-i\delta^e} & 
b_1 \left(\th^\nu, \th^e, \delta^e\right) & b_2 \left(\th^\nu, \th^e, \delta^e\right) \\[0.1cm]
2\, i\, c^e + \sqrt2\, s^e e^{-\frac{i\pi}{6}} e^{i\delta^e} & 
b_3 \left(\th^\nu, \th^e, \delta^e\right) & b_4 \left(\th^\nu, \th^e, \delta^e\right) 
\end{pmatrix}\,,
\label{eq:UPMNSB}
\ee
%
with
\begin{align}
b_1 \left(\th^\nu, \th^e, \delta^e\right) &= \left(3s^\nu - c^\nu\right)c^e e^{-\frac{i\pi}{6}} 
- 2 \sqrt2\, i\, c^\nu s^e e^{-i\delta^e}\,,  \\
b_2 \left(\th^\nu, \th^e, \delta^e\right) &= - \left(3c^\nu + s^\nu\right) c^e e^{-\frac{i\pi}{6}} 
- 2\sqrt2\, i\, s^\nu s^e e^{-i\delta^e}\,,  \\
b_3 \left(\th^\nu, \th^e, \delta^e\right) &= - 2\sqrt2\, i\, c^\nu c^e
- \left(3s^\nu - c^\nu\right) s^e e^{-\frac{i\pi}{6}} e^{i\delta^e}\,,  \\
b_4 \left(\th^\nu, \th^e, \delta^e\right) &= - 2\sqrt2\, i\, s^\nu c^e
+ \left(3c^\nu + s^\nu\right) s^e e^{-\frac{i\pi}{6}} e^{i\delta^e}\,.
\end{align}
%
Equation~\eqref{eq:UPMNSB} implies that the absolute value of one element of the PMNS 
matrix is predicted to be $1/\sqrt{2}$. Thus, we have 
four potentially viable 
cases.
\bigskip

\begin{figure}
\centering
\includegraphics[width=\textwidth]{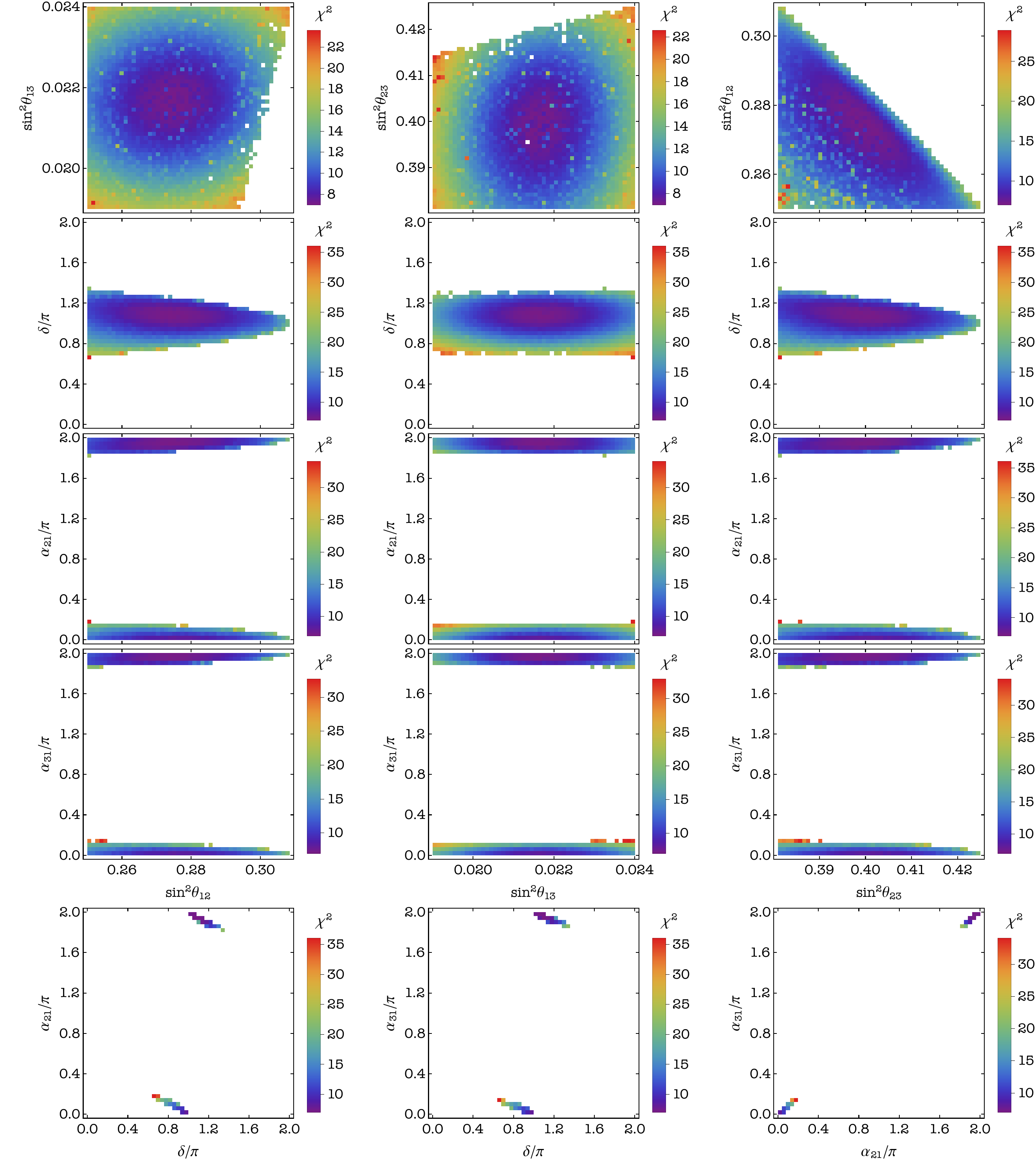}
\caption{Correlations between the neutrino mixing parameters in case B1. 
The values of all the three mixing angles are required to lie 
in their respective $3\sigma$ ranges. 
Colour represents values of $\chi^2$ for the NO 
neutrino mass spectrum.}
\label{fig:caseB1}
\end{figure}

$\bullet$ \textbf{Case B1: $\bo{|(U_{\rm PMNS})_{\mu2}| = 1/\sqrt{2}~%
(P_e = P_\nu = P_{213})}$.} 
Note that from eqs.~\eqref{eq:PMNS3sigmaNO} and 
\eqref{eq:PMNS3sigmaIO} it follows 
that this magnitude of the fixed element is inside 
its $3\sigma$ range for NO, but 
slightly outside the corresponding range for IO. 
Hence, we will focus on the results for NO. 
The characteristic feature of this case is the 
following sum rule for $\cos\delta$:
\be
\cos\delta = \frac{2\cos^2\theta_{12} \cos^2\theta_{23} 
+ 2\sin ^2\theta_{12} \sin^2\theta_{23} \sin^2 \theta_{13} - 1}
{\sin2\th_{12} \sin2\th_{23} \sin\th_{13}}\,,
\label{eq:cosdeltaB1}
\ee
%
which arises from the equality of $|(U_{\rm PMNS})_{\mu2}|$ to $1/\sqrt{2}$.
The pair correlations between the mixing parameters in this case are summarised 
in Fig.~\ref{fig:caseB1}. 
The colour palette corresponds to values of $\chi^2$ for NO. 
As can be seen, while all values of $\sin^2\th_{13}$ in its $3\sigma$ 
range are allowed, the parameters $\sin^2\th_{12}$ 
and $\sin^2\th_{23}$ are found to lie in 
$[0.250,0.308]$ and $[0.381,0.425)$ intervals, respectively. 
The predicted values of $\delta$ span the range $[0.68\pi,1.32\pi]$.
Thus, CPV effects in neutrino oscillations 
due to the phase $\delta$ can be suppressed. 
The Majorana phases instead are distributed in relatively narrow regions 
around $0$, so the magnitude of the neutrinoless double beta decay 
effective Majorana mass (see Section \ref{sec:bb0nu} and, e.g., 
\cite{Bilenky:2001rz,Bilenky:1987ty,bb0nuth})
is predicted (for $k_1 = k_2 = 0$) to have a value close to 
the maximal possible for the  NO spectrum.
Namely, $\alpha_{21} \in [-0.16\pi,0.16\pi]$ and
$\alpha_{31} \in (-0.13\pi,0.13\pi)$. 
In addition, $\delta$ is strongly correlated with 
$\alpha_{21}$ and $\alpha_{31}$, 
which in turn exhibit a strong correlation between themselves. 
Finally, $\chi^2_{\rm min} \approx 7$ for both NO and IO, i.e., 
this case is compatible with the global data at less than $3\sigma$~%
\footnote{The apparent contradiction between the obtained value of 
$\chi^2_{\rm min} \approx 7$, 
which suggests compatibility also for IO, 
and the expectation of $\chi^2_{\rm min} \gtap 9$, 
according to eq.~\eqref{eq:PMNS3sigmaIO}, 
arises from the way we construct the $\chi^2$ 
function (see eq.~\eqref{eq:chisq}), 
which does not explicitly include covariances 
between the oscillation parameters.
\label{foot:chisq}}. 
\bigskip

\begin{figure}
\centering
\includegraphics[width=\textwidth]{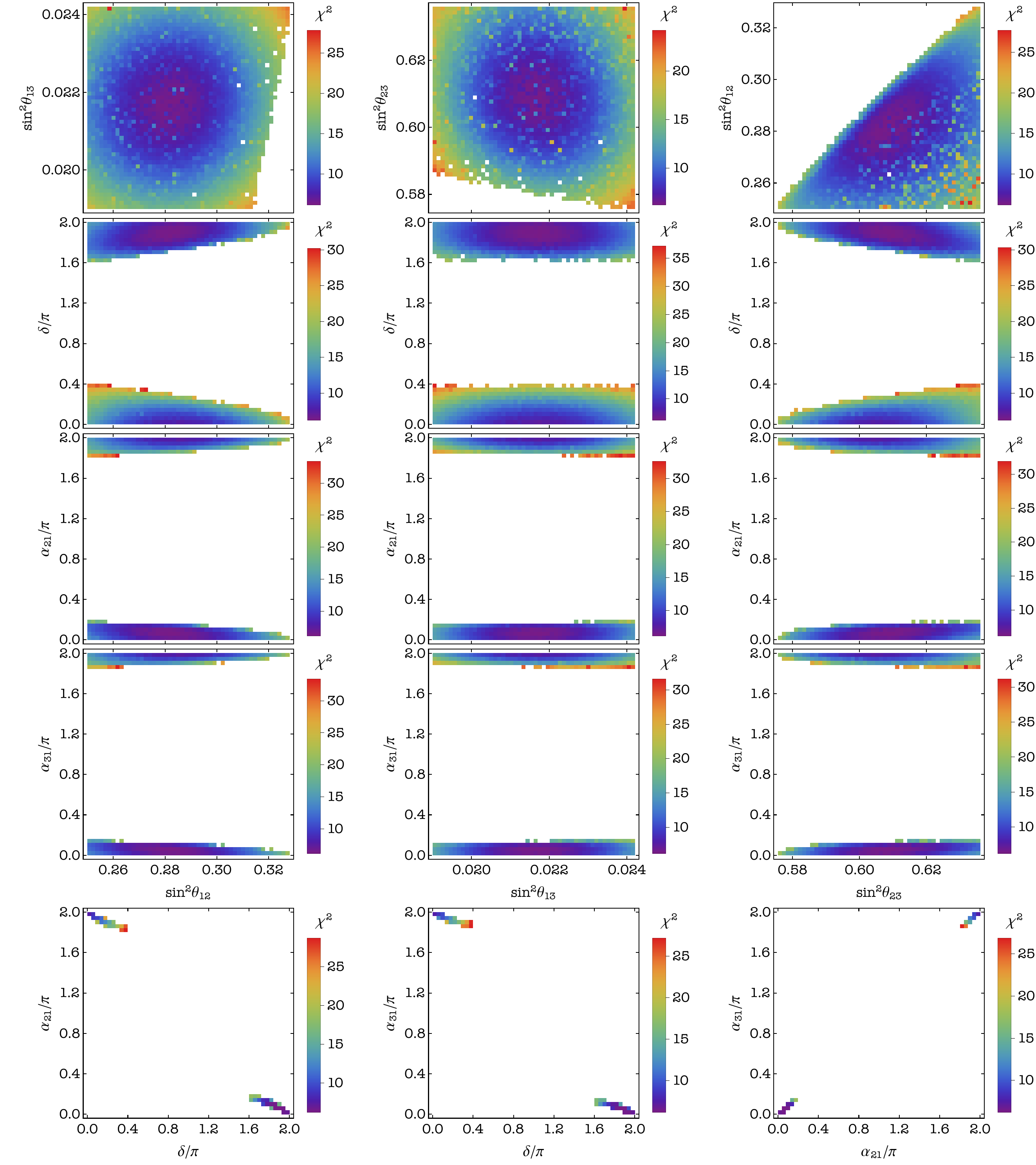}
\caption{Correlations between the neutrino mixing parameters in case B2. 
The values of all the three mixing angles are required to lie 
in their respective $3\sigma$ ranges. 
Colour represents values of $\chi^2$ for the IO 
neutrino mass spectrum.}
\label{fig:caseB2}
\end{figure}

$\bullet$ \textbf{Case B2: $\bo{|(U_{\rm PMNS})_{\tau2}| = 1/\sqrt{2}~%
(P_e = P_{321},~P_\nu = P_{213})}$.} 
Note that this value of $|(U_{\rm PMNS})_{\tau2}|$ is compatible at 
$3\sigma$ with the global data in the case of IO spectrum,
but not in the case of NO spectrum, as can be seen 
from eqs.~\eqref{eq:PMNS3sigmaNO} and \eqref{eq:PMNS3sigmaIO}. 
Thus, below we present results for the IO spectrum only. 
As in case B1, the whole $3\sigma$ range for $\sin^2\th_{13}$ is allowed. 
The obtained ranges of values of $\alpha_{21}$ and $\alpha_{31}$ are 
the same of the preceding case. The range for $\sin^2\th_{12}$ differs somewhat 
from that  obtained in case B1, and it reads $\sin^2\th_{12} \in [0.250,0.328]$~%
\footnote{This difference is related to the fact that the current $3\sigma$ range of 
$\sin^2\th_{23}$ for IO, which reads $[0.384,0.636]$, is not symmetric with respect to $0.5$. 
The asymmetry of $0.02$ translates to increase of the allowed range of 
$\sin^2\th_{12}$ by approximately $0.02$. 
This can be better understood from the top right plots 
in Figs.~\ref{fig:caseB1} and \ref{fig:caseB2}.}. 
The predictions for $\sin^2\th_{23}$ and $\delta$ are different. Now the following 
sum rule, derived from $|(U_{\rm PMNS})_{\tau2}| = 1/\sqrt{2}$, holds:
\be
\cos\delta = \frac{1 - 2\cos^2\theta_{12} \sin^2\theta_{23} 
- 2\sin ^2\theta_{12} \cos^2\theta_{23} \sin^2 \theta_{13}}
{\sin2\th_{12} \sin2\th_{23} \sin\th_{13}}\,.
\label{eq:cosdeltaB2}
\ee
The values of $\delta$ are concentrated in $[-0.38\pi,0.38\pi]$. 
For $\sin^2\th_{23}$ we find the range $(0.575,0.636]$. 
The correlations between the phases are of the same type as in case B1. 
We summarise the results in Fig.~\ref{fig:caseB2}. 
Finally, $\chi^2_{\rm min} \approx 6$ in the case of IO and $\chi^2_{\rm min} \approx 12.5$ 
for NO, 
which reflects incompatibility of this case at more than $3\sigma$ for the NO spectrum.
This occurs mainly due to the predicted values of $\sin^2\th_{23}$, which are outside its 
current $2\sigma$ range for NO.
\bigskip

\begin{figure}
\centering
\includegraphics[width=\textwidth]{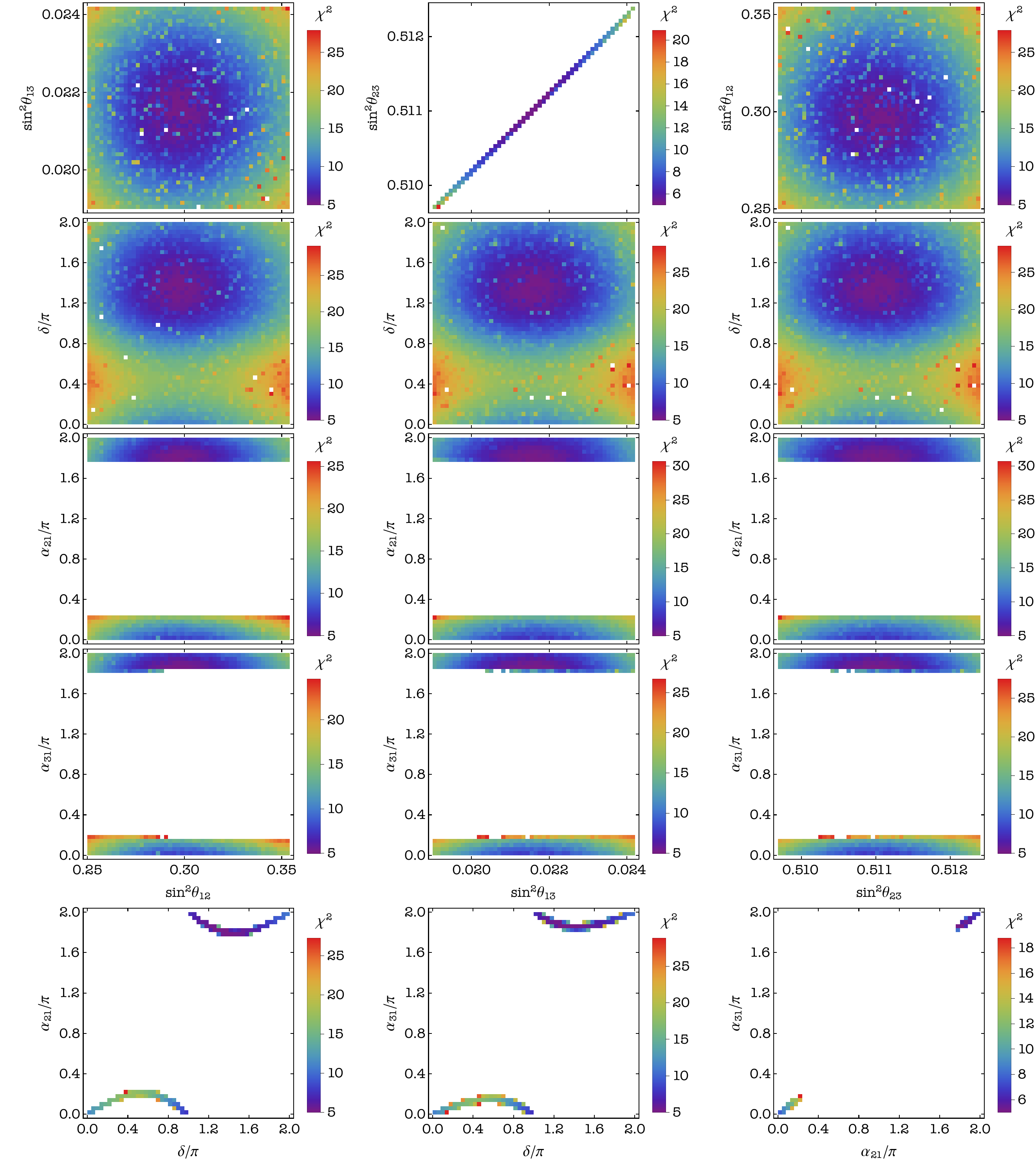}
\caption{Correlations between the neutrino mixing parameters in case B3. 
The values of all the three mixing angles are required to lie 
in their respective $3\sigma$ ranges. 
Colour represents values of $\chi^2$ for the IO 
neutrino mass spectrum. 
Note that this case is not compatible with the $2\sigma$ range of $\sin^2\th_{23}$.}
\label{fig:caseB3}
\end{figure}

$\bullet$ \textbf{Case B3: $\bo{|(U_{\rm PMNS})_{\mu3}| = 1/\sqrt{2}~%
(P_e = P_{213},~P_\nu = P_{321})}$.} 
Since $|(U_{\rm PMNS})_{\mu3}| = 1/\sqrt{2}$, the angles $\th_{13}$ and $\th_{23}$ 
are correlated as in case A1, i.e., according to eq.~\eqref{eq:ss23A1}. 
For IO this leads to 
$\sin^2\th_{23} \in [0.5097,0.5124]$ due to the fact that the whole $3\sigma$ range of 
$\sin^2\th_{13}$ is found to be allowed, as can be seen from Fig.~\ref{fig:caseB3}. 
Note that this range is outside the current $2\sigma$ range of $\sin^2\th_{23}$. 
In addition, we find that the whole $3\sigma$ range 
of the values of $\sin^2\th_{12}$ can be reproduced.
In contrast to case A1, $|(U_{\rm PMNS})_{\mu1}|$ does not equal $1/2$, but 
depends on $\th^\nu$ in the following way: 
\be
|(U_{\rm PMNS})_{\mu1}|^2 = \frac{1 - \sin2\th^\nu}{4}.
\ee
From this equation we find
\be
\cos\delta = \frac{1 - 4 \sin ^2\theta_{12} \cos^2\theta_{23} 
- 4 \cos^2\theta_{12} \sin^2\theta_{23} \sin^2\th_{13} - \sin2\th^\nu}
{2\sin2\th_{12} \sin2\th_{23} \sin\th_{13}}\,, 
\label{eq:cosdeltaB3}
\ee
i.e., $\cos\delta$  depends on $\th^\nu$ explicitly 
(not only via $\th_{12}$, $\th_{23}$ and $\th_{13}$).
With this relation, any value of $\delta$ between $0$ and $2\pi$ is allowed 
(see Fig.~\ref{fig:caseB3}). The Majorana phases, however, are constrained to 
lie around $0$ in the following intervals: 
$\alpha_{21} \in [-0.23\pi,0.23\pi]$ and 
$\alpha_{31} \in (-0.18\pi,0.18\pi)$.
Moreover, both phases $\alpha_{21}$ and $\alpha_{31}$ are correlated 
in one and the same peculiar way with the phase $\delta$. 
The correlation between $\alpha_{21}$ and $\alpha_{31}$ is similar to those 
in cases B1 and B2 
(cf. Figs.~\ref{fig:caseB1} and \ref{fig:caseB2}).
Due to the predicted values of $\sin^2\th_{23}$, which belong to the upper octant, IO is preferred over NO, the corresponding $\chi^2_{\rm min}$ being approximately $5$ and $8.5$. 
\bigskip

\begin{figure}
\centering
\includegraphics[width=\textwidth]{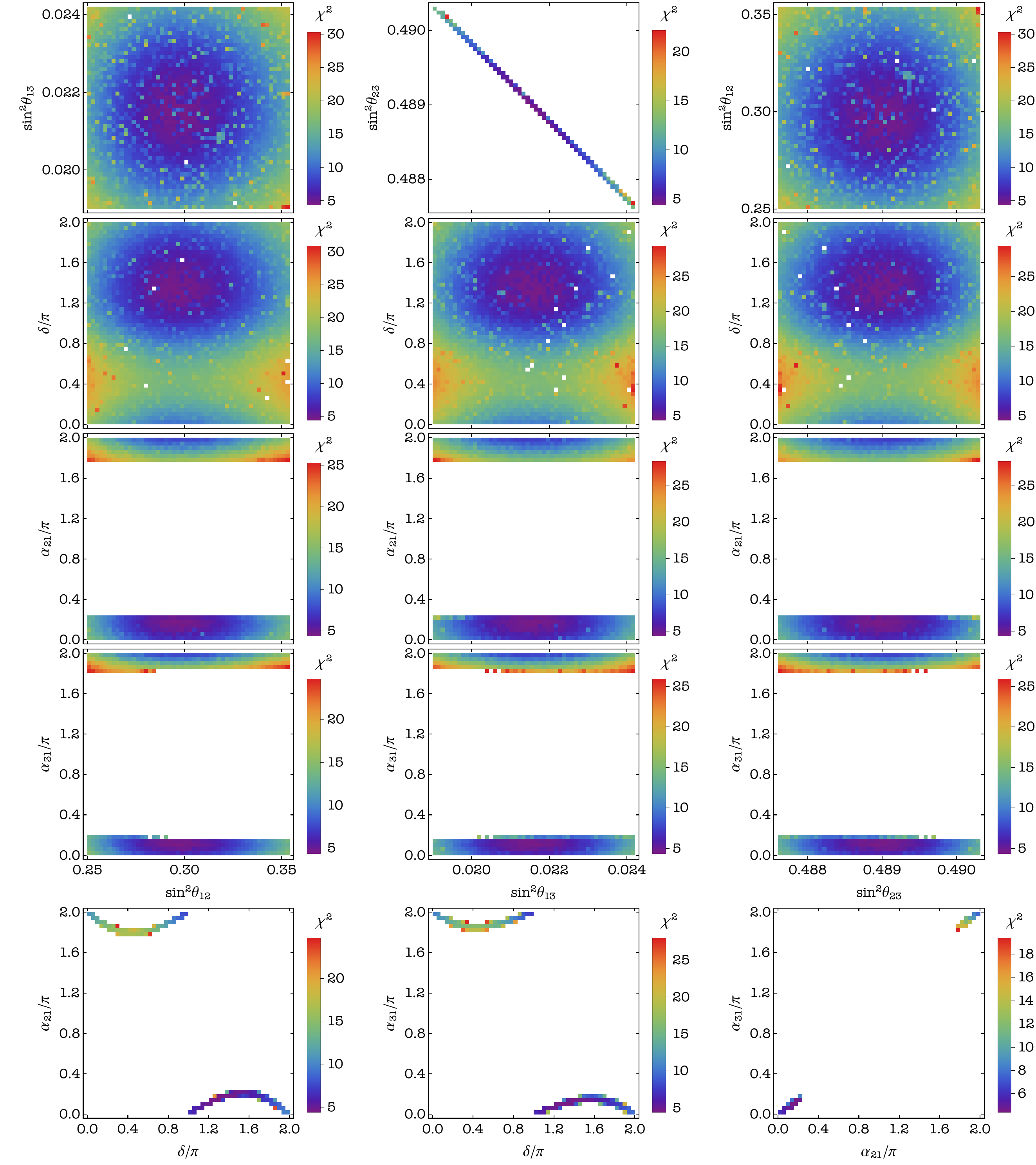}
\caption{Correlations between the neutrino mixing parameters in case B4. 
The values of all the three mixing angles are required to lie 
in their respective $3\sigma$ ranges. 
Colour represents values of $\chi^2$ for the IO 
neutrino mass spectrum. 
Note that this case is not compatible with the $2\sigma$ range of $\sin^2\th_{23}$.}
\label{fig:caseB4}
\end{figure}

$\bullet$ \textbf{Case B4: $\bo{|(U_{\rm PMNS})_{\tau3}| = 1/\sqrt{2}~%
(P_e = P_\nu = P_{321})}$.} 
The predicted ranges of all the mixing parameters are the same of case B3, 
except for $\sin^2\th_{23}$, which respects the
relation in eq.~\eqref{eq:ss23A2}, 
and thus belongs to $[0.4876,0.4903]$ in the case of IO spectrum. 
As in the previous case, this interval falls outside the $2\sigma$
range of $\sin^2\th_{23}$. 
The results obtained in this case for the IO spectrum 
are presented in Fig.~\ref{fig:caseB4}.
Similarly to the preceding case, we find 
\be
|(U_{\rm PMNS})_{\tau1}|^2 = \frac{1 - \sin2\th^\nu}{4}\,,
\ee
which leads to
\be
\cos\delta = \frac{\sin2\th^\nu + 4 \sin ^2\theta_{12} \sin^2\theta_{23} 
+ 4 \cos^2\theta_{12} \cos^2\theta_{23} \sin^2\th_{13} - 1}
{2\sin2\th_{12} \sin2\th_{23} \sin\th_{13}}\,.
\label{eq:cosdeltaB4}
\ee
%
The correlation between the Majorana phases is similar to that in the previous case. 
Also in this case, $\chi^2_{\rm min} \approx 4.5$ for IO is lower than that of 
approximately $6.5$ for NO, the reason being again the predicted range of
$\sin^2\theta_{23}$.
\bigskip
\bigskip

\noindent\textbf{Group C: $\bo{\{G_e, G_\nu\} = \{Z_2^{TU}, Z_2^{U} \times H^\nu_{\rm CP}\}}$ with 
$\bo{H^\nu_{\rm CP} = \{1,U\}}$.}
Using the correspon\-ding matrices $\O_e$ and $\O_\nu$ 
given in Table~\ref{tab:OeOnu} and eq.~\eqref{eq:UPMNS}, 
we obtain the following form of the PMNS matrix 
(up to permutations of rows and columns and the phases in the matrix $Q_\nu$):
\be
U_{\rm PMNS}^{\rm C} = \frac{1}{2}\begin{pmatrix}
e^{i\frac{\pi}{3}} & \sqrt3\, c^\nu e^{-\frac{i\pi}{6}} & \sqrt3\, s^\nu e^{-\frac{i\pi}{6}} \\[0.1cm]
\sqrt3\, c^e e^{-\frac{i\pi}{6}} & c^\nu c^e e^{\frac{i\pi}{3}} - 2\, s^\nu s^e e^{-i \delta^e} 
& s^\nu c^e e^{\frac{i\pi}{3}} + 2\, c^\nu s^e e^{-i \delta^e} \\[0.1cm]
- \sqrt3\, s^e e^{-\frac{i\pi}{6}} e^{i\delta^e} & - 2\, s^\nu c^e - c^\nu s^e e^{\frac{i\pi}{3}} e^{i\delta^e} 
& 2\, c^\nu c^e - s^\nu s^e e^{\frac{i\pi}{3}} e^{i\delta^e}
\end{pmatrix}\,.
\label{eq:UPMNSC}
\ee
%
Thus, this
pair of residual symmetries leads the absolute value of the fixed element to be $1/2$. 
Taking into account the current uncertainties in the values of the neutrino mixing parameters, 
eqs.~\eqref{eq:PMNS3sigmaNO} and \eqref{eq:PMNS3sigmaIO}, 
we have to consider five potentially viable cases corresponding to 
$(U_{\rm PMNS})_{e2}$, $(U_{\rm PMNS})_{\mu1}$, $(U_{\rm PMNS})_{\tau1}$, 
$(U_{\rm PMNS})_{\mu2}$ or $(U_{\rm PMNS})_{\tau2}$ 
being the fixed element.
\bigskip

\begin{figure}
\centering
\includegraphics[width=\textwidth]{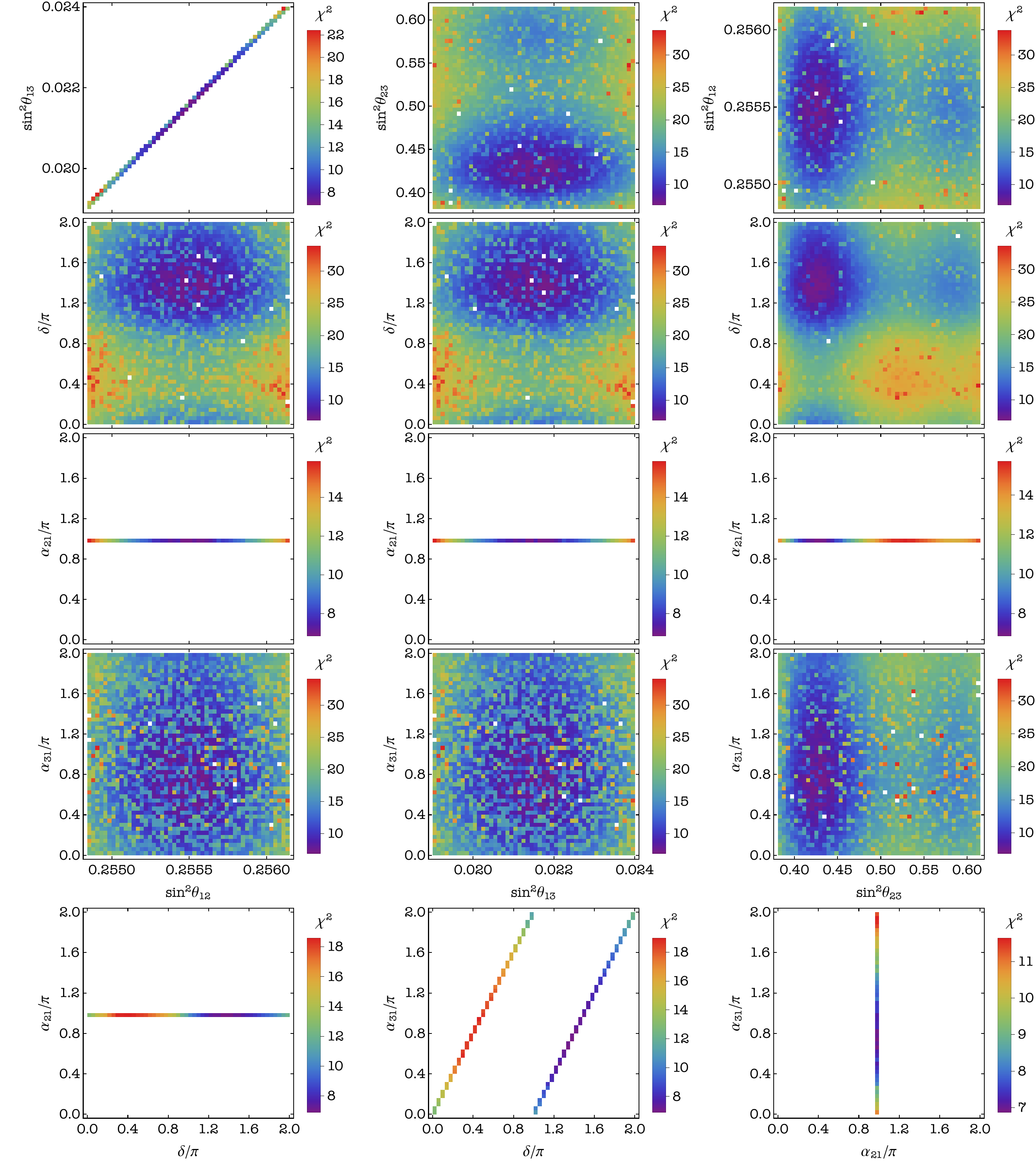}
\caption{Correlations between the neutrino mixing parameters in case C1. 
The values of all the three mixing angles are required to lie 
in their respective $3\sigma$ ranges. 
Colour represents values of $\chi^2$ for the NO 
neutrino mass spectrum. 
Note that this case is not compatible with the $2\sigma$ range of $\sin^2\th_{12}$.}
\label{fig:caseC1}
\end{figure}

$\bullet$ \textbf{Case C1: $\bo{|(U_{\rm PMNS})_{e2}| = 1/2~%
(P_e = P_{123},~P_\nu = P_{213})}$.}
Fixing $(U_{\rm PMNS})_{e2}$ leads to the following relation between $\sin^2\th_{13}$ 
and $\sin^2\th_{12}$:
\begin{align}
\sin^2\th_{12} & = \frac{1}{4\left(1 - \sin^2\th_{13}\right)} 
\label{eq:ss12C1}\\
& = \frac{1}{4} \left(1 + \sin^2\th_{13}\right) + \mathcal{O}\left(\sin^4\th_{13}\right)\,.
\end{align}
Since this case allows for the whole $3\sigma$ range of $\sin^2\th_{13}$ 
(see Fig.~\ref{fig:caseC1}), we find $\sin^2\th_{12} \in (0.2548,0.2562)$.
Note that this narrow interval is outside the current $2\sigma$ range of $\sin^2\th_{12}$. 
At the same time, this case reproduces the whole $3\sigma$ range of 
the values of $\sin^2\th_{23}$. From 
\be
|(U_{\rm PMNS})_{\mu2}|^2 = \frac{3\cos^2\th^e}{4}\,,
\ee
we obtain
\be
\cos\delta = \frac{4\cos^2\theta_{12} \cos^2\theta_{23} 
+ 4\sin ^2\theta_{12} \sin^2\theta_{23} \sin^2 \theta_{13} - 3\cos^2\th^e}
{2\sin2\th_{12} \sin2\th_{23} \sin\th_{13}}\,,
\label{eq:cosdeltaC1}
\ee
i.e., $\cos\delta$ explicitly depends on $\th^e$, and eventually this relation 
does not constrain $\delta$. 
Instead the Majorana phase $\alpha_{21}$ is predicted to be exactly $\pi$ 
(exactly $0$) for $k_1 = 0$ ($k_1 = 1$). 
While the second Majorana phase $\alpha_{31}$ itself remains unconstrained, 
the difference $\alpha_{31} - 2\delta = 0~(\pi)$ for $k_2 = 0$ ($k_2 = 1$), 
i.e., we have a strong linear correlation 
between $\delta$ and $\alpha_{31}$ (see Fig.~\ref{fig:caseC1}). 
The reason for these trivial values of $\alpha_{21}$ and $\alpha_{31} - 2\delta$ 
is the following.
In the standard parametrisation of the PMNS matrix,
$\alpha_{21}$ and the combination $(\alpha_{31}-2\delta)$
may be extracted from the phases of the first row of the PMNS matrix,
as can be seen from eqs.~\eqref{eq:inv}\,--\,\eqref{eq:invstd}.
In case C1, 
none of the phases of the first row elements of the PMNS matrix depend (mod $\pi$) 
on the free parameters $\theta^\nu$, $\theta^e$ and $\delta^e$. 
Namely, the phases of $(U_{\rm PMNS})_{e1}$, 
$(U_{\rm PMNS})_{e2}$ and $(U_{\rm PMNS})_{e3}$ are fixed 
(mod $\pi$ and up to a global phase) to be 
$-\pi/6$, $\pi/3$ and $-\pi/6$, respectively. 
Notice that only in groups B and C the relative phases of the first row can
be predicted (mod $\pi$) to be independent of $\theta^\nu$, $\theta^e$ and $\delta^e$.
Furthermore, case C1 stands out since it is, out of these relevant cases,
the only one which survives the constraints on the magnitudes of the PMNS matrix elements
given in eqs.~\eqref{eq:PMNS3sigmaNO} and \eqref{eq:PMNS3sigmaIO}.
Finally, $\chi^2_{\rm min} \approx 7$ for both mass orderings. 
\bigskip

\begin{figure}
\centering
\includegraphics[width=\textwidth]{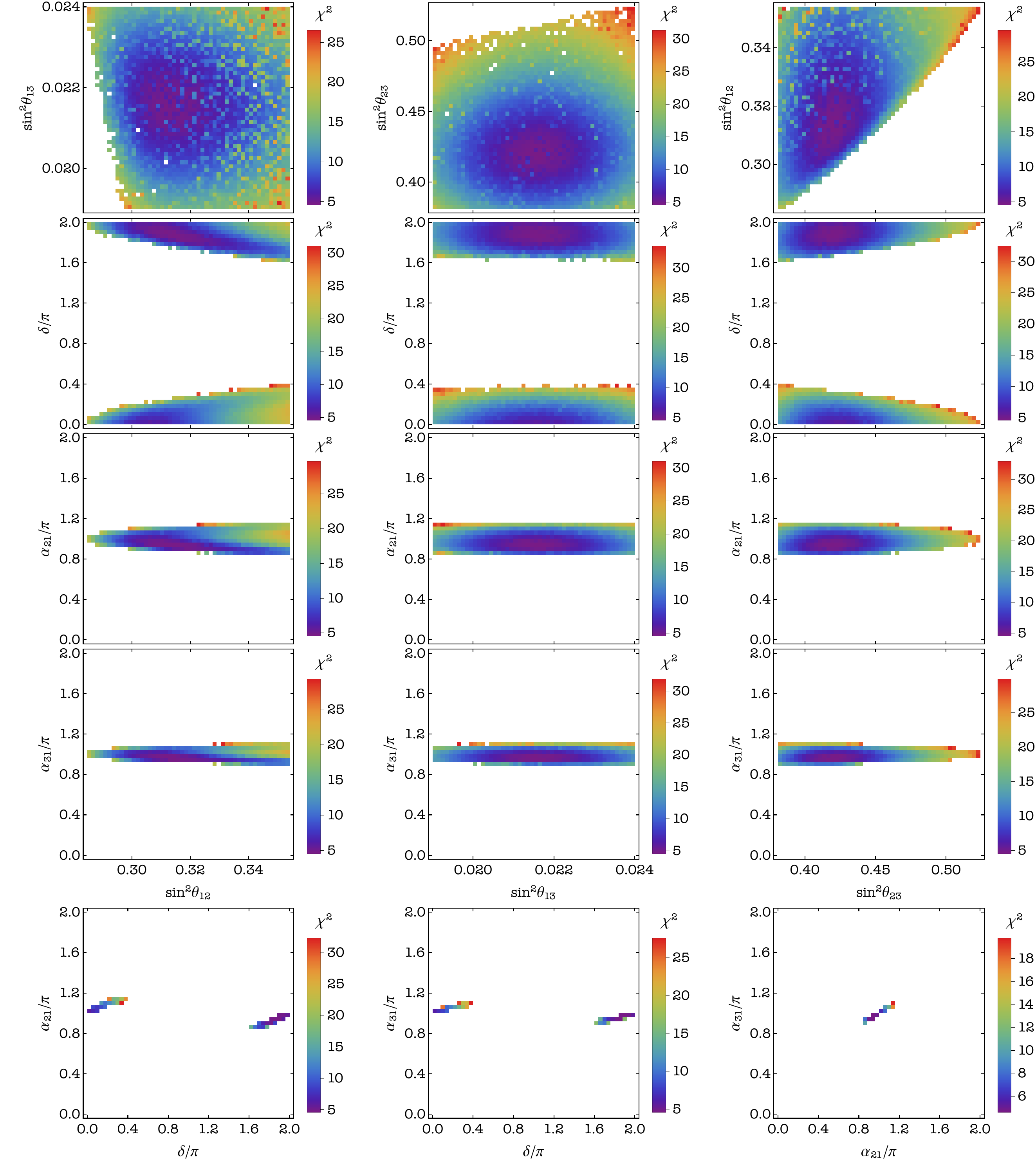}
\caption{Correlations between the neutrino mixing parameters in case C2. 
The values of all the three mixing angles are required to lie 
in their respective $3\sigma$ ranges. 
Colour represents values of $\chi^2$ for the NO 
neutrino mass spectrum.}
\label{fig:caseC2}
\end{figure}

$\bullet$ \textbf{Case C2: $\bo{|(U_{\rm PMNS})_{\mu1}| = 1/2~%
(P_e = P_{213},~P_\nu = P_{123})}$.} 
The correlations between the mixing parameters obtained in this case for NO 
are summarised in Fig.~\ref{fig:caseC2} (the results for IO are very similar). 
This case accounts for the whole $3\sigma$ range of 
$\sin^2\th_{13}$, but constrains the values of the two other angles. 
Namely, we find $\sin^2\th_{12} \in [0.285,0.354]$ and $\sin^2\th_{23} \in [0.381,0.524]$.
This case enjoys the sum rule for $\cos\delta$ given in eq.~\eqref{eq:cosdeltaA1}, since 
$|(U_{\rm PMNS})_{\mu1}| = 1/2$ as it was also in case A1. 
As a consequence, we find $\delta$ to be constrained: 
$\delta \in (-0.38\pi,0.38\pi)$. 
Both Majorana phases are distributed in relatively narrow
intervals around $\pi$: 
$\alpha_{21} \in (0.85\pi,1.15\pi)$ and $\alpha_{31} \in [0.91\pi,1.09\pi]$. 
The phase $\delta$ is correlated with each of the two 
Majorana phases in a similar way.
The latter in turn are correlated linearly between themselves. 
Overall, NO is slightly preferred over IO in this case.
The corresponding values of 
$\chi^2_{\rm min}$ read $4.5$ and $5.5$, respectively.
\bigskip

\begin{figure}
\centering
\includegraphics[width=\textwidth]{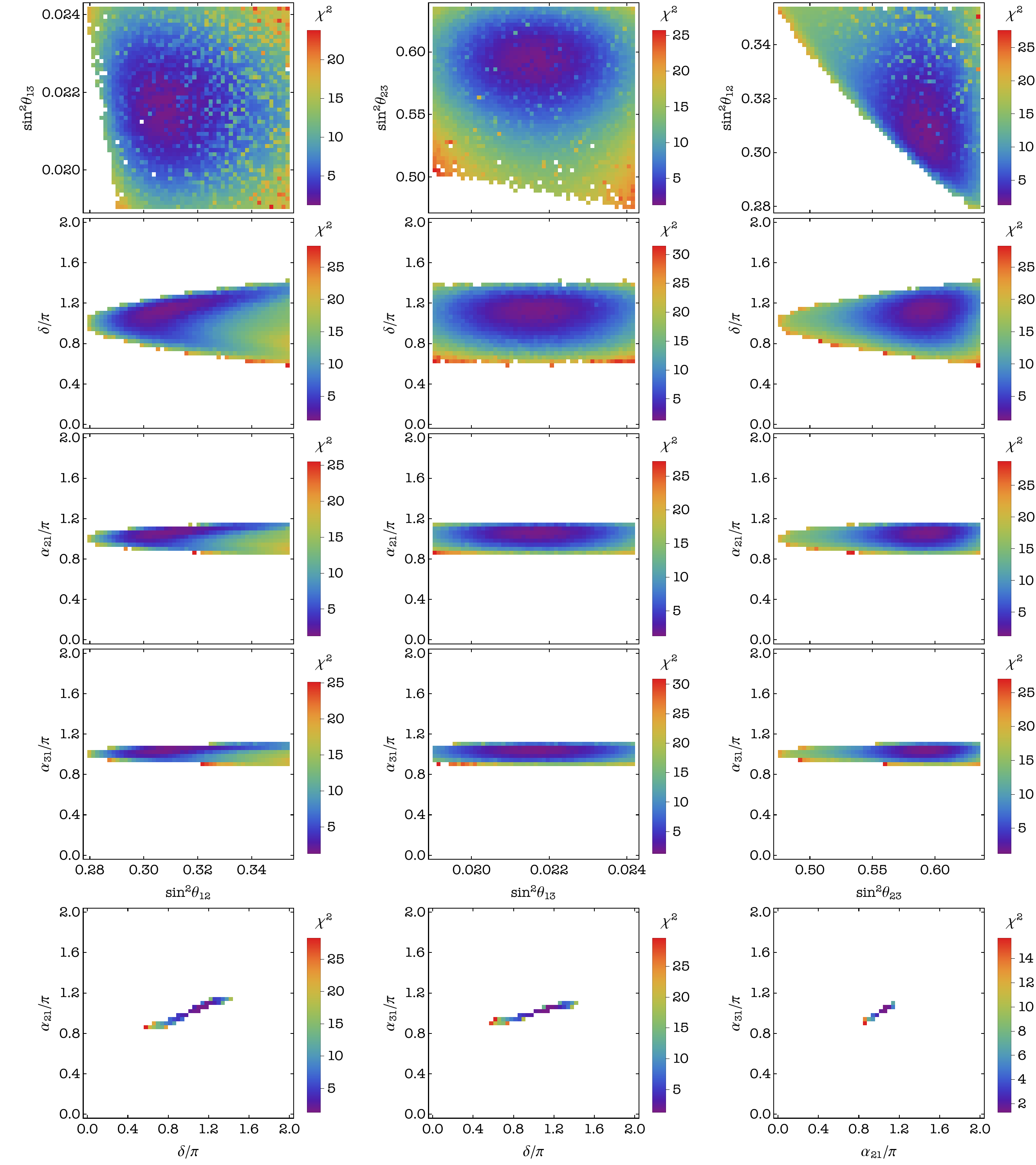}
\caption{Correlations between the neutrino mixing parameters in case C3. 
The values of all the three mixing angles are required to lie 
in their respective $3\sigma$ ranges. 
Colour represents values of $\chi^2$ for the IO 
neutrino mass spectrum.}
\label{fig:caseC3}
\end{figure}

$\bullet$ \textbf{Case C3: $\bo{|(U_{\rm PMNS})_{\tau1}| = 1/2~%
(P_e = P_{321},~P_\nu = P_{123})}$.}
This case shares some of the predictions of case C2.
Namely, the whole $3\sigma$ range of $\sin^2\th_{13}$ is allowed, 
and the ranges of $\alpha_{21}$ and $\alpha_{31}$ are 
the same as in the preceding case, as can be seen from Fig.~\ref{fig:caseC3}, 
in which we present the results for the IO neutrino mass spectrum. 
The interval of values of $\sin^2\th_{12}$ differs somewhat from that of case C2  
and reads $\sin^2\th_{12} \in [0.279,0.354]$. 
The predictions for $\sin^2\th_{23}$ and $\delta$, however, are very different 
from those of case C2. 
The allowed values of $\sin^2\th_{23}$ are concentrated mostly in the upper octant, 
$\sin^2\th_{23} \in [0.475,0.636]$. The sum rule for $\cos\delta$ in eq.~\eqref{eq:cosdeltaA2} 
is valid in this case, since $|(U_{\rm PMNS})_{\tau1}| = 1/2$, and we find the values 
of $\delta$ to be symmetrically distributed around $\pi$ in the interval $[0.60\pi,1.40\pi]$. 
The pairwise correlations between the CPV phases are of the same type as in case C2 
(taking into account an approximate shift of $\delta$ by $\pi$, as
suggested by Figs. \ref{fig:caseC2} and \ref{fig:caseC3}).
Due to the predicted range of $\sin^2\th_{23}$, this case is
favoured by the data for IO, 
for which $\chi^2_{\rm min} \approx 1.5$, while for
NO we find $\chi^2_{\rm min} \approx 8.5$.
\bigskip

\begin{figure}
\centering
\includegraphics[width=\textwidth]{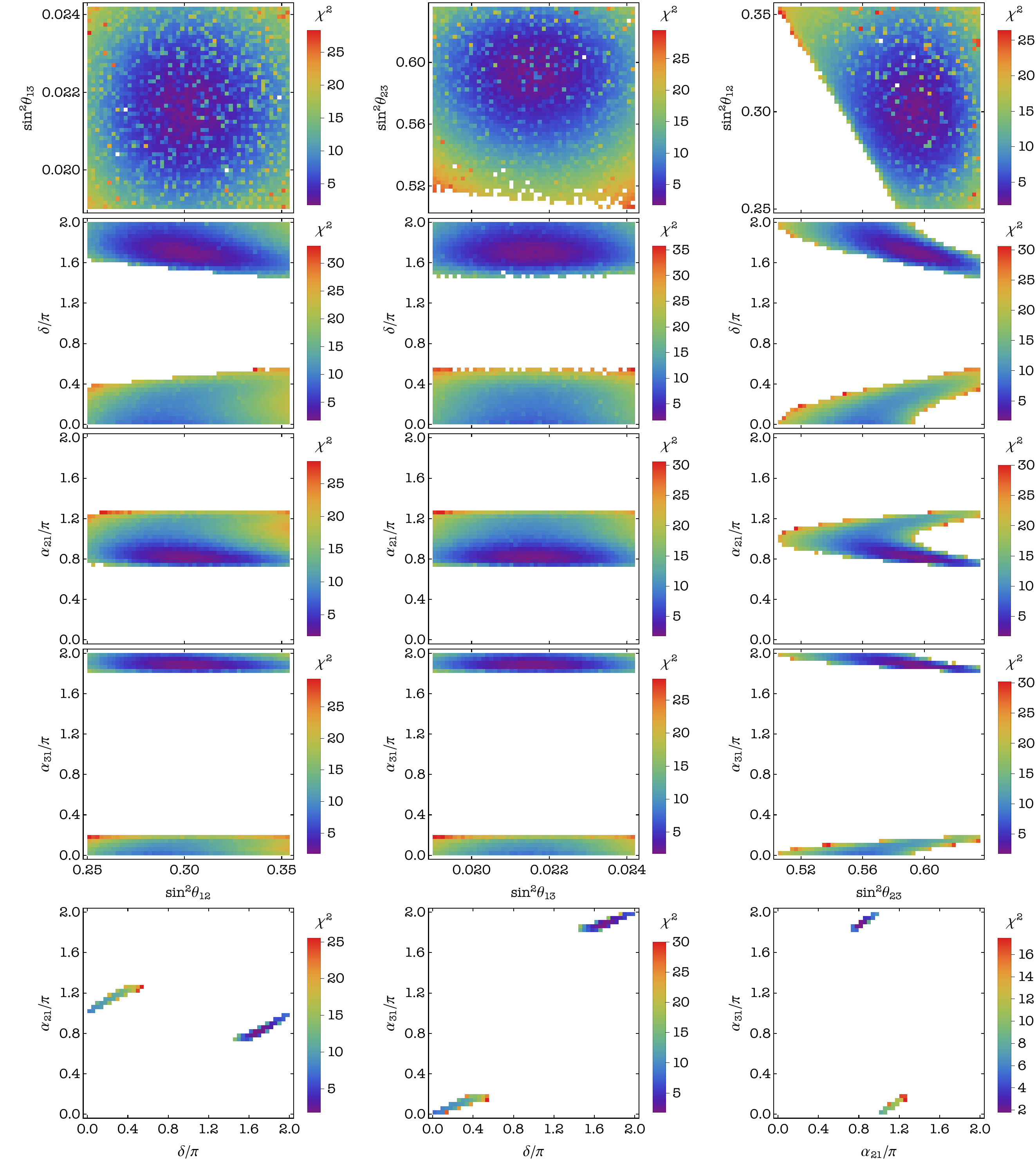}
\caption{Correlations between the neutrino mixing parameters in case C4. 
The values of all the three mixing angles are required to lie 
in their respective $3\sigma$ ranges. 
Colour represents values of $\chi^2$ for the IO 
neutrino mass spectrum.}
\label{fig:caseC4}
\end{figure}

$\bullet$ \textbf{Case C4: $\bo{|(U_{\rm PMNS})_{\mu2}| = 1/2~%
(P_e = P_\nu = P_{213})}$.}
From eqs.~\eqref{eq:PMNS3sigmaNO} and \eqref{eq:PMNS3sigmaIO} it follows that 
the value of $|(U_{\rm PMNS})_{\mu2}| = 1/2$ is allowed at $3\sigma$ only for IO. 
Thus, below we present results obtained in the IO case. 
In the case under consideration there are 
no constraints on the ranges of 
$\sin^2\th_{12}$ and $\sin^2\th_{13}$. 
The atmospheric angle is, in turn, found to lie in the upper octant, 
$\sin^2\th_{23} \in (0.505,0.636]$. As can be seen in Fig.~\ref{fig:caseC4}, 
$\delta \in [-0.54\pi,0.54\pi]$, 
which is a consequence of the following correlation 
between $\cos\delta$ and the mixing angles: 
\be
\cos\delta = \frac{4\cos^2\theta_{12} \cos^2\theta_{23} 
+ 4\sin ^2\theta_{12} \sin^2\theta_{23} \sin^2 \theta_{13} - 1}
{2\sin2\th_{12} \sin2\th_{23} \sin\th_{13}}\,,
\label{eq:cosdeltaC4}
\ee
obtained from $|(U_{\rm PMNS})_{\mu2}| = 1/2$. 
There is also a peculiar correlation between $\sin^2\th_{23}$ and $\delta$. 
The phases $\alpha_{21} \in [0.73\pi,1.27\pi]$ and 
$\alpha_{31} \in [-0.18\pi,0.18\pi]$. 
The values of all the three phases are highly correlated among themselves. 
The predicted values of $\sin^2\th_{23}$ in the upper octant lead to 
$\chi^2_{\rm min} \approx 8.5$ for NO (see footnote~\ref{foot:chisq}), which is bigger than that of $\chi^2_{\rm min} \approx 2$ for IO. 
\bigskip

\begin{figure}
\centering
\includegraphics[width=\textwidth]{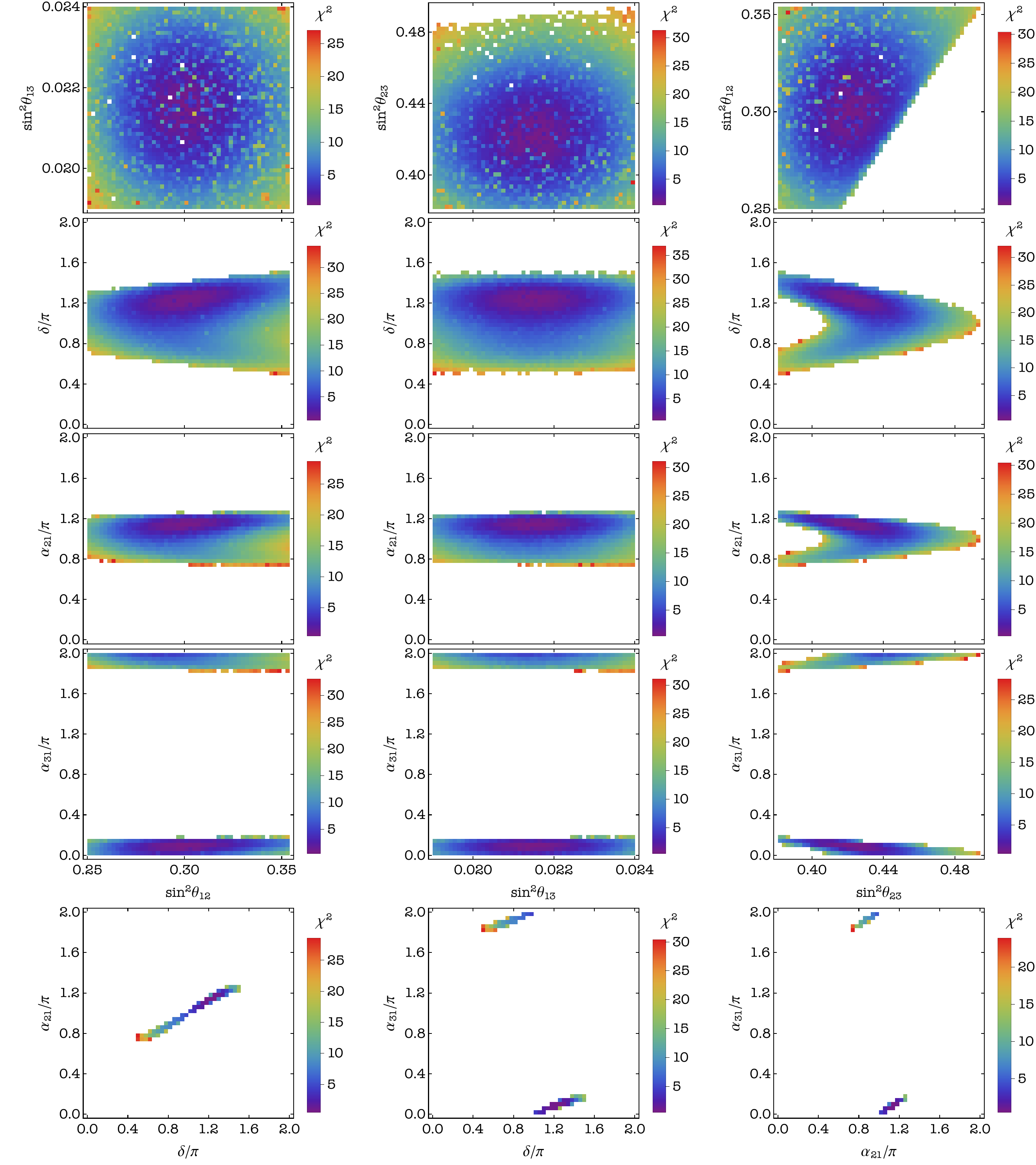}
\caption{Correlations between the neutrino mixing parameters in case C5. 
The values of all the three mixing angles are required to lie 
in their respective $3\sigma$ ranges. 
Colour represents values of $\chi^2$ for the NO 
neutrino mass spectrum.}
\label{fig:caseC5}
\end{figure}

$\bullet$ \textbf{Case C5: $\bo{|(U_{\rm PMNS})_{\tau2}| = 1/2~%
(P_e = P_{321},~P_\nu = P_{213})}$.}
The last case of this group, analogously to case C4, does not constrain the ranges of 
$\sin^2\th_{12}$ and $\sin^2\th_{13}$. Moreover, it leads to almost the same 
allowed ranges of $\alpha_{21}$ and $\alpha_{31}$ as in the previous case, 
$\alpha_{21} \in (0.74\pi,1.26\pi)$ and $\alpha_{31} \in [-0.16\pi,0.16\pi]$. 
The differences are in predictions for $\sin^2\th_{23}$ and $\delta$. 
Now the atmospheric angle lies in the lower octant, namely, for NO we find 
$\sin^2\th_{23} \in [0.381,0.494]$. 
The condition $|(U_{\rm PMNS})_{\tau2}| = 1/2$ gives rise to the following sum rule: 
\be
\cos\delta = \frac{1 - 4\cos^2\theta_{12} \sin^2\theta_{23} 
- 4\sin ^2\theta_{12} \cos^2\theta_{23} \sin^2 \theta_{13}}
{2\sin2\th_{12} \sin2\th_{23} \sin\th_{13}}\,.
\label{eq:cosdeltaC5}
\ee
The allowed values of $\delta$ span the range $[0.51\pi,1.49\pi]$. 
The correlations between the mixing parameters in this case are summarised in 
Fig.~\ref{fig:caseC5}. 
Finally, we have $\chi^2_{\rm min} \approx 0.5$ for both NO and IO. 
\bigskip
\bigskip

\noindent\textbf{Group D: $\bo{\{G_e, G_\nu\} = \{Z_2^{TU}, Z_2^{U} \times H^\nu_{\rm CP}\}}$ with 
$\bo{H^\nu_{\rm CP} = \{S,SU\}}$.}
For this last group of cases, 
we find that the PMNS matrix takes the following form 
(up to permutations of rows and columns and the phases in the matrix $Q_\nu$):
\be
U_{\rm PMNS}^{\rm D} = \frac{1}{2\sqrt3}\begin{pmatrix}
-\sqrt3\, e^{-\frac{i\pi}{6}} & \sqrt3 \left(\sqrt2\, c^\nu + i\, s^\nu\right) e^{-\frac{i\pi}{6}} & 
\sqrt3 \left(\sqrt2\, s^\nu - i\, c^\nu\right) e^{-\frac{i\pi}{6}} \\[0.1cm]
3\, c^e e^{\frac{i\pi}{3}} & d_1 \left(\th^\nu, \th^e, \delta^e\right) & 
d_2 \left(\th^\nu, \th^e, \delta^e\right)\\[0.1cm]
-3\, s^e e^{\frac{i\pi}{3}} e^{i\delta^e} & d_3 \left(\th^\nu, \th^e, \delta^e\right) & 
d_4 \left(\th^\nu, \th^e, \delta^e\right)
\end{pmatrix}\,,
\label{eq:UPMNSD}
\ee
%
where
\begin{align}
d_1 \left(\th^\nu, \th^e, \delta^e\right) &= \left(\sqrt2\, c^\nu + i\, s^\nu\right) c^e 
e^{\frac{i\pi}{3}} + 2 \left(c^\nu - i \sqrt2\, s^\nu\right) s^e e^{-i\delta^e} \,, \\
d_2 \left(\th^\nu, \th^e, \delta^e\right) &= \left(\sqrt2\, s^\nu - i\, c^\nu\right) c^e 
e^{\frac{i\pi}{3}} + 2 \left(s^\nu + i \sqrt2\, c^\nu\right) s^e e^{-i\delta^e}\,, \\
d_3 \left(\th^\nu, \th^e, \delta^e\right) &= 2 \left(c^\nu - i \sqrt2\, s^\nu\right) c^e 
- \left(\sqrt2\, c^\nu + i\, s^\nu\right) s^e e^{\frac{i\pi}{3}} e^{i\delta^e}\,, \\
d_4 \left(\th^\nu, \th^e, \delta^e\right) &= 2 \left(s^\nu + i \sqrt2\, c^\nu\right) c^e
- \left(\sqrt2\, s^\nu - i\, c^\nu\right) s^e e^{\frac{i\pi}{3}} e^{i\delta^e}\,.
\end{align}
%
Therefore,
the absolute value of the fixed element of the neutrino 
mixing matrix yields $1/2$. Thus, we have again five potentially viable cases.
\bigskip

$\bullet$ \textbf{Case D1: $\bo{|(U_{\rm PMNS})_{e2}| = 1/2~%
(P_e = P_{123},~P_\nu = P_{213})}$.}
In this case we find 
\be
\sin^2\th_{13} = \frac{3 - \cos2\th^\nu}{8}\,,
\ee
which implies that $\sin^2\th_{13}$ can have values between $1/4$ and $1/2$. 
Thus, this case is ruled out.
\bigskip

\begin{figure}
\centering
\includegraphics[width=\textwidth]{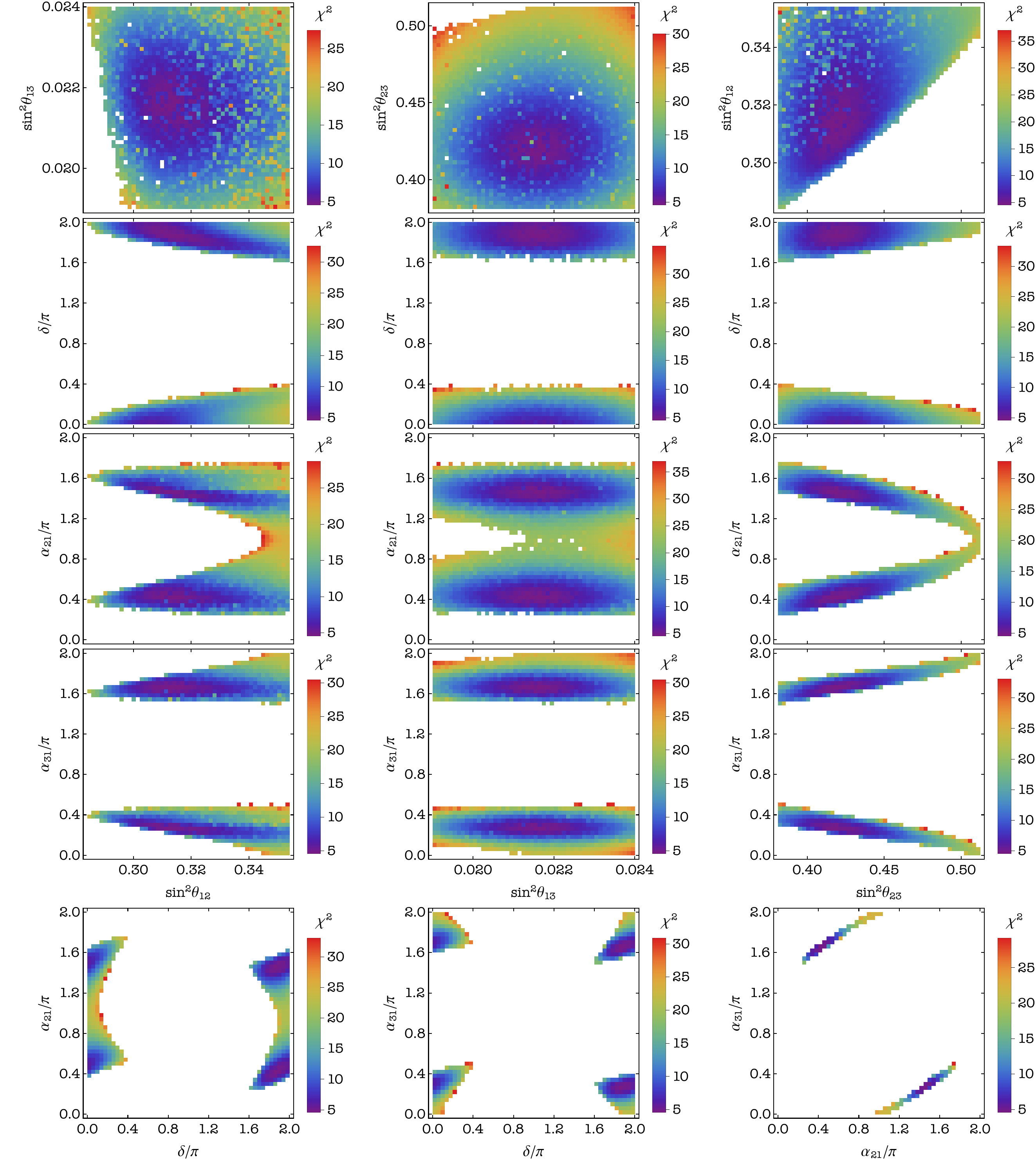}
\caption{Correlations between the neutrino mixing parameters in case D2. 
The values of all the three mixing angles are required to lie 
in their respective $3\sigma$ ranges. 
Colour represents values of $\chi^2$ for the NO 
neutrino mass spectrum.}
\label{fig:caseD2}
\end{figure}

$\bullet$ \textbf{Case D2: $\bo{|(U_{\rm PMNS})_{\mu1}| = 1/2~%
(P_e = P_{213},~P_\nu = P_{123})}$.}
This case allows for the whole $3\sigma$ range of $\sin^2\th_{13}$ 
and, in the case of NO, for the following ranges of $\sin^2\th_{12}$ and $\sin^2\th_{23}$:
$\sin^2\th_{12} \in [0.284,0.354]$ and $\sin^2\th_{23} \in [0.381,0.512]$. 
The sum rule for $\cos\delta$ in eq.~\eqref{eq:cosdeltaA1} holds, 
since $|(U_{\rm PMNS})_{\mu1}| = 1/2$. We find $\delta \in [-0.37\pi,0.37\pi]$.
What concerns the Majorana phases, $\alpha_{21}$ spans 
a relatively broad interval $[0.25\pi,1.75\pi]$, while 
$\alpha_{31} \in [-0.48\pi,0.48\pi]$.
There are very particular correlations between $\alpha_{21(31)}$ and all the other 
mixing parameters in this case, as can be seen in Fig.~\ref{fig:caseD2}, 
in which we summarise the results for NO. 
Finally, $\chi^2_{\rm min} \approx 4.5$ for NO, and 
it is slightly higher, $\chi^2_{\rm min} \approx 5.5$, for IO.
\bigskip

\begin{figure}
\centering
\includegraphics[width=\textwidth]{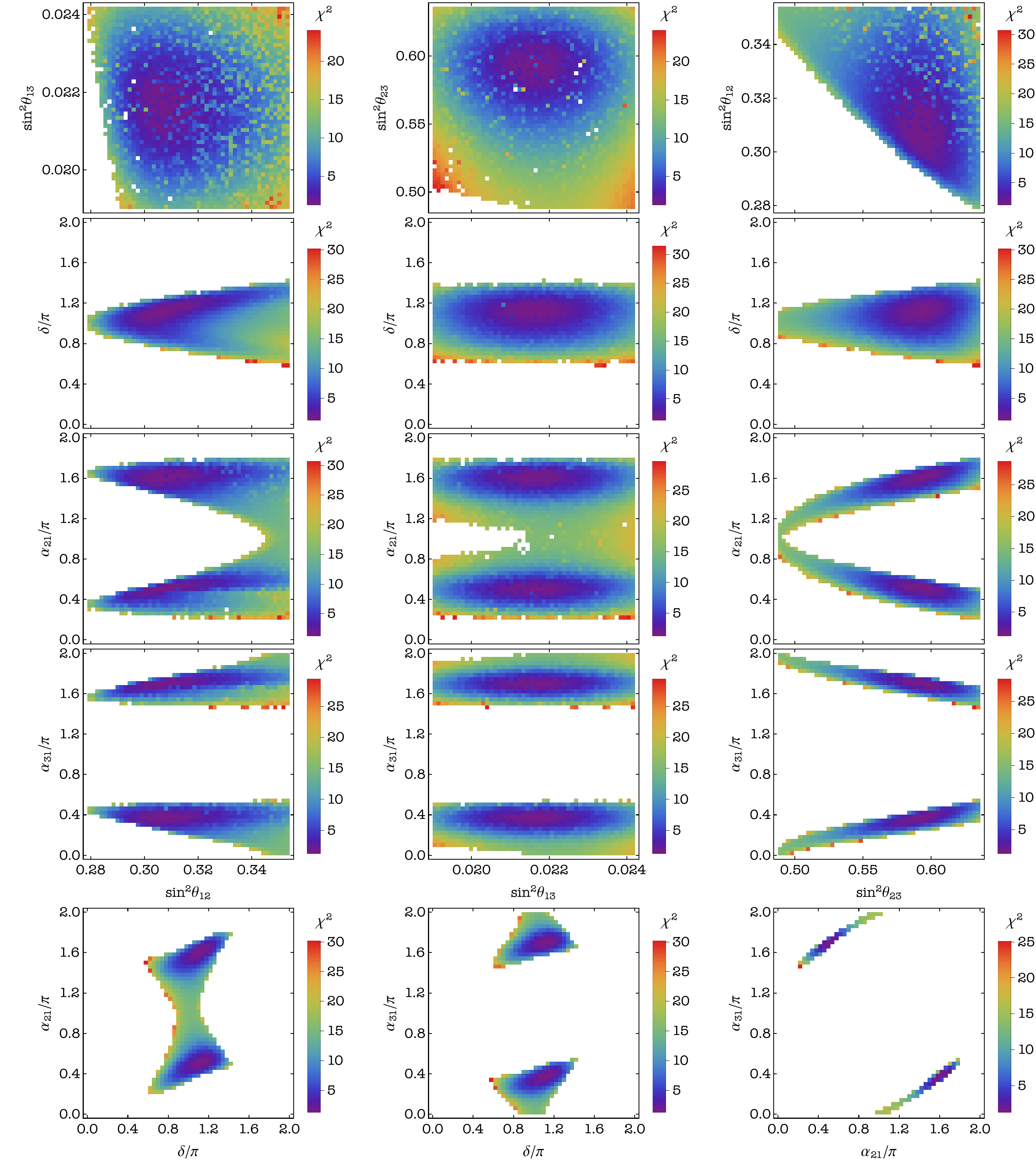}
\caption{Correlations between the neutrino mixing parameters in case D3. 
The values of all the three mixing angles are required to lie 
in their respective $3\sigma$ ranges. 
Colour represents values of $\chi^2$ for the IO 
neutrino mass spectrum.}
\label{fig:caseD3}
\end{figure}

$\bullet$ \textbf{Case D3: $\bo{|(U_{\rm PMNS})_{\tau1}| = 1/2~%
(P_e = P_{321},~P_\nu = P_{123})}$.} 
As in the previous case, the whole $3\sigma$ range of $\sin^2\th_{13}$
gets reproduced. 
The allowed ranges of $\sin^2\th_{12}$,
$\alpha_{21}$ and $\alpha_{31}$ are very similar to 
those of case D2. Namely, in the case of IO spectrum 
we have $\sin^2\th_{12} \in [0.279,0.354]$, 
$\alpha_{21} \in [0.21\pi,1.79\pi]$ and $\alpha_{31} \in (-0.53\pi,0.53\pi)$. 
Instead, the values of $\sin^2\th_{23}$ occupy mostly the upper octant, 
$\sin^2\th_{23} \in [0.488,0.636]$. The sum rule in eq.~\eqref{eq:cosdeltaA2}, which 
holds in this case since $|(U_{\rm PMNS})_{\tau1}| = 1/2$, leads to the values 
of $\delta$ distributed around $\pi$ in a rather broad range of $(0.59\pi,1.41\pi)$. 
The correlations between the Majorana phases and $\delta$ are as in the previous case, 
but again with an approximate shift of $\delta$ by $\pi$ (see Fig.~\ref{fig:caseD3}). 
The minimal value $\chi^2_{\rm min} \approx 1.5$ in the IO case,
while for the NO spectrum we get 
approximately $8.5$. This difference is due
to the allowed values of $\sin^2\th_{23}$. 
\bigskip

$\bullet$ \textbf{Case D4: $\bo{|(U_{\rm PMNS})_{\mu2}| = 1/2~%
(P_e = P_\nu = P_{213})}$.} 
This case can account only for a part of the
$3\sigma$ range of $\sin^2\th_{13}$, 
namely, $\sin^2\th_{13} \in [0.0214,0.0240(2)]$ for NO (IO) spectrum.
The constraints on two other angles are more severe.
We find that only a narrow region 
of the values of $\sin^2\th_{23}$, which falls outside
its present $2\sigma$ range, 
is allowed, namely, $\sin^2\th_{23} \in [0.505,0.512]$. 
For the solar mixing angle we have
$\sin^2\th_{12} \in [0.345,0.354]$, which is also outside 
the current $2\sigma$ range of this parameter.
The sum rule in eq.~\eqref{eq:cosdeltaC4}, 
which is also valid in this case,
constrains $\delta$ to lie in a narrow interval around $0$: 
$\delta \in [-0.11\pi,0.11\pi]$.
 The Majorana phases instead are distributed in 
narrow intervals around $\pi$. Namely, $\alpha_{21} \in (0.83\pi,1.17\pi)$ and 
$\alpha_{31} \in [0.92\pi,1.08\pi]$. 
However, the global minimum of $\chi^2$ is somewhat large in this case
for both NO and IO orderings. 
Namely, we find $\chi^2_{\rm min} \approx 22~(19)$ for NO (IO), i.e.,
this case is disfavoured 
at more than $4\sigma$ by the current global data.
\bigskip

$\bullet$ \textbf{Case D5: $\bo{|(U_{\rm PMNS})_{\tau2}| = 1/2~%
(P_e = P_{321},~P_\nu = P_{213})}$.}
This last case shares the predicted
ranges for $\sin^2\th_{12}$, $\sin^2\th_{13}$, 
$\alpha_{21}$ and $\alpha_{31}$ with case D4. 
Therefore, this case is also not compatible with the $2\sigma$ range of 
the values of $\sin^2\th_{12}$.
For $\sin^2\th_{23}$ instead we find the narrow interval 
in the lower octant, $\sin^2\th_{23} \in [0.488,0.495]$, which lies  
outside the $2\sigma$ range of $\sin^2\th_{23}$. 
We find $\cos\delta$ to satisfy the sum rule
in eq.~\eqref{eq:cosdeltaC5}, 
which in this case gives us the values of
$\delta$ in a narrow interval around $\pi$, 
$\delta \in [0.89\pi,1.11\pi]$. Thus, all the three CPV phases
are concentrated in narrow ranges around $\pi$. 
Finally, we find $\chi^2_{\rm min} \approx 18.5~(15)$ for NO (IO), 
which implies that this case is also disfavoured by
the latest global neutrino oscillation data. 
\bigskip\bigskip

 The PMNS matrix in case A2 is related with that in case A1 by 
the permutation matrix $P_{312}$ as 
$U_{\rm PMNS}^{\rm A2} = P_{312}\,U_{\rm PMNS}^{\rm A1}$. 
Given that $P_{312} = P_{132} P_{321}$, one can see that these
matrices are related by $\mu-\tau$ interchange,
after an unphysical exchange of
the first and third rows of $U_{\rm PMNS}^{\rm A1}$ has been performed
(which amounts to a redefinition of the free parameter $\theta^e$,
as shown in eq.~\eqref{eq:unph2}).
The same also holds for the following pairs of cases:  
(B1, B2), (B3, B4), (C2, C3), (C4, C5), (D2, D3) and (D4, D5). 
As can be seen from the discussion above and 
Figs.~\ref{fig:caseB1}\,--\,\ref{fig:caseB4} and 
\ref{fig:caseC2}\,--\,\ref{fig:caseD3}, 
cases inside a pair share some qualitative features. 
Namely, i) the predicted ranges of $\sin^2\th_{12}$, $\sin^2\th_{13}$, 
$\alpha_{21}$ and $\alpha_{31}$ are approximately the same; 
ii) the predicted range of $\sin^2\th_{23}$ gets approximately reflected around $1/2$, 
i.e., $\sin^2\th_{23} \rightarrow 1 - \sin^2\th_{23}$; 
iii) the predicted range of the CPV phase $\delta$ experiences an 
approximate shift by $\pi$, i.e., $\delta \rightarrow \delta + \pi$. 

 In Tables~\ref{tab:rangesAB} and \ref{tab:rangesCD} we summarise 
the predicted ranges of the mixing parameters obtained in all the 
phenomenologically viable cases discussed above.  
The corresponding best fit values together with $\chi^2_{\rm min}$ 
are presented in Tables~\ref{tab:bestfitAB} and \ref{tab:bestfitCD}.
Finally, in Table~\ref{tab:checks} we show whether the cases compatible with 
the $3\sigma$ ranges of the three mixing angles 
are also compatible with their corresponding $2\sigma$ ranges.

 The results shown in Tables~%
\ref{tab:rangesAB}\,--\,\ref{tab:bestfitCD} allow to assess 
the possibilities to critically test the predictions 
of the viable cases of the model and to distinguish between 
them. We recall that the current $1\sigma$ uncertainties 
on the measured values of  $\sin^2\th_{12}$, 
$\sin^2\th_{13}$ and  $\sin^2\th_{23}$ are \cite{Capozzi:2017ipn}
5.8\%, 4.0\% and 9.6\%, respectively.
These uncertainties are foreseen to be further reduced 
by the currently active and/or future planned experiments.
The Daya Bay collaboration plans to determine 
$\sin^2\th_{13}$ with  $1\sigma$ uncertainty of 3\% \cite{Ling:2016wgq}.
The uncertainties on  $\sin^2\th_{12}$ and  $\sin^2\th_{23}$
are planned to be reduced significantly.
The parameter $\sin^2\theta_{12}$  is foreseen to be measured 
with $1\sigma$ relative error of 0.7\% in the JUNO 
experiment \cite{JUNO}.
In the proposed upgrading of the currently taking 
data T2K experiment \cite{T2K20152016}, for example, 
$\theta_{23}$ is estimated to be determined with a 
$1\sigma$ error of $1.7^\circ$, $0.5^\circ$ and $0.7^\circ$ 
if the best fit value of $\sin^2\theta_{23} = 0.50$, $0.43$ and $0.60$, 
respectively. This implies that for these three values of 
 $\sin^2\theta_{23}$ the absolute $1\sigma$ error 
would be 0.0297, 0.0086 and 0.0120.
This error on $\sin^2\theta_{23}$ will 
be further reduced in the future planned T2HK \cite{T2HK2015} 
 and DUNE \cite{DUNE2016} experiments.
If $\delta = 3\pi/2$, the CP-conserving case of $\sin\delta = 0$ 
would be disfavoured for the NO mass 
spectrum in the same experiment at least at $3\sigma$ C.L.
Higher precision measurements of $\delta$ 
are planned to be performed in  the T2HK and DUNE 
experiments.

  We turn now to the possibilities to discriminate 
experimentally between the different cases listed in Tables%
~\ref{tab:rangesAB}\,--\,\ref{tab:bestfitCD} using 
the prospective data on  $\sin^2\theta_{12}$,  
$\sin^2\theta_{13}$,  $\sin^2\theta_{23}$ and $\delta$.
The first thing to notice is that the 
predicted ranges for  $\sin^2\th_{12}$,  $\sin^2\theta_{13}$, 
$\sin^2\th_{23}$ and $\delta$ 
in cases A1 and A2 practically coincide with the predictions 
respectively in cases D4 and D5. However,
cases  A1, D4 and cases A2, D5
are strongly disfavoured by the current data: 
for the NO (IO) neutrino mass spectrum 
A1 and D4 are disfavoured at $4.7\sigma$ ($4.4\sigma$), 
while A2 and D5 are disfavoured at $4.3\sigma$ ($3.9\sigma$).
In all these cases  $\sin^2\th_{12}$, in particular, is predicted to 
lie in the interval (0.345,0.354) compatible with the 
current $3\sigma$ range
 and, given the current best fit value 
of  $\sin^2\th_{12}$ and prospective JUNO precision 
on  $\sin^2\th_{12}$, it is very probable that future 
more precise data on  $\sin^2\th_{12}$ 
will rule out completely these scenarios. 
We will not discuss them further in this subsection.

  It follows also from Tables~\ref{tab:bestfitAB} and \ref{tab:bestfitCD} that 
the combined results on the best fit values of
$\sin^2\theta_{12}$, $\sin^2\theta_{23}$ and $\delta$ 
we have obtained in the different viable cases 
(excluding A1, A2, D4 and D5) differ significantly. 
Assuming, for example, that the experimentally determined 
best fit values of $\sin^2\theta_{12}$ and $\sin^2\theta_{23}$
will coincide with those found by us for 
a given viable case, it is not difficult 
to convince oneself inspecting 
Tables \ref{tab:bestfitAB} and \ref{tab:bestfitCD} 
that the cited prospective $1\sigma$ errors 
on  $\sin^2\theta_{12}$ and $\sin^2\theta_{23}$ 
will allow to discriminate between 
the different viable cases identified 
in our study. 
More specifically, considering as an example only the 
case of NO neutrino mass spectrum, 
the prospective high precision measurement of 
 $\sin^2\theta_{12}$ will allow to discriminate between
case C1 and all other cases 
B1\,--\,B4, C2\,--\,C5, D2 and D3. The same measurement 
will make it possible to distinguish 
i)~between case B1 and 
all the other cases except B2, 
ii)  between case B2 and all the other cases except 
B1, B3 and B4, and similarly 
iii) between case B3 and all the other cases except 
B2, B4, C4 and C5.
However, the differences between the best fit values 
of  $\sin^2\theta_{23}$ in cases B1, B2 and B3 (or B4) 
are sufficiently large, which would permit to 
distinguish between these three cases if 
$\sin^2\theta_{23}$ were measured with 
the prospective precision. It follows from 
Table \ref{tab:bestfitAB}, however, that 
it would be very challenging to discriminate between 
cases B3 and B4: it will require extremely high 
precision measurement of $\sin^2\theta_{23}$. 
These two cases would be ruled out, however, 
if the experimentally determined best fit value of 
$\sin^2\theta_{23}$ differs significantly from 
the results for $\sin^2\theta_{23}$, namely, 0.511 and 
0.489, we have obtained for  $\sin^2\theta_{23}$ 
in the B3 and B4 cases. 

 In the remaining cases C2\,--\,C5 and D2\,--\,D3, the results we have obtained for 
$\sin^2\theta_{12}$, as Table~ \ref{tab:bestfitCD} shows, 
are very similar. However, the predictions for the pair 
$\sin^2\theta_{23}$ and $\delta$ differ significantly 
in cases C2 or D2, and C3 or D3. 
The cases within each pair would be ruled out if 
the experimentally determined values of 
$\sin^2\theta_{23}$ and $\delta$ differ significantly 
from the predicted best fit values.

 Thus, the planned future high precision measurements of 
$\sin^2\theta_{12}$ and $\sin^2\theta_{23}$, together 
with more precise data on the Dirac phase $\delta$,  
will make it possible to critically test 
the predictions of the cases listed in Tables 
\ref{tab:rangesAB}\,--\,\ref{tab:bestfitCD}.
A comprehensive analysis of the possibilities 
to distinguish between the different viable cases 
found in our work in the considered $S_4$ model 
can only be done when more precise data 
first of all on $\sin^2\theta_{12}$ and  
$\sin^2\theta_{23}$, and then on $\delta$, 
will be available.

\begin{figure}[t]
\centering
\includegraphics[width=10cm]{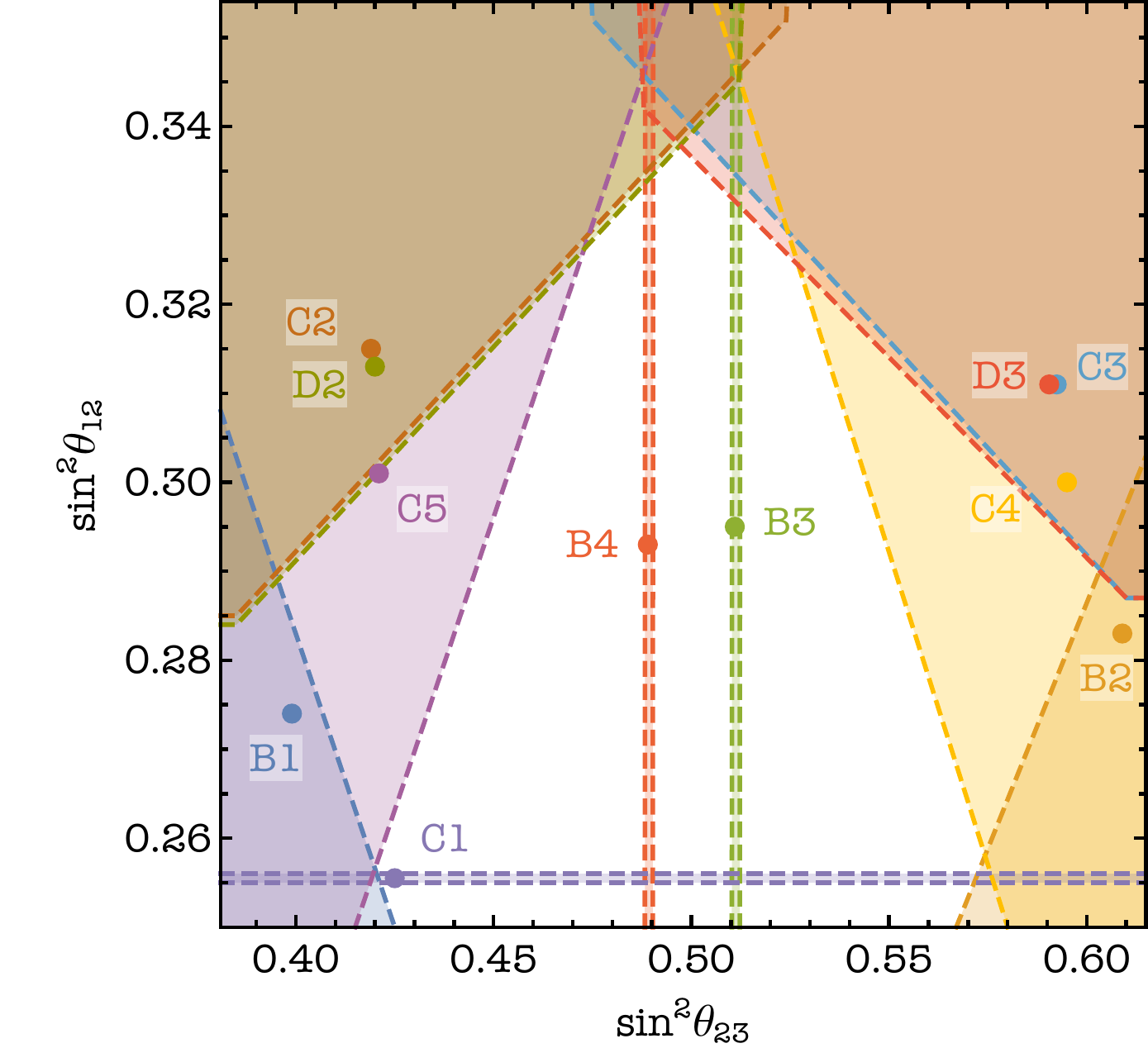}
\caption{Summary of the predicted allowed regions in the 
$(\sin^2\th_{23},\sin^2\th_{12})$ plane and the corresponding 
best fit points in cases B1\,--\,B4, C1\,--\,C5, D2 and D3 
for the NO neutrino mass spectrum. 
The values of all the three mixing angles are required to lie 
in their respective current $3\sigma$ ranges.}
\label{fig:regions}
\end{figure}

 We schematically summarise in Fig.~\ref{fig:regions} 
the predicted $3\sigma$
allowed regions in the plane $(\sin^2\th_{23},\sin^2\th_{12})$
for all currently viable cases from 
Figs.~\ref{fig:caseB1}\,--\,\ref{fig:caseD3}. 
In this figure we also present the best fit point in each case 
used in the preceding discussion.  
When future more precise data on $\sin^2\th_{23}$ and 
$\sin^2\th_{12}$ become available,
the experimentally allowed region in 
the $(\sin^2\th_{23},\sin^2\th_{12})$ plane will shrink, 
and only a limited number of cases, 
if any,  will remain viable. 
It will be possible to distinguish further               
between some or all of the remaining viable cases 
with a high precision measurement of $\delta$.
%
 
 Finally, we note that the sum rules for $\sin^2\th_{23}$ 
($\sin^2\th_{12}$ in case C1) and/or $\cos\delta$ obtained in the present study 
follow from those derived in \cite{Girardi:2015rwa} 
for certain values of the parameters $\sin^2\th^\circ_{ij}$, 
fixed by $G_f = S_4$ and the residual $Z_2^{g_e}$ and $Z_2^{g_\nu}$ 
flavour symmetries, 
and the additional constraints provided by the GCP symmetry $H^\nu_{\rm CP}$.
Note that in \cite{Girardi:2015rwa} only flavour symmetry, 
without imposing a GCP symmetry, has been considered. 
As we have seen in subsection~\ref{subsec:PMNS}, a GCP symmetry 
does not allow for a free phase $\delta^\nu$ coming from the neutrino sector, 
which is present otherwise. 
This, in turn, leads to the fact that in certain cases 
the parameter $\sin\hat\th^\nu_{ij}$ (see eq.~(213) in \cite{Girardi:2015rwa}), 
which is free in \cite{Girardi:2015rwa}, gets fixed by the GCP symmetry. 
Thus, we find additional correlations between 
$\th_{ij}$ and between $\th_{ij}$ and $\cos\delta$ in these cases.
We provide the correspondence between the phenomenologically viable cases 
of the present study and the cases considered in \cite{Girardi:2015rwa} 
in Appendix~\ref{app:sumrulescorrespondence}.

\begin{landscape}
\thispagestyle{empty}
\begin{table}
\centering
\renewcommand{\arraystretch}{1.2}
\begin{tabular}{cccccccc} 
\toprule
 \multirow{2}{*}{$H^\nu_{\rm CP}$}
& Case
& \multirow{2}{*}{$\dfrac{\sin^2\theta_{12}}{10^{-1}}$}
& \multirow{2}{*}{$\dfrac{\sin^2\theta_{13}}{10^{-2}}$}
& \multirow{2}{*}{$\dfrac{\sin^2\theta_{23}}{10^{-1}}$}
& \multirow{2}{*}{$\delta/\pi$}
&  $\alpha_{21}/\pi$ &  $\alpha_{31}/\pi$ 
\\ 
& (p.f.e.) & & & & &(mod 1) &(mod 1)  \\ 
\midrule
 \multirow{4}{*}[-1mm]{$\{1,S\}$}
& A1
& $3.45 - 3.54$
& $2.13 - 2.40$
& $5.11 - 5.12$
& $0 - 0.11 \oplus 1.89 - 2$
& $0 - 0.07 \oplus 0.93 - 1$
& $0-1$
\\ 
& $(\mu 3)$
& $3.44-3.54$ 
& $2.13 - 2.42$
& $5.11 - 5.12$
& $0 - 0.12 \oplus 1.88 - 2$
& $0 - 0.07 \oplus 0.93 - 1$
& $0-1$
\\
\cmidrule(lr){2-8}
& A2
& $3.45 - 3.54$
& $2.13 - 2.40$
& $4.88 - 4.89$
& $0.89 - 1.11$
& $0 - 0.07 \oplus 0.93 - 1$
& $0-1$
\\ 
& $(\tau 3)$
& $3.44 - 3.54$ 
& $2.13 - 2.42$
& $4.88 - 4.89$
& $0.88 - 1.12$
& $0 - 0.07 \oplus 0.93 - 1$
& $0-1$
\\
\midrule
 \multirow{8}{*}[-2.7mm]{$\{U,SU\}$}
& B1
& $2.50 - 3.08$
& Full $3\sigma$
& $3.81 - 4.25$
& $0.68 - 1.32$
& $0 - 0.16 \oplus 0.84 - 1$
& $0 - 0.13 \oplus 0.88 - 1$
\\ 
& $(\mu 2)$
& $2.50 - 3.06$
& Full $3\sigma$
& $3.84 - 4.25$
& $0.69 - 1.31$
& $0 - 0.16 \oplus 0.84 - 1$
& $0 - 0.12 \oplus 0.88 - 1$
\\
\cmidrule(lr){2-8}
& B2
& $2.50 - 3.03$
& Full $3\sigma$
& $5.76 - 6.15$
& $0 - 0.30 \oplus 1.70 - 2$
& $0 - 0.16 \oplus 0.84 - 1$
& $0 - 0.12 \oplus 0.88 - 1$
\\ 
& $(\tau 2)$
& $2.50-3.28$ 
& Full $3\sigma$
& $5.76 - 6.36$
& $0 - 0.38 \oplus 1.61 - 2$
& $0 - 0.17 \oplus 0.83 - 1$
& $0 - 0.13 \oplus 0.87 - 1$
\\
\cmidrule(lr){2-8}
& B3
& Full $3\sigma$
& Full $3\sigma$
& $5.10 - 5.12$
& $0 - 2$
& $0 - 0.23 \oplus 0.77 - 1$
& $0 - 0.18 \oplus 0.83 - 1$
\\ 
& $(\mu 3)$
& Full $3\sigma$
& Full $3\sigma$
& $5.10 - 5.12$
& $0 - 2$
& $0 - 0.23 \oplus 0.77 - 1$
& $0 - 0.18 \oplus 0.82 - 1$
 \\
\cmidrule(lr){2-8}
& B4
& Full $3\sigma$
& Full $3\sigma$
& $4.88 - 4.90$
& $0 - 2$
& $0 - 0.23 \oplus 0.77 - 1$
& $0 - 0.17 \oplus 0.83 - 1$
\\ 
& $(\tau 3)$
& Full $3\sigma$ 
& Full $3\sigma$
& $4.88 - 4.90$
& $0 - 2$
& $0 - 0.23 \oplus 0.77 - 1$
& $0 - 0.18 \oplus 0.82 - 1$\\
\bottomrule
\end{tabular}
\caption{Ranges of the mixing parameters 
for the viable cases, i.e., 
those cases for which 
the predicted values of all the three mixing angles
lie inside their respective $3\sigma$ allowed ranges.
The cases presented here
correspond to 
$G_e = Z_2^{g_e}$ and $G_\nu = Z_2^{g_\nu} \times H^\nu_{\rm CP}$ 
with $\{g_e,g_\nu\} = \{TU,S\}$,
for which the magnitude of the fixed element is $1/\sqrt{2}$ 
(p.f.e.~denotes its position in $U_{\rm PMNS}$).  
For each case, the upper and lower rows
refer to NO and IO, respectively.}
\label{tab:rangesAB}
\end{table}
%
%
%
%
\clearpage
\thispagestyle{empty}
\begin{table}
\centering
\renewcommand{\arraystretch}{1.2}
\begin{tabular}{cccccccc} 
\toprule
 \multirow{2}{*}{$H^\nu_{\rm CP}$}
& Case
& \multirow{2}{*}{$\dfrac{\sin^2\theta_{12}}{10^{-1}}$}
& \multirow{2}{*}{$\dfrac{\sin^2\theta_{13}}{10^{-2}}$}
& \multirow{2}{*}{$\dfrac{\sin^2\theta_{23}}{10^{-1}}$}
& \multirow{2}{*}{$\delta/\pi$}
&  $\alpha_{21}/\pi$ &  $\alpha_{31}/\pi$ 
\\ 
& (p.f.e.) & & & & &(mod 1) &(mod 1) \\ 
\midrule
\multirow{10}{*}[-3.5mm]{$\{1,U\}$}
& C1
& $2.55 - 2.56$
& Full $3\sigma$
& Full $3\sigma$
& $0 - 2$
& $0$ (exactly)
& $0 - 1$
\\ 
& $(e 2)$
& $2.55 - 2.56$
& Full $3\sigma$
& Full $3\sigma$
& $0 - 2$
& $0$ (exactly)
& $0 - 1$
\\
\cmidrule(lr){2-8}
& C2
& $2.85 - 3.54$
& Full $3\sigma$
& $3.81 - 5.24$
& $0 - 0.38 \oplus 1.62 - 2$
& $0 - 0.15 \oplus 0.85 - 1$
& $0 - 0.09 \oplus 0.91 - 1$
\\ 
& $(\mu 1)$
& $2.86 - 3.54$
& Full $3\sigma$
& $3.84 - 5.25$
& $0 - 0.37 \oplus 1.63 - 2$
& $0 - 0.15 \oplus 0.85 - 1$
& $0 - 0.09 \oplus 0.91 - 1$
\\
\cmidrule(lr){2-8}
& C3
& $2.87 - 3.54$
& Full $3\sigma$
& $4.75 - 6.15$
& $0.63 - 1.37$
& $0 - 0.15 \oplus 0.86 - 1$
& $0 - 0.09 \oplus 0.91 - 1$
\\ 
& $(\tau 1)$
& $2.79 - 3.54$
& Full $3\sigma$
& $4.75 - 6.36$

& $0.60 - 1.40$
& $0 - 0.15 \oplus 0.85 - 1$
& $0 - 0.09 \oplus 0.91 - 1$
\\
\cmidrule(lr){2-8}
& C4
& Full $3\sigma$
& Full $3\sigma$
& $5.06 - 6.15$
& $0 - 0.48 \oplus 1.52 - 2$
& $0 - 0.25 \oplus 0.75 - 1$
& $0 - 0.16 \oplus 0.84 - 1$
\\ 
& $(\mu 2)$
& Full $3\sigma$
& Full $3\sigma$
& $5.05 - 6.36$
& $0 - 0.54 \oplus 1.45 - 2$
& $0 - 0.27 \oplus 0.73 - 1$
& $0 - 0.18 \oplus 0.82 - 1$
\\
\cmidrule(lr){2-8}
& C5
& Full $3\sigma$
& Full $3\sigma$
& $3.81 - 4.94$
& $0.51 - 1.49$
& $0 - 0.26 \oplus 0.74 - 1$
& $0 - 0.17 \oplus 0.84 - 1$
\\ 
& $(\tau 2)$
& Full $3\sigma$
& Full $3\sigma$
& $3.84 - 4.94$
& $0.52 - 1.48$
& $0 - 0.25 \oplus 0.74 - 1$
& $0 - 0.16 \oplus 0.84 - 1$
\\
\midrule
\multirow{8}{*}[-3mm]{$\{S,SU\}$}
& D2
& $2.84 - 3.54$
& Full $3\sigma$
& $3.81 - 5.12$
& $0 - 0.38 \oplus 1.63 - 2$
& $0 - 1$
& $0 - 0.48 \oplus 0.52 - 1$
\\ 
& $(\mu 1)$
& $2.85 - 3.54$
& Full $3\sigma$
& $3.84 - 5.12$
& $0 - 0.37 \oplus 1.63 - 2$
& $0 - 1$
& $0 - 0.48 \oplus 0.52 - 1$
\\
\cmidrule(lr){2-8}
& D3
& $2.87 - 3.54$
& Full $3\sigma$
& $4.88 - 6.15$
& $0.63 - 1.37$
& $0 - 1$
& $0 - 0.47 \oplus 0.52 - 1$
\\ 
& $(\tau 1)$
& $2.79 - 3.54$
& Full $3\sigma$
& $4.88 - 6.36$
& $0.59 - 1.41$
& $0 - 1$
& $0 - 1$
\\
\cmidrule(lr){2-8}
& D4
& $3.45 - 3.54$
& $2.14 - 2.40$
& $5.05 - 5.12$
& $0 - 0.11 \oplus 1.89 - 2$
& $0 - 0.16 \oplus 0.83 - 1$
& $0 - 0.08 \oplus 0.92 - 1$
\\ 
& $(\mu 2)$
& $3.45 - 3.54$
& $2.14 - 2.42$
& $5.05 - 5.12$
& $0 - 0.11 \oplus 1.89 - 2$
& $0 - 0.17 \oplus 0.83 - 1$
& $0 - 0.08 \oplus 0.91 - 1$
\\
\cmidrule(lr){2-8}
& D5
& $3.45 - 3.54$
& $2.13 - 2.40$
& $4.88 - 4.95$
& $0.89 - 1.11$
& $0 - 0.16 \oplus 0.83 - 1$
& $0 - 0.08 \oplus 0.92 - 1$
\\ 
& $(\tau 2)$
& $3.45 - 3.54$
& $2.13 - 2.42$
& $4.88 - 4.95$
& $0.88 - 1.11$
& $0 - 0.17 \oplus 0.83 - 1$
& $0 - 0.09 \oplus 0.91 - 1$
\\
\bottomrule
\end{tabular}
\caption{The same as in Table~\ref{tab:rangesAB}, but for 
$G_e = Z_2^{g_e}$ and $G_\nu = Z_2^{g_\nu} \times H^\nu_{\rm CP}$ with 
$\{g_e,g_\nu\} = \{TU,U\}$. In this case the magnitude of the fixed element is $1/2$.}
\label{tab:rangesCD}
\end{table}
\end{landscape}
%
%
%
%
\begin{table}
\centering
\renewcommand{\arraystretch}{1.2}
\begin{tabular}{ccccccccc}
\toprule
 \multirow{2}{*}{$H^\nu_{\rm CP}$}
& Case
& \multirow{2}{*}{$\dfrac{\sin^2\theta_{12}}{10^{-1}}$}
& \multirow{2}{*}{$\dfrac{\sin^2\theta_{13}}{10^{-2}}$}
& \multirow{2}{*}{$\dfrac{\sin^2\theta_{23}}{10^{-1}}$}
& \multirow{2}{*}{$\delta/\pi$}
&  $\alpha_{21}/\pi$ &  $\alpha_{31}/\pi$ 
& \multirow{2}{*}{$\chi^2_{\rm min}$} \\ 
& (p.f.e.) & & & & &(mod 1) &(mod 1) & \\ 
\midrule
 \multirow{4}{*}[-1mm]{$\{1,S\}$}
& A1
& $3.54$
& $2.18$
& $5.11$
& $1.96$
& $0.97$
& $0.43$
& $22.0$ \\ 
& $(\mu 3)$
& $3.53$ 
& $2.19$
& $5.11$
& $1.95$
& $0.97$
& $0.89$
& $19.0$ \\
\cmidrule(lr){2-9}
& A2 
& $3.54$
& $2.18$
& $4.89$
& $1.05$
& $0.03$
& $0.01$
& $18.5$ \\ 
& $(\tau 3)$
& $3.53$
& $2.20$
& $4.89$
& $1.04$
& $0.02$
& $0.67$
& $15.0$ \\
\midrule
 \multirow{8}{*}[-2.7mm]{$\{U,SU\}$}
& B1
& $2.74$
& $2.17$
& $3.99$
& $1.09$
& $0.94$
& $0.96$
& $7.0$ \\ 
& $(\mu 2)$
& $2.75$
& $2.18$
& $4.01$
& $1.07$
& $0.96$
& $0.97$
& $7.0$ \\
\cmidrule(lr){2-9}
& B2
& $2.83$
& $2.17$
& $6.09$
& $1.89$
& $0.07$
& $0.05$
& $12.5$ \\ 
& $(\tau 2)$
& $2.83$
& $2.17$
& $6.09$
& $1.89$
& $0.07$
& $0.05$
& $6.0$ \\
\cmidrule(lr){2-9}
& B3
& $2.95$
& $2.15$
& $5.11$
& $1.36$
& $0.80$
& $0.85$
& $8.5$ \\ 
& $(\mu 3)$
& $2.95$
& $2.15$
& $5.11$
& $1.36$
& $0.80$
& $0.85$
& $5.0$ \\
\cmidrule(lr){2-9}
& B4
& $2.93$
& $2.16$
& $4.89$
& $1.38$
& $0.19$
& $0.13$
& $6.5$ \\ 
& $(\tau 3)$
& $2.97$
& $2.16$
& $4.89$
& $1.31$
& $0.16$
& $0.11$
& $4.5$ \\
\bottomrule
\end{tabular}
\caption{Best fit values of the mixing parameters 
and the corresponding value of the $\chi^2$ function, 
$\chi^2_{\rm min}$, for the viable cases, i.e., 
those cases for which the predicted values of all the three mixing angles
lie inside their respective $3\sigma$ allowed ranges.
The cases presented here
correspond to  
$G_e = Z_2^{g_e}$ and $G_\nu = Z_2^{g_\nu} \times H^\nu_{\rm CP}$
with $\{g_e,g_\nu\} = \{TU,S\}$,
for which the magnitude of the fixed element is $1/\sqrt{2}$ 
(p.f.e.~denotes its position in $U_{\rm PMNS}$).  
For each case, the upper and lower rows
refer to NO and IO, respectively.}
\label{tab:bestfitAB}
\end{table}
%
%
%
%
\begin{table}
\centering
\renewcommand{\arraystretch}{1.2}
\begin{tabular}{ccccccccc}
\toprule
 \multirow{2}{*}{$H^\nu_{\rm CP}$}
& Case
& \multirow{2}{*}{$\dfrac{\sin^2\theta_{12}}{10^{-1}}$}
& \multirow{2}{*}{$\dfrac{\sin^2\theta_{13}}{10^{-2}}$}
& \multirow{2}{*}{$\dfrac{\sin^2\theta_{23}}{10^{-1}}$}
& \multirow{2}{*}{$\delta/\pi$}
&  $\alpha_{21}/\pi$ &  $\alpha_{31}/\pi$ 
& \multirow{2}{*}{$\chi^2_{\rm min}$} \\ 
& (p.f.e.) & & & & &(mod 1) &(mod 1) & \\ 
\midrule
\multirow{10}{*}[-3.5mm]{$\{1,U\}$}
& C1
& $2.56$
& $2.16$
& $4.25$
& $1.32$
& $0$
& $0.64$
& $7.0$ \\ 
& $(e 2)$
& $2.56$
& $2.16$
& $5.85$
& $1.36$
& $0$
& $0.73$
& $7.0$ \\
\cmidrule(lr){2-9}
& C2
& $3.15$
& $2.16$
& $4.19$
& $1.86$
& $0.93$
& $0.96$
& $4.5$ \\ 
& $(\mu 1)$
& $3.14$
& $2.16$
& $4.24$
& $1.88$
& $0.94$
& $0.96$
& $5.5$ \\
\cmidrule(lr){2-9}
& C3
& $3.11$
& $2.16$
& $5.92$
& $1.15$
& $0.07$
& $0.05$
& $8.5$ \\ 
& $(\tau 1)$
& $3.08$
& $2.17$
& $5.93$
& $1.13$
& $0.06$
& $0.04$
& $1.5$ \\
\cmidrule(lr){2-9}
& C4
& $3.00$
& $2.14$
& $5.95$
& $1.69$
& $0.81$
& $0.88$
& $8.5$ \\ 
& $(\mu 2)$
& $3.00$
& $2.14$
& $5.95$
& $1.69$
& $0.81$
& $0.88$
& $2.0$ \\
\cmidrule(lr){2-9}
& C5
& $3.01$
& $2.15$
& $4.21$
& $1.25$
& $0.15$
& $0.10$
& $0.5$ \\ 
& $(\tau 2)$
& $2.99$
& $2.17$
& $4.26$
& $1.22$
& $0.13$
& $0.09$
& $0.5$ \\
\midrule
\multirow{8}{*}[-3mm]{$\{S,SU\}$}
& D2
& $3.13$
& $2.15$
& $4.20$
& $1.88$
& $0.43$
& $0.65$
& $4.5$ \\ 
& $(\mu 1)$
& $3.15$
& $2.17$
& $4.23$
& $1.87$
& $0.43$
& $0.66$
& $5.5$ \\
\cmidrule(lr){2-9}
& D3
& $3.11$
& $2.17$
& $5.91$
& $1.14$
& $0.61$
& $0.38$
& $8.5$ \\ 
& $(\tau 1)$
& $3.06$
& $2.16$
& $5.96$
& $1.12$
& $0.50$
& $0.69$
& $1.5$ \\
\cmidrule(lr){2-9}
& D4
& $3.54$
& $2.18$
& $5.11$
& $1.96$
& $0.97$
& $0.98$
& $22.0$ \\ 
& $(\mu 2)$
& $3.53$
& $2.20$
& $5.11$
& $1.95$
& $0.97$
& $0.98$
& $19.0$ \\
\cmidrule(lr){2-9}
& D5
& $3.54$
& $2.19$
& $4.89$
& $1.05$
& $0.03$
& $0.02$
& $18.5$ \\ 
& $(\tau 2)$
& $3.53$
& $2.19$
& $4.89$
& $1.04$
& $0.03$
& $0.01$
& $15.0$ \\
\bottomrule
\end{tabular}
\caption{The same as in Table~\ref{tab:bestfitAB}, but for 
$G_e = Z_2^{g_e}$ and $G_\nu = Z_2^{g_\nu} \times H^\nu_{\rm CP}$ with
$\{g_e,g_\nu\} = \{TU,U\}$. In this case the magnitude of the fixed element is $1/2$.}
\label{tab:bestfitCD}
\end{table}
%
%
%
%
\begin{table}
\centering
\renewcommand{\arraystretch}{1.2}
\begin{tabular}{ccccccccccccccccc}
\toprule
  &  & A1 & A2 & B1 & B2 & B3 & B4 & C1 & C2 & C3 & C4 & C5 & D2 & D3 & D4 & D5 \\ 
\midrule
\multirow{2}{*}{\text{3$\sigma $}} & \text{NO} & \cmark & \cmark & \cmark & \cmark & \cmark & \cmark & \cmark & \cmark & \cmark & \cmark & \cmark & \cmark & \cmark & \cmark & \cmark \\
& \text{IO} & \cmark & \cmark & \cmark & \cmark & \cmark & \cmark & \cmark & \cmark & \cmark & \cmark & \cmark & \cmark & \cmark & \cmark & \cmark \\ 
\midrule
\multirow{2}{*}{\text{2$\sigma $}} & \text{NO} & \xmark & \xmark & \cmark & \xmark & \xmark & \xmark & \xmark & \cmark & \xmark & \xmark & \cmark & \cmark & \xmark & \xmark & \xmark \\
 & \text{IO} & \xmark & \xmark & \cmark & \cmark & \xmark & \xmark & \xmark & \cmark & \cmark & \cmark & \cmark & \cmark & \cmark & \xmark & \xmark \\  
\bottomrule
\end{tabular}
\caption{Compatibility of the cases under consideration with the $3\sigma$ and $2\sigma$ 
experimentally allowed ranges of the three neutrino mixing angles for both 
types of the neutrino mass spectrum.}
\label{tab:checks}
\end{table}

\clearpage
\section{Neutrinoless Double Beta Decay}
\label{sec:bb0nu}

 As we have seen, in the class of models investigated 
in the present article the Dirac and Majorana CPV phases, 
$\delta$ and $\alpha_{21}$, $\alpha_{31}$, 
are (statistically) predicted to lie in specific, 
in most cases relatively narrow, 
intervals and their values are 
strongly correlated. 
The only exception is case C1, in which 
the exact predictions $\alpha_{21} = 0$ or $\pi$ and 
$(\alpha_{31} - 2\delta) = 0$ or $\pi$ hold.

   These results make it possible to derive predictions 
for the absolute value of the 
neutrinoless double beta ($\betabeta$-) decay 
effective Majorana mass, $\mefff$
(see, e.g., refs.~\cite{Olive:2016xmw,Bilenky:1987ty,bb0nuth}), 
as a function of the lightest neutrino mass.
As is well known, information about $\meff$ is provided
by the experiments on $\betabeta$-decay
of even-even nuclei $^{48}Ca$, $^{76}Ge$, 
$^{82}Se$, $^{100}Mo$, $^{116}Cd$, $^{130}Te$, 
$^{136}Xe$, $^{150}Nd$, etc., 
$(A,Z) \rightarrow (A,Z+2) + e^- + e^-$,
in which the total lepton charge
changes by two units, and through
the observation of which the possible
Majorana nature of massive neutrinos can be revealed.
If the light neutrinos with definite mass $\nu_j$ 
are Majorana fermions, their exchange
between two neutrons of the initial nucleus $(A,Z)$ 
can trigger the process of  $\betabeta$-decay.
In  this case the $\betabeta$-decay amplitude has the 
following general form (see, e.g., refs.~\cite{Bilenky:1987ty,bb0nuth}):
$A(\betabeta) = G^2_{\rm F}\, \mefff\,M(A,Z)$, with 
$G_{\rm F}$,  $\mefff$ and $M(A,Z)$ being respectively  
the Fermi constant, 
the  $\betabeta$-decay effective Majorana mass and 
the nuclear matrix element (NME) of the process.
All the dependence of  $A(\betabeta)$ on
the neutrino mixing parameters is contained in $\mefff$. 
The current best limits on  $\meff$ have been obtained  
by the KamLAND-Zen \cite{KamLAND-Zen:2016pfg}
and GERDA Phase II \cite{Agostini:2017iyd} 
experiments searching for $\betabeta$-decay 
of $^{136}Xe$ and  $^{76}Ge$, respectively:
\begin{equation}
\meff < (0.061 - 0.165)~{\rm eV}~\text{\cite{KamLAND-Zen:2016pfg}}\,\quad
{\rm and}\quad
\meff < (0.15 - 0.33)~{\rm eV}~\text{\cite{Agostini:2017iyd}}\,, 
\label{eq:mefflimits}
\end{equation}
%
both at $90\%$ C.L., 
where the intervals reflect the
estimated uncertainties in the relevant
NMEs used to extract the
limits on $\meff$ from the experimentally obtained lower
bounds on the $^{136}Xe$ and  $^{76}Ge$ $\betabeta$-decay
half-lives (for a review of
the limits on $\meff$ obtained in other $\betabeta$-decay 
experiments and a detailed discussion of the 
NME calculations for $\betabeta$-decay and
their uncertainties see, e.g., \cite{Vergados:2016hso}).
It is important to note that a large number of 
experiments of a new generation aims at a 
sensitivity to $\meff \sim (0.01 - 0.05)$~eV,
which will allow to probe the whole
range of the predictions for $\meff$ in the case of 
IO neutrino mass spectrum \cite{Pascoli:2002xq} 
(see, e.g., \cite{Vergados:2016hso,DellOro:2016tmg}  for 
reviews of the currently running and future planned
$\betabeta$-decay experiments and their prospective
sensitivities).

The predictions for $\meff$ 
(see, e.g., \cite{Bilenky:1987ty,Bilenky:2001rz,bb0nuth}), 
\begin{align}
\meff &= \left | \sum_{i=1}^3 m_i U_{ei}^2 \right| \nonumber \\
&= \left| m_1 \cos^2\theta_{12} \cos^2\theta_{13} 
+ m_2 \sin^2\theta_{12} \cos^2\theta_{13} e^{i \alpha_{21}}
+ m_3 \sin^2 \theta_{13} e^{i \left(\alpha_{31} - 2 \delta\right)} \right|\,,
\label{meff}
\end{align}
%
$m_{1,2,3}$ being the light Majorana neutrino masses,
depend on the values of the Majorana phase $\alpha_{21}$ 
and on the Majorana-Dirac phase difference 
$(\alpha_{31} - 2\delta)$.
For the normal hierarchical (NH),
inverted hierarchical (IH) and quasi-degenerate (QD), 
neutrino mass spectra
$\meff$ is given by (see, e.g., \cite{Olive:2016xmw,Petcov:2016ovu}):
\begin{align}
\label{NH}
&\meff\cong \left|\sqrt{\Delta m^2_{21}}\, \sin^2\th_{12} \cos^2\th_{13}\, e^{i\alpha_{21}}
+ \sqrt{\Delta m^2_{31}}\, \sin^2\th_{13}\, e^{i(\alpha_{31}-2\delta)}\right|
\quad
\text{(NH)}\,,\\
\label{IH}
&\meff \cong \sqrt{\Delta m^2_{23}}\, \cos^2\th_{13}
\left|\cos^2\th_{12} + \sin^2\th_{12}\,e^{i\alpha_{21}}\right|
\quad
\text{(IH)},\\[0.30cm]
\label{QD}
&\meff \cong m_0\,
\left|\cos^2\th_{12} + \sin^2\th_{12}\,e^{i \alpha_{21}}\right|
\quad
\text{(QD)}\,,
\end{align}
%
where $m_0 \cong m_{1,2,3}$.
We recall that the NH spectrum corresponds to 
$m_1 \ll m_2 < m_3$, and thus, $m_2 = (m_1^2 + \Delta m^2_{21})^{1\over{2}} 
\cong (\Delta m^2_{21})^{1\over{2}}
\cong 8.6\times 10^{-3}$~eV,
$m_3 =  (m_3^2 + \Delta m^2_{31})^{1\over{2}}
 \cong (\Delta m^2_{31})^{1\over{2}} \cong 0.0506$~eV.
The IH spectrum corresponds to 
$m_3 \ll m_1 < m_2$, and therefore, 
$m_1 = (m_3^2 + \Delta m^2_{23} - \Delta m^2_{21})^{1\over{2}}
 \cong (\Delta m^2_{23} -  \Delta m^2_{21})^{1\over{2}} \cong 0.0497$~eV,
$m_{2}= (m_3^2 + \Delta m^2_{23})^{1\over{2}}\cong 
(\Delta m^2_{23})^{1\over{2}}\cong 0.0504$ eV.
In the case of QD spectrum we have: 
$m_{1} \cong m_{2}\cong m_{3}\cong m_0$,  
$m_j^2 \gg \Delta m^2_{31(23)}$, $m_0 \gtap 0.10$ eV.
In eqs.~(\ref{NH}) and (\ref{IH}) we have 
assumed that the contributions respectively 
$\propto m_1$ and $\propto m_3$
are negligible, while in eq.~(\ref{QD}) we have neglected 
corrections $\propto \sin^2\theta_{13}$~%
\footnote{The term  $\propto \sin^2\theta_{13}$ gives a 
subleading contribution because even in the case of 
$\alpha_{21} = \pi$, when the leading term  
$\propto (\cos^2\th_{12} - \sin^2\th_{12})$ has a minimal value, 
$\sin^2\theta_{13} \ll \cos2\theta_{12}$ since 
$\sin^2\theta_{13} \leq 0.0242 $ while 
$\cos2\theta_{12} \geq 0.29$ at $3\sigma$. 
} 
and $\propto \Delta m^2_{31(23)}/m^2_0$. 
Clearly, the values of the phases $(\alpha_{31} - \alpha_{21} - 2\delta)$ 
and  $\alpha_{21}$ determine the ranges of possible values of 
$\meff$ in the cases of NH and IH (QD) spectra, respectively.
Using the $3\sigma$ ranges of the allowed values of 
the neutrino oscillation parameters 
from Table~\ref{tab:parameters}, we find that:
\begin{enumerate}[i)]
\item 
$0.79\times 10^{-3}~{\rm eV} \ltap \meff \ltap 4.33\times 10^{-3}$~eV 
in the case of NH spectrum;
\item 
$\sqrt{\Delta m^2_{23}}\,\cos^2\th_{13} \cos2\theta_{12} \ltap \meff \ltap \sqrt{\Delta m^2_{23}}\,\cos^2\th_{13}$, or 
$1.4\times 10^{-2}~{\rm eV} \ltap \meff  
\ltap 5.1\times 10^{-2}$~eV in the case of IH spectrum; 
\item
$m_0\,\cos2\theta_{12}\ltap \meff \ltap m_0$, or 
$2.9\times 10^{-2}~{\rm eV} \ltap \meff \ltap m_0$~eV, $m_0\gtap 0.10$~eV,
in the case of QD spectrum, 
where we have used the fact that at $3\sigma$ C.L., 
$\cos2\theta_{12} \geq 0.29$. 
\end{enumerate}
%

\begin{figure}
\vspace{-2.5cm}
\centering
\includegraphics[width=0.49\textwidth]{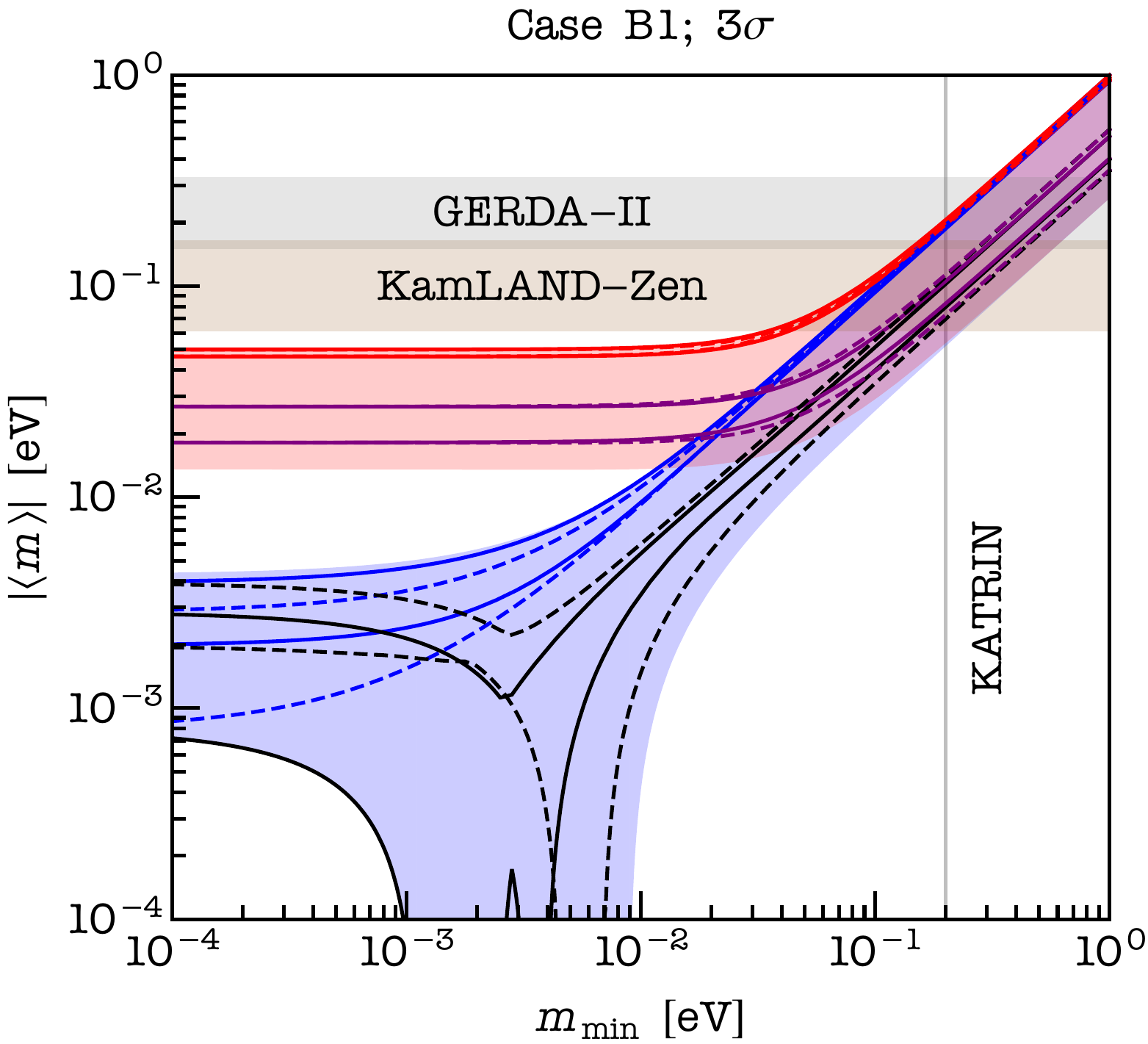}
\includegraphics[width=0.49\textwidth]{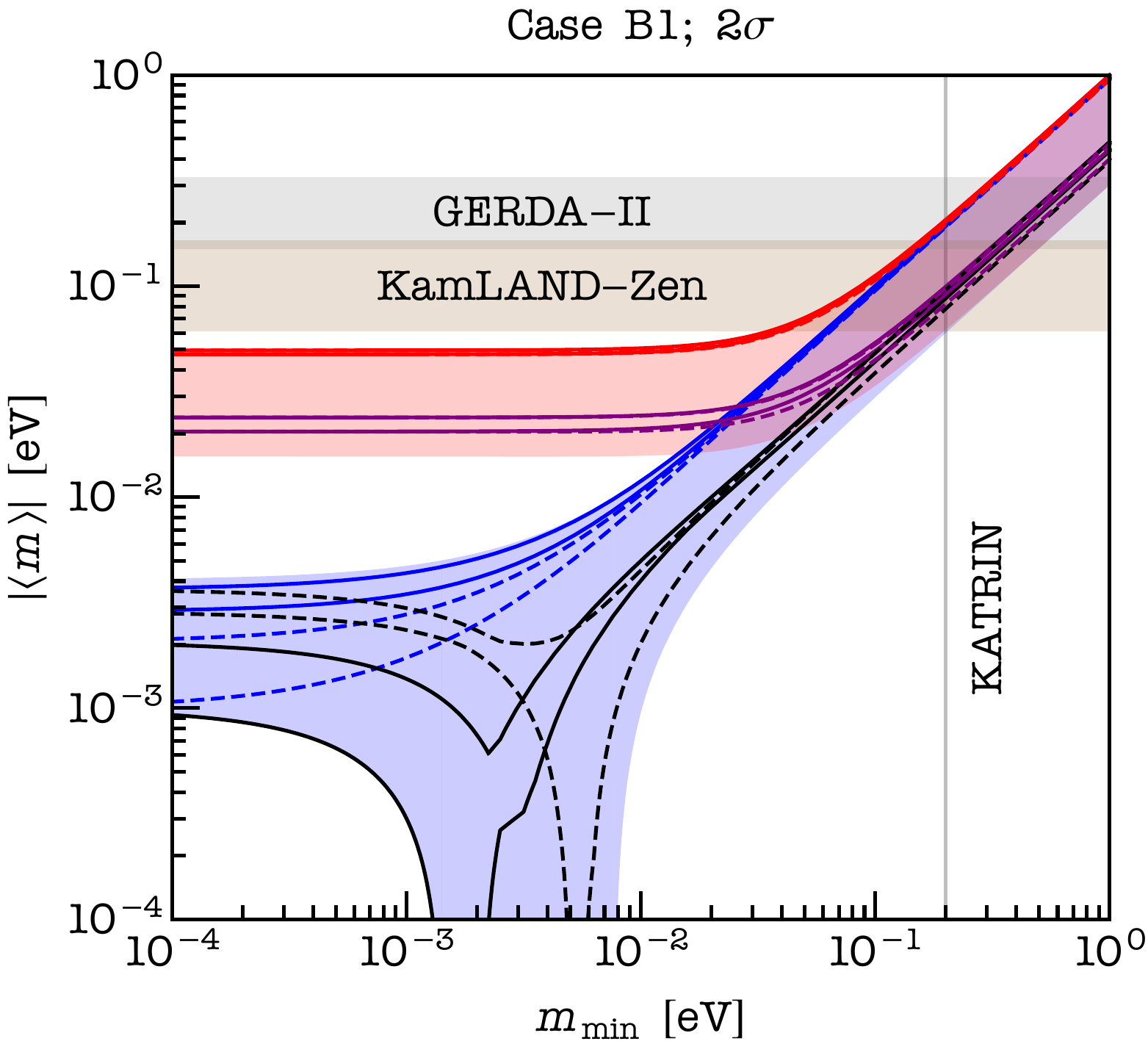}\\[6mm]
\includegraphics[width=0.49\textwidth]{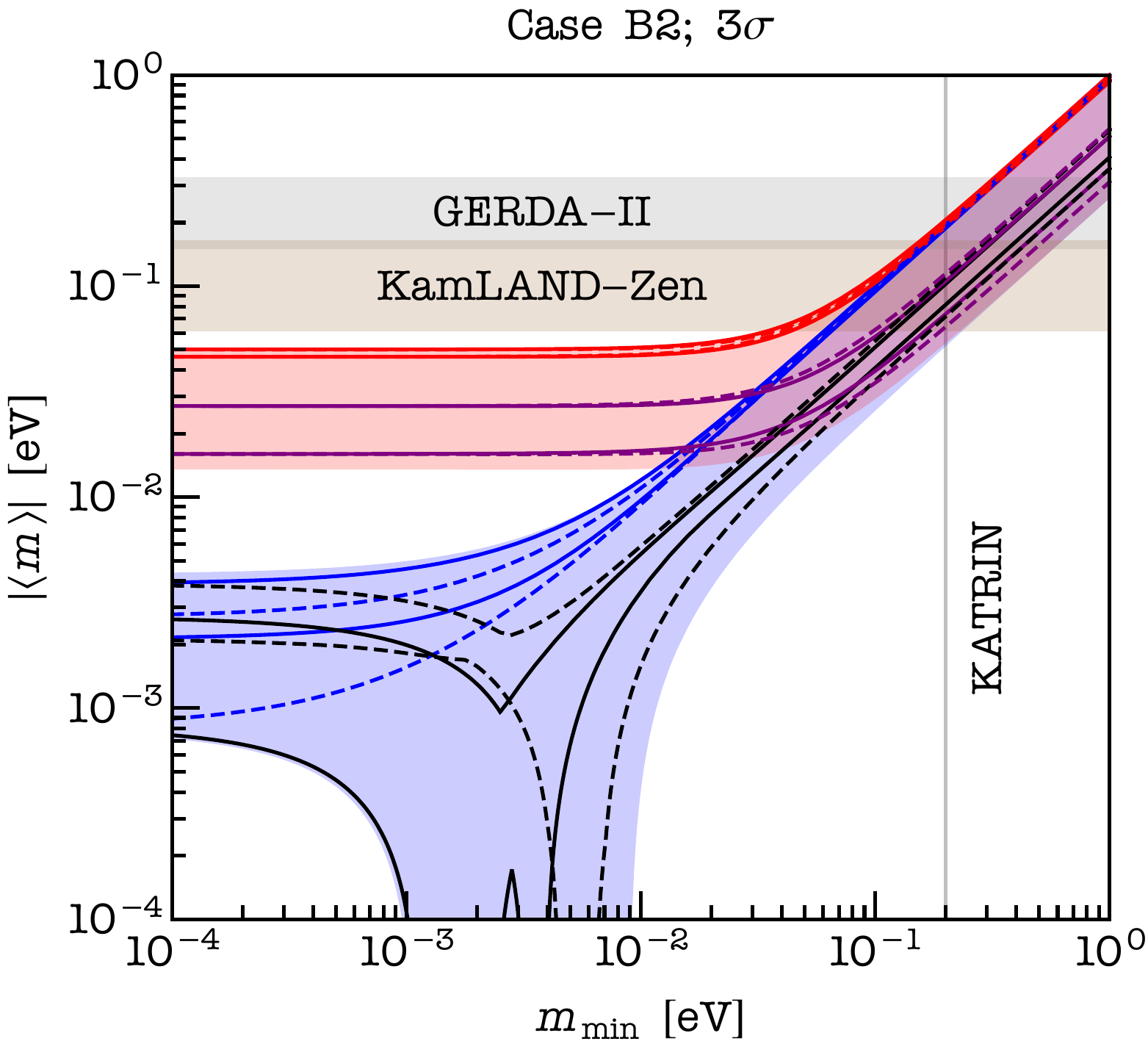}
\includegraphics[width=0.49\textwidth]{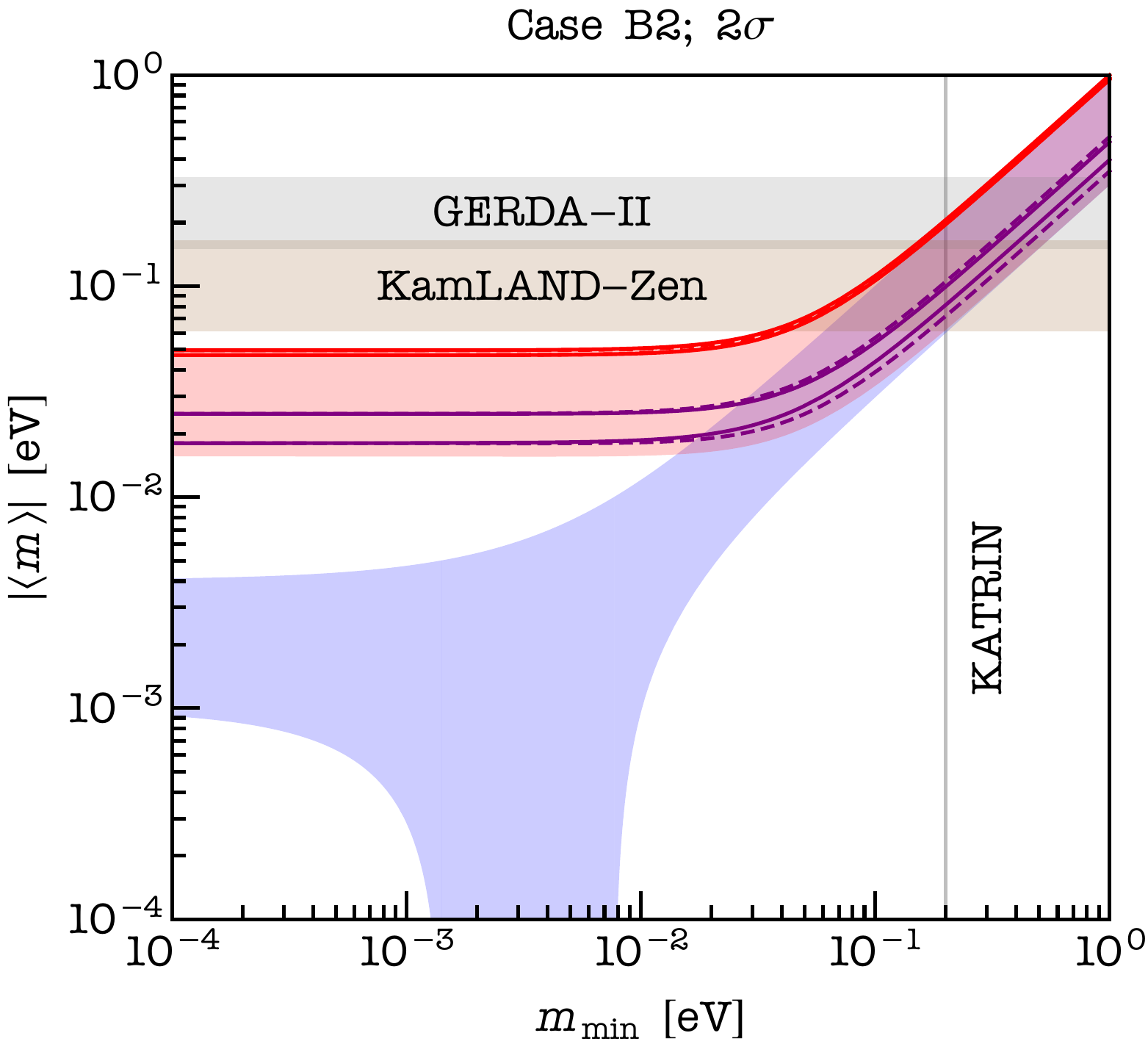}\\[6mm]
\includegraphics[width=0.49\textwidth]{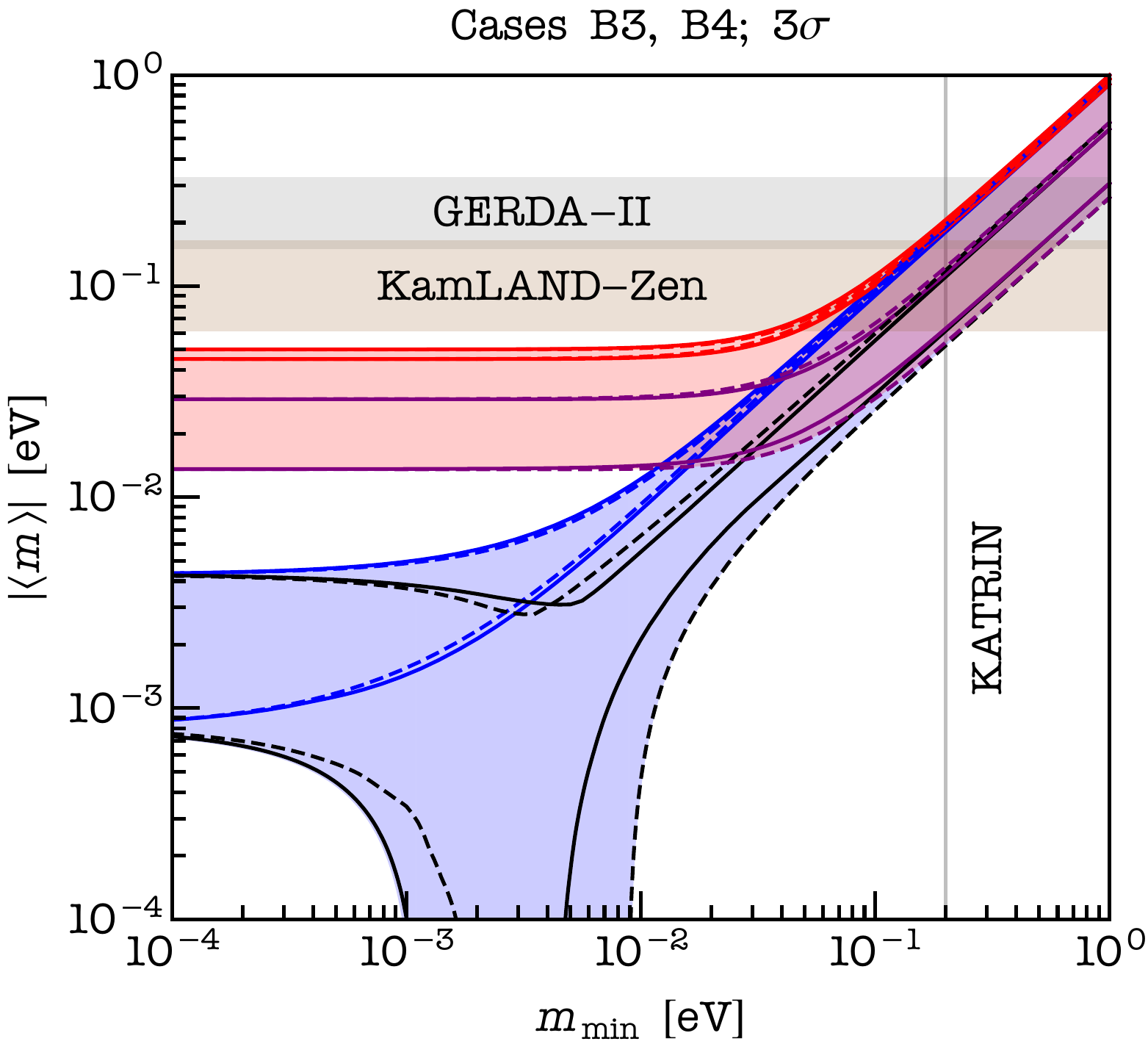}
\includegraphics[width=0.49\textwidth]{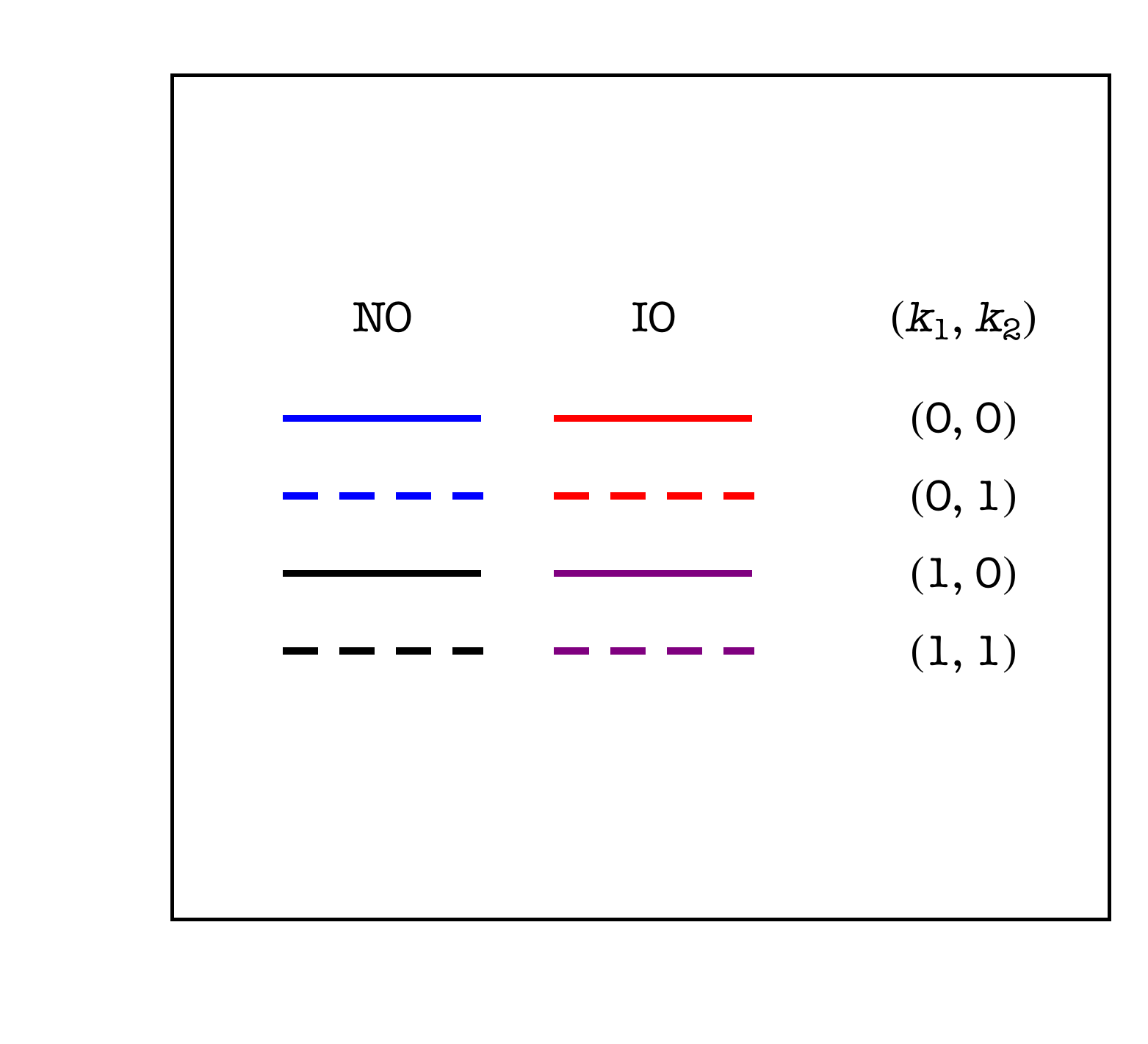}
\caption{The magnitude of the effective Majorana mass $\meff$ 
versus the lightest neutrino mass $m_{\rm min}$. 
The {\it lines} limit the allowed regions of 
$\meff$ calculated using the predictions for the relevant mixing angles 
and the CPV phases obtained in cases B1\,--\,B4 
and compatible with the $3\sigma$ ({\it left panels}) and $2\sigma$ ({\it right panels}) 
ranges of all the three mixing angles. 
The {\it light-blue} ({\it light-red}) {\it areas} are obtained varying 
the neutrino oscillation parameters 
$\theta_{12}$, $\theta_{13}$, $\Delta m^2_{21}$ and $\Delta m^2_{31(23)}$ for NO (IO)
in their allowed $3\sigma$ and $2\sigma$ ranges in 
the left and right panels, respectively,   
and the phases $\alpha_{21}$ and $(\alpha_{31} - 2\delta)$ in the interval $[0,2\pi)$. 
The {\it horizontal brown} and {\it grey bands} indicate the 
current upper bounds on $\meff$ quoted in eq.~\eqref{eq:mefflimits} 
set by KamLAND-Zen \cite{KamLAND-Zen:2016pfg} and 
GERDA Phase II \cite{Agostini:2017iyd}, respectively. 
The {\it vertical grey line} represents the prospective 
upper limit on $m_{\rm min} \protect\ltap 0.2$~eV from KATRIN \cite{Eitel:2005hg}. 
Cases B3 and B4 are compatible with the $3\sigma$ ranges of the mixing angles, 
but not with their $2\sigma$ ranges.}
\label{fig:bb0nuB}
\end{figure}

\begin{figure}
\centering
\includegraphics[width=0.49\textwidth]{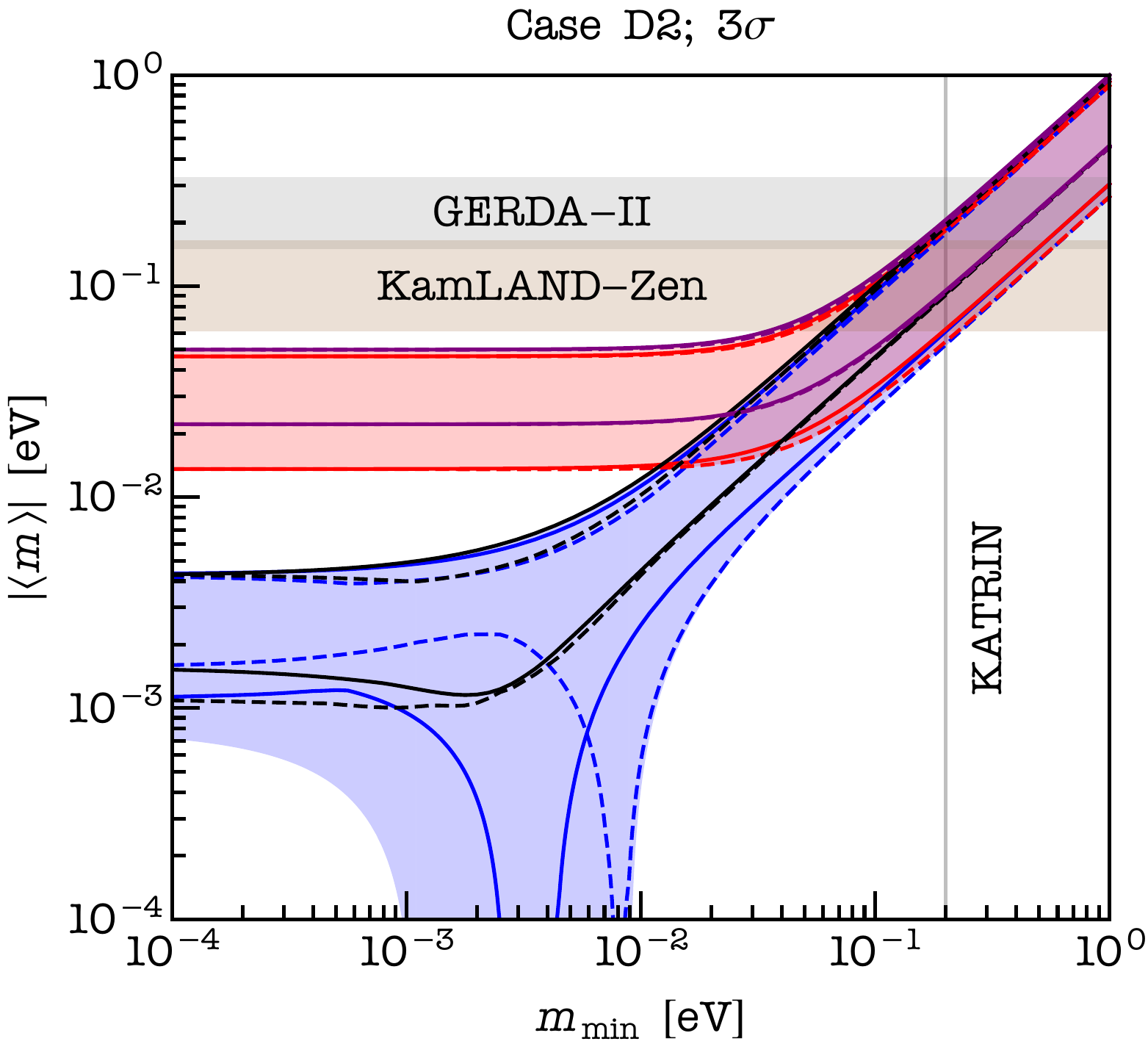}
\includegraphics[width=0.49\textwidth]{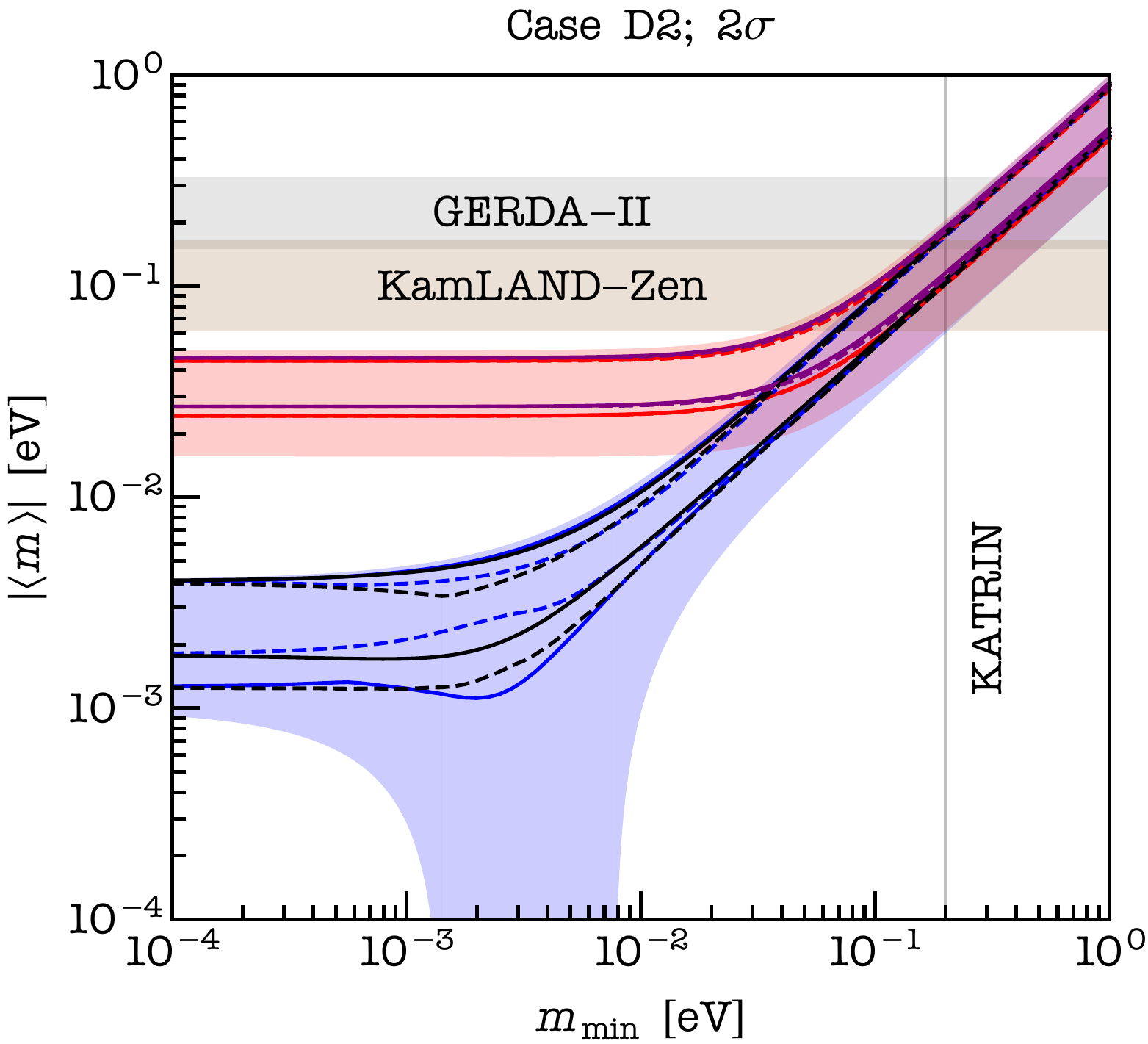}\\[6mm]
\includegraphics[width=0.49\textwidth]{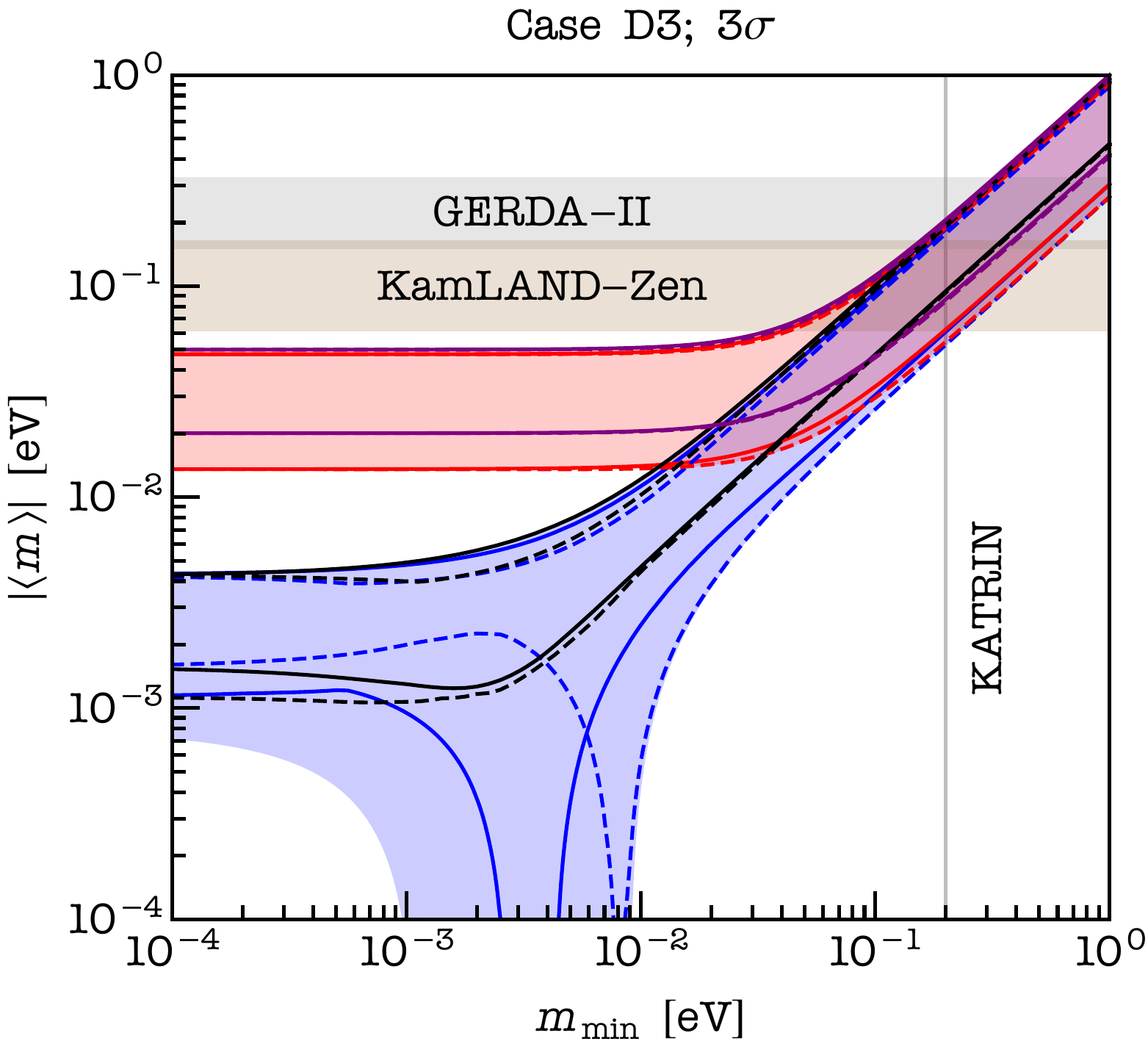}
\includegraphics[width=0.49\textwidth]{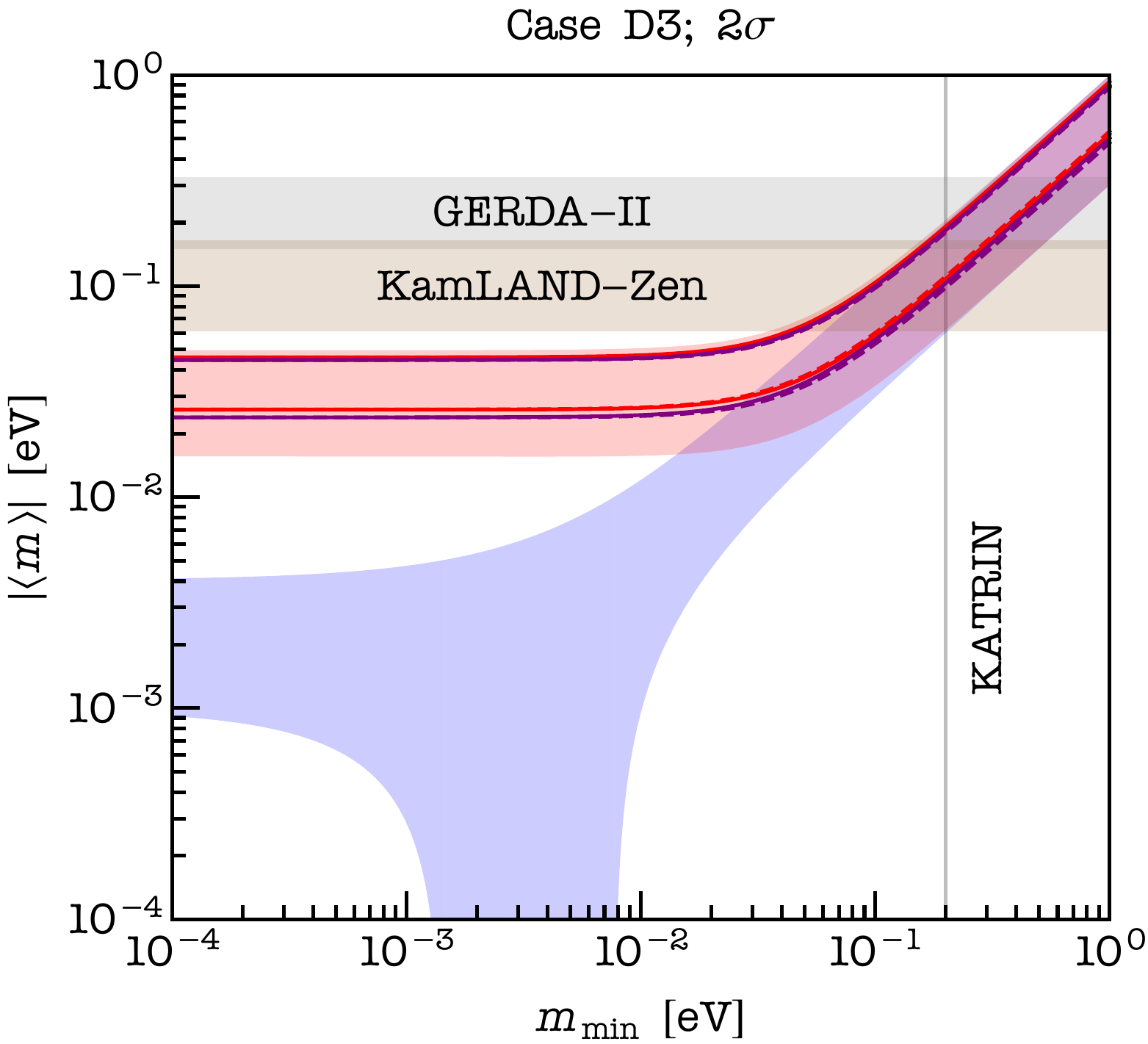}
\caption{The same as in Fig.~\ref{fig:bb0nuB}, but for cases D2 and D3.}
\label{fig:bb0nuD}
\end{figure}

  In what follows, we obtain predictions for $\meff$ using 
the phenomenologically viable neutrino mixing patterns found
in subsection~\ref{subsec:results}. 
In Figs.~\ref{fig:bb0nuB}\,--\,\ref{fig:bb0nuD} we present $\meff$ 
as a function of the lightest neutrino mass $m_{\rm min}$ 
($m_{\rm min} = m_1$ for the NO spectrum and $m_{\rm min} = m_3$ for 
the IO spectrum) 
in cases B1\,--\,B4, C1\,--\,C3, C4 and C5, and D2 and D3.  
The solid and dashed lines limit the found allowed regions of
$\meff$ calculated using the predicted ranges for 
$\th_{12}$, $\th_{13}$, $\alpha_{21}$, $(\alpha_{31} - 2\delta)$. 
In the left panels we require the predicted values of 
$\sin^2\th_{12}$, $\sin^2\th_{13}$ and $\sin^2\th_{23}$ to lie in 
their corresponding experimentally allowed $3\sigma$ intervals, 
while in the right panels we require them to be inside the 
corresponding $2\sigma$ ranges. The mass squared differences
$\Delta m^2_{21}$ and $\Delta m^2_{31(23)}$ in the case of 
NO (IO) spectrum  are varied in their appropriate ranges given 
in Table~\ref{tab:parameters}. 
The light-blue (light-red) areas in the left and right panels 
are obtained varying the neutrino oscillation parameters 
$\theta_{12}$, $\theta_{13}$, $\Delta m^2_{21}$ and 
$\Delta m^2_{31(23)}$ in their full $3\sigma$ and 
$2\sigma$ NO (IO) ranges, respectively,   
and varying the phases $\alpha_{21}$ and 
$(\alpha_{31} - 2\delta)$ in the interval $[0,2\pi)$. 
The horizontal brown and grey bands indicate the 
current most stringent upper limits on 
$\meff$, given in eq.~\eqref{eq:mefflimits},
set by KamLAND-Zen and GERDA Phase II, respectively. 
The vertical grey line represents the prospective 
upper limit on $m_{\rm min} \ltap 0.2$~eV from 
the KATRIN experiment \cite{Eitel:2005hg}. 

 Several comments are in order. 
Firstly, for given values of $(k_1,k_2)$ and a given ordering we find 
$\meff$ to be inside of a band, which 
occupies a certain part of the allowed parameter space. 
Secondly, we note that most cases are compatible with both 
$3\sigma$ and $2\sigma$ ranges of all the mixing angles 
for both neutrino mass orderings 
(see Table~\ref{tab:checks}). 
There are several exceptions. 
Namely, cases B2, C3, C4 and D3, in which, due to the 
correlations imposed by the employed symmetry, 
the predictions for $\sin^2\th_{23}$ for the NO spectrum 
are not compatible with its $2\sigma$ allowed range 
(see Tables~\ref{tab:rangesAB} and \ref{tab:rangesCD}). 
Moreover, there is incompatibility for both orderings of 
cases B3 and B4 with the allowed $2\sigma$ ranges of $\sin^2\th_{23}$ 
(see Table~\ref{tab:rangesAB}), 
and of case C1 with the $2\sigma$ range of $\sin^2\th_{12}$ 
(see Table~\ref{tab:rangesCD}).
Thirdly, the predictions for $\meff$ compatible with the $3\sigma$ ranges 
of all the mixing angles are almost the same for the following pairs of cases: 
(B1, B2), (B3, B4), (C2, C3), (C4, C5) and (D2, D3). 
As discussed at the end of subsection~\ref{subsec:results}, 
the cases in each pair share some qualitative features, in particular, 
the allowed ranges of $\th_{12}$, $\th_{13}$, $\alpha_{21}$
and $(\alpha_{31} - 2\delta)$ are approximately equal.
We note also that case C1 stands out by having relatively 
narrow bands for $\meff$  due to the predicted values of 
$\alpha_{21} = k_1\,\pi$ and $(\alpha_{31} -2\delta) = k_2\,\pi$.
Finally, the results shown in 
Figs.~\ref{fig:bb0nuB}\,--\,\ref{fig:bb0nuD} and derived    
using the predictions for the CPV phases and the mixing angles 
$\theta_{12}$ and $\theta_{13}$ in the case 
when the predicted values of all the 
three mixing angles $\theta_{12}$, $\theta_{13}$ 
and $\theta_{23}$ are compatible with their respective 
$3\sigma$ experimentally allowed ranges, 
can be obtained analytically in the limiting cases of NH, IH and QD 
spectra using eqs.~(\ref{NH})\,--\,(\ref{QD}), the values 
of $\Delta m^2_{21}$ and $\Delta m^2_{31(23)}$
quoted in Table~\ref{tab:parameters} and 
the results on  $\sin^2\theta_{12}$, $\sin^2\theta_{13}$, 
$\delta$, $\alpha_{21}$ and $\alpha_{31}$ given in   
Tables~\ref{tab:rangesAB} and \ref{tab:rangesCD}.

\clearpage
\thispagestyle{empty}
%
%
%
%
\clearpage
\thispagestyle{empty}
\begin{figure}
\centering
\includegraphics[width=0.49\textwidth]{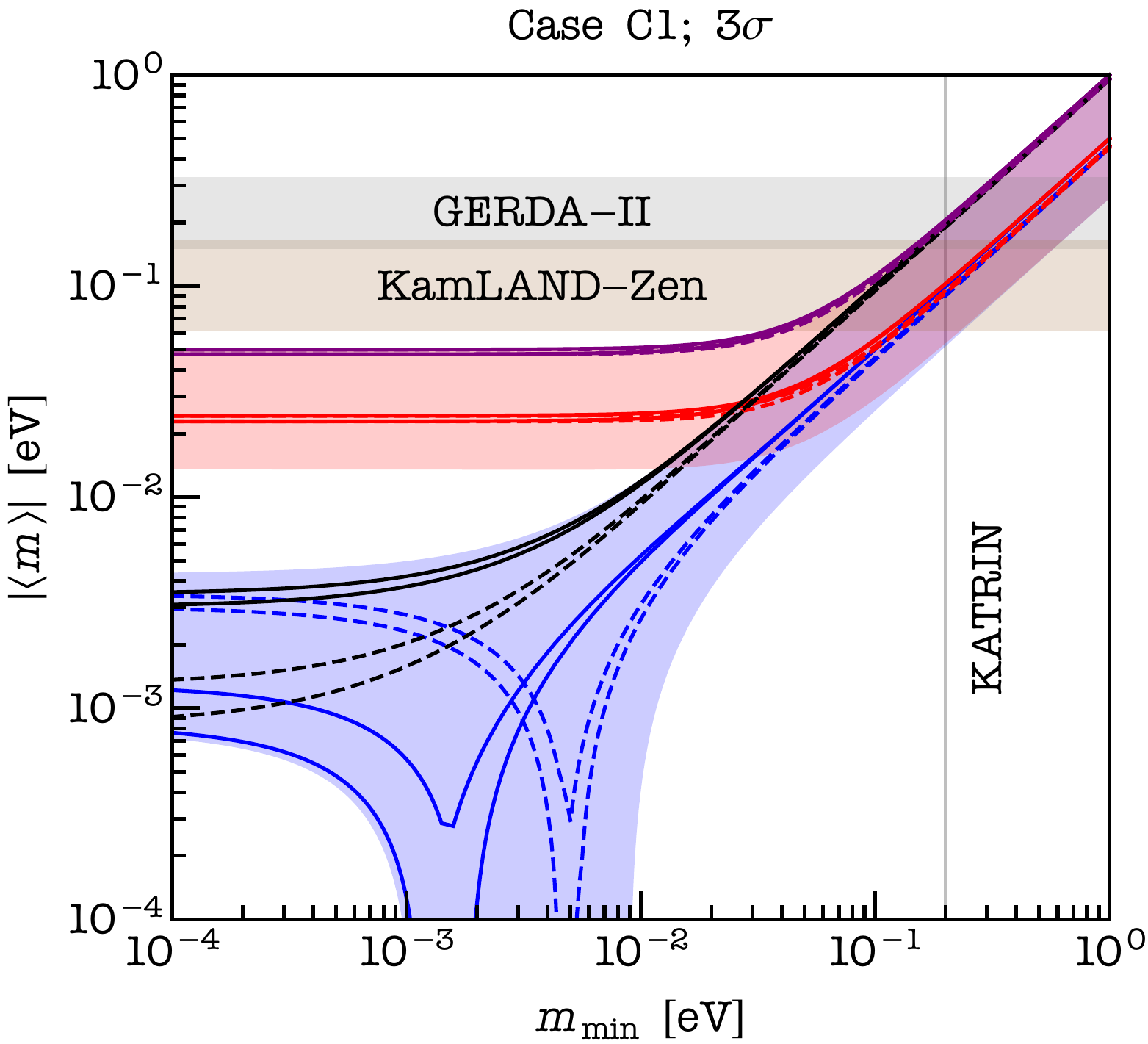}
\includegraphics[width=0.49\textwidth]{plots/bb0nu/legend.pdf}\\[6mm]
\includegraphics[width=0.49\textwidth]{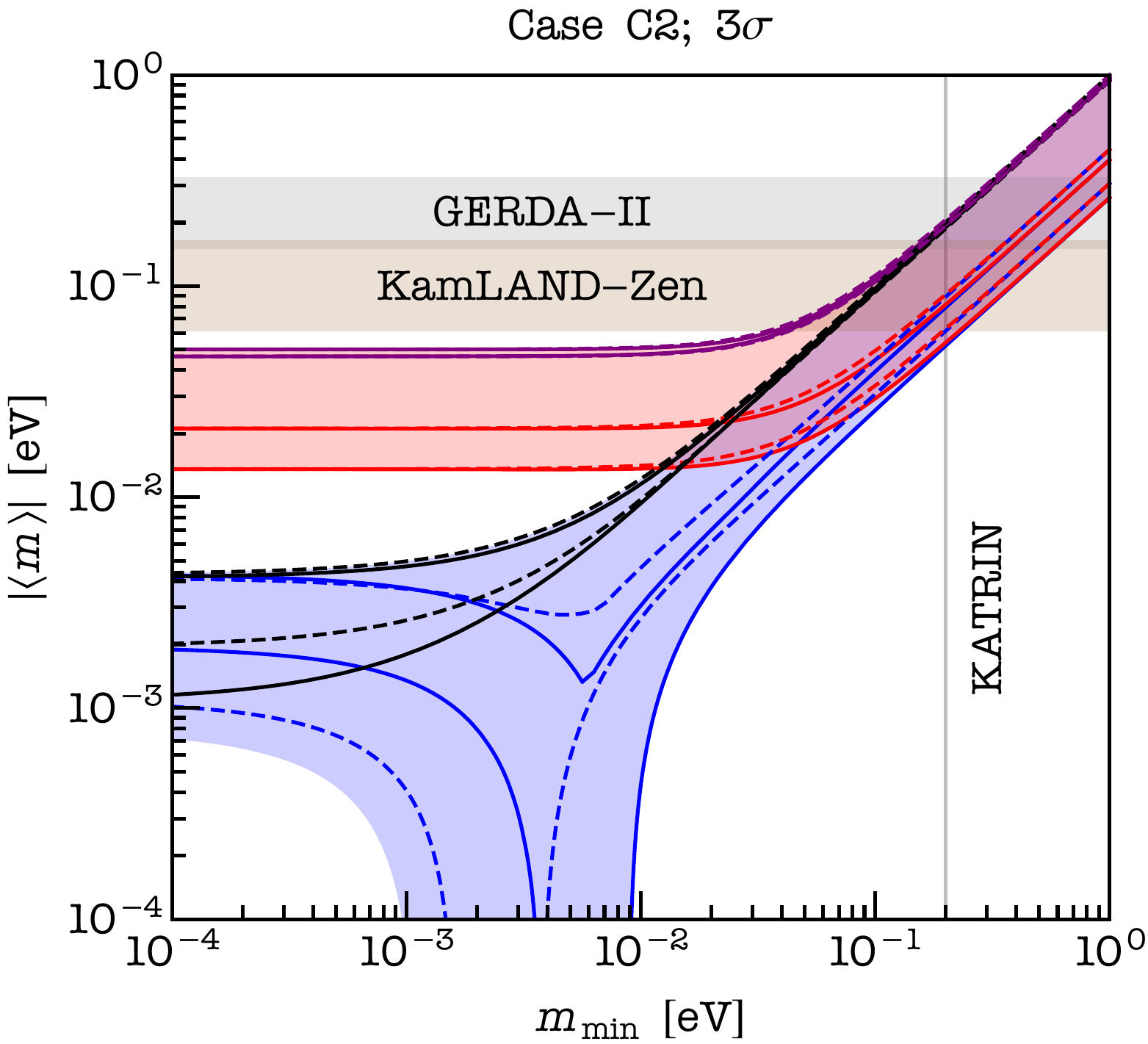}
\includegraphics[width=0.49\textwidth]{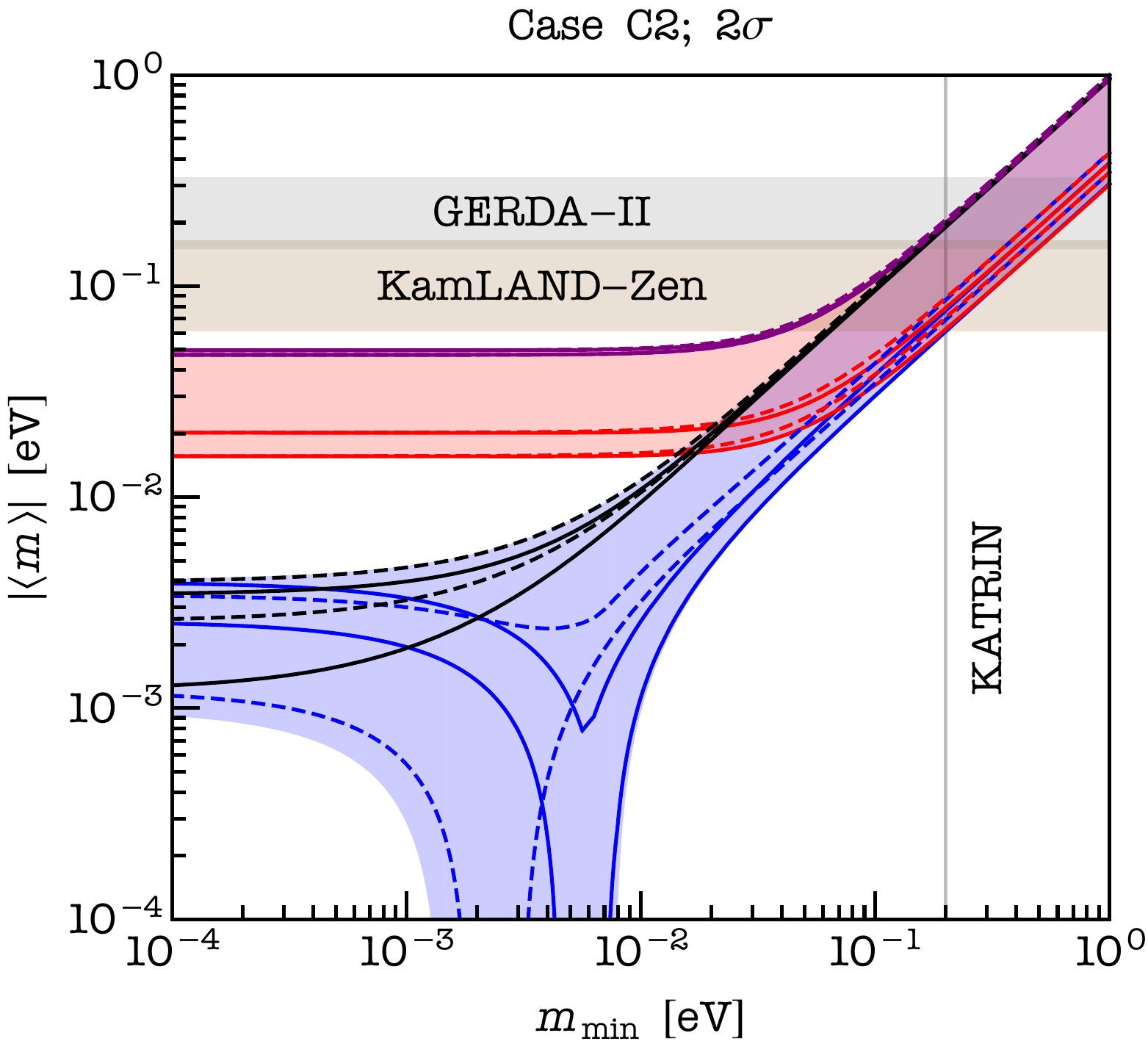}\\[6mm]
\includegraphics[width=0.49\textwidth]{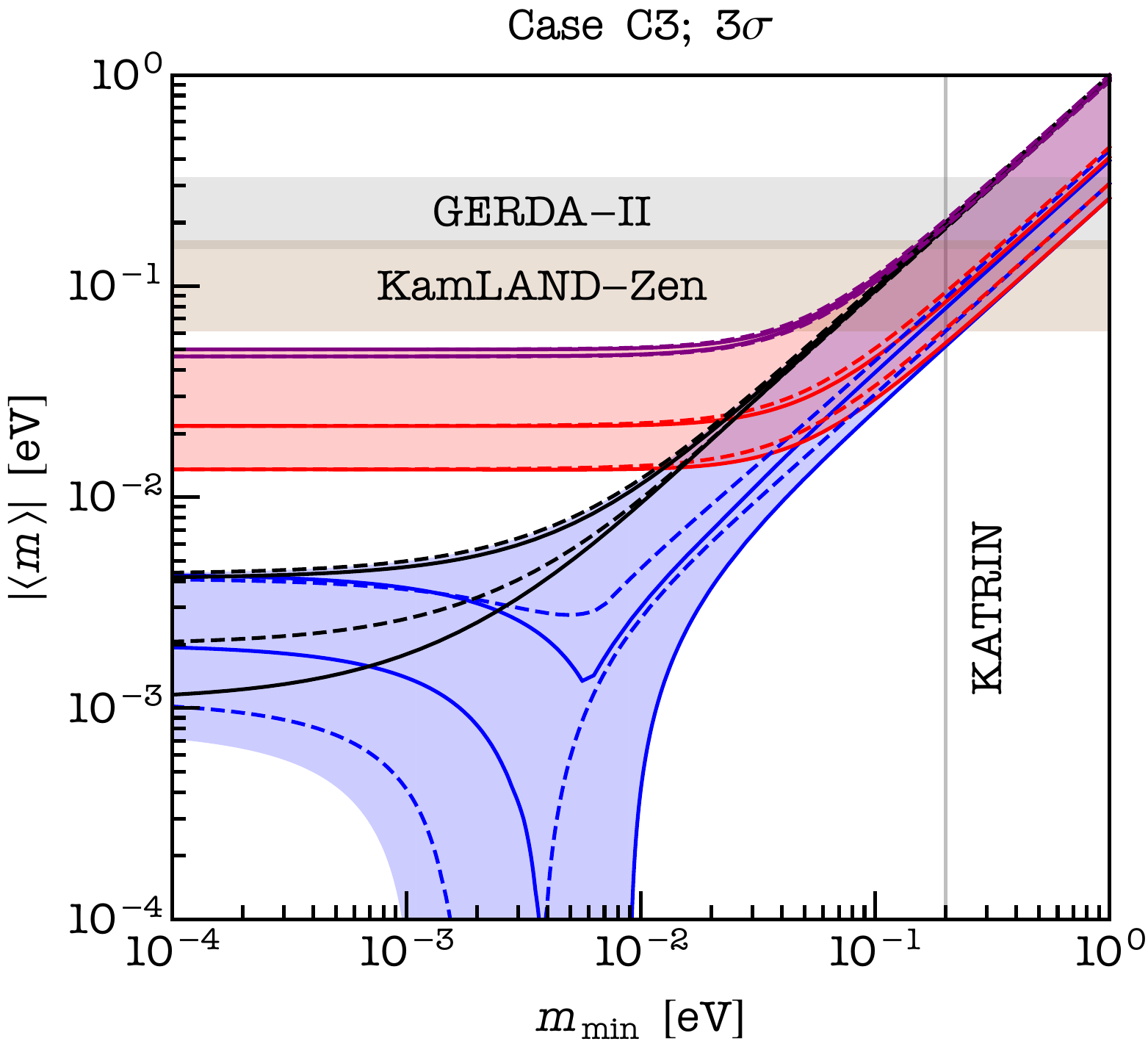}
\includegraphics[width=0.49\textwidth]{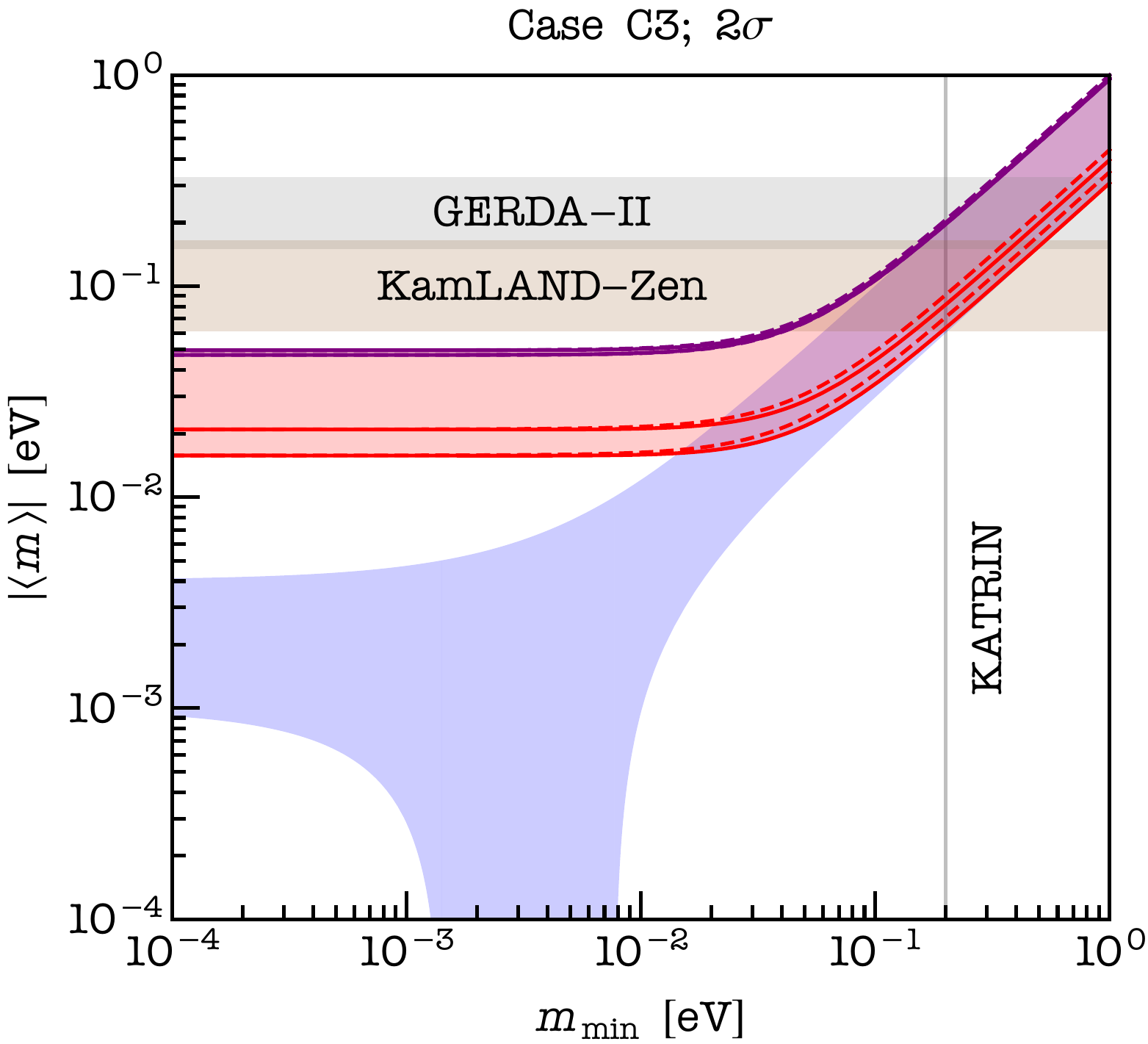}
\caption{The same as in Fig.~\ref{fig:bb0nuB}, but for cases C1\,--\,C3. 
Case C1 is compatible with the $3\sigma$ ranges of the mixing angles, 
but not with their $2\sigma$ ranges.}
\label{fig:bb0nuC123}
\end{figure}
%
%
%
%
\begin{figure}
\centering
\includegraphics[width=0.49\textwidth]{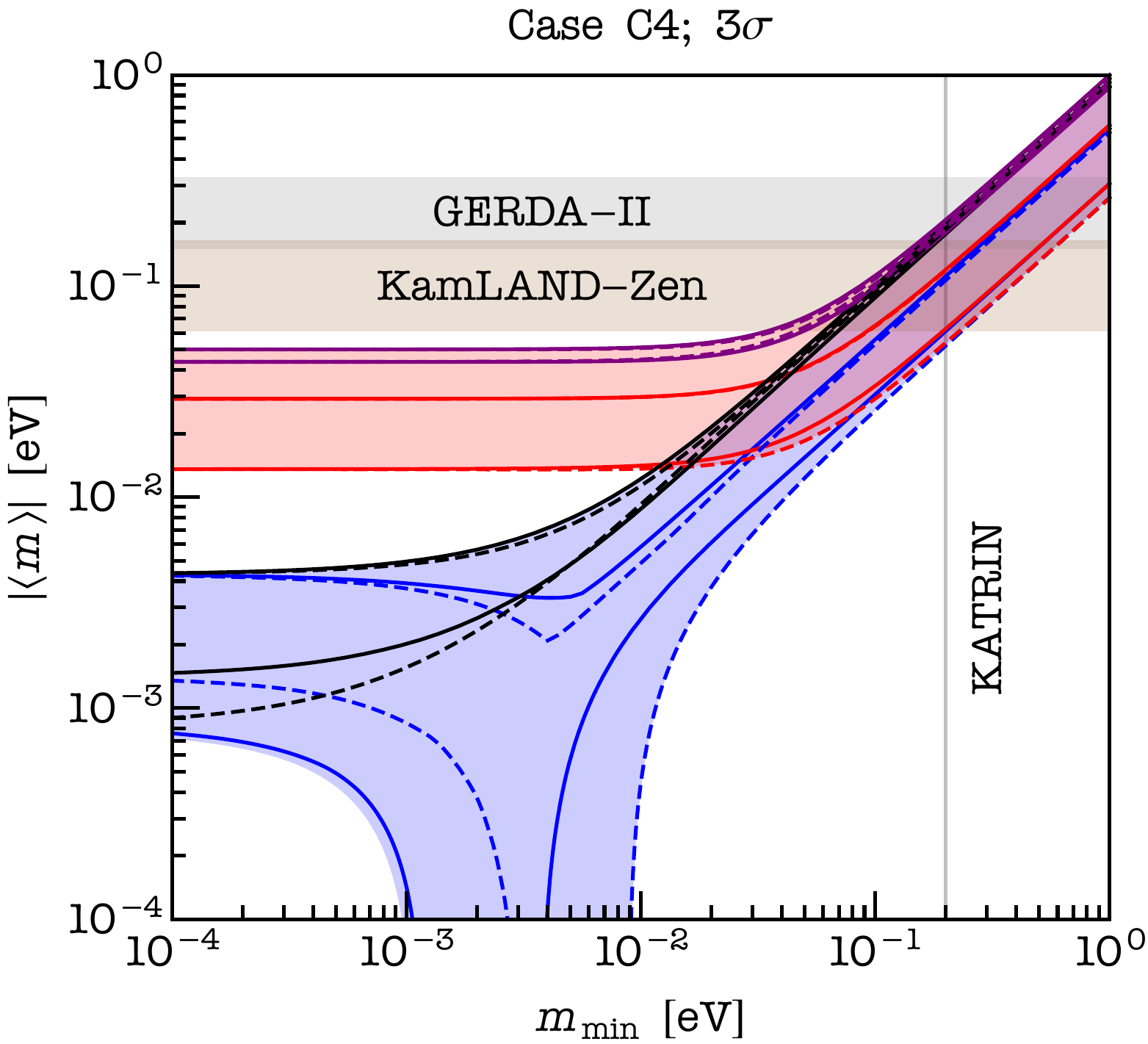}
\includegraphics[width=0.49\textwidth]{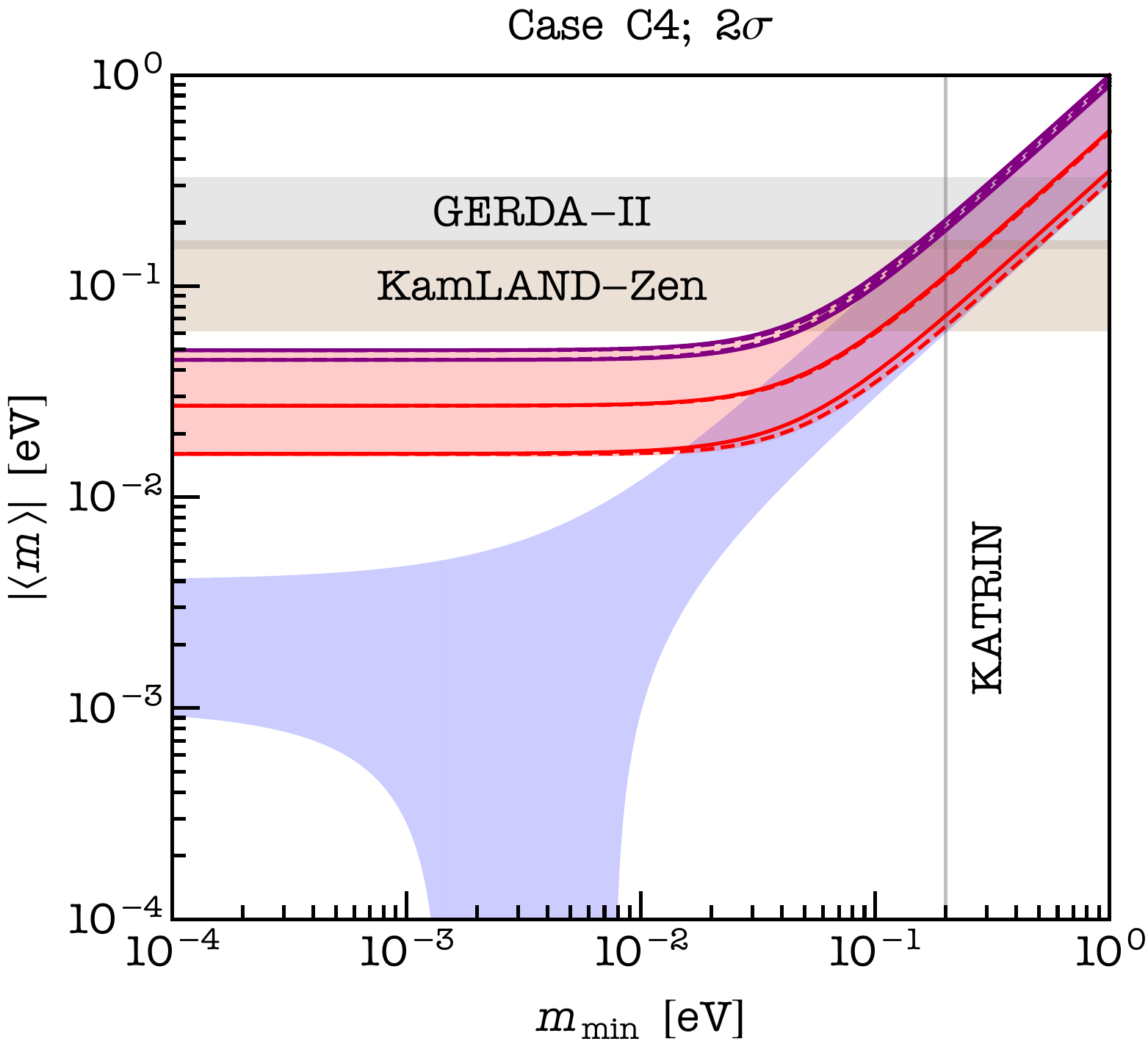}\\[6mm]
\includegraphics[width=0.49\textwidth]{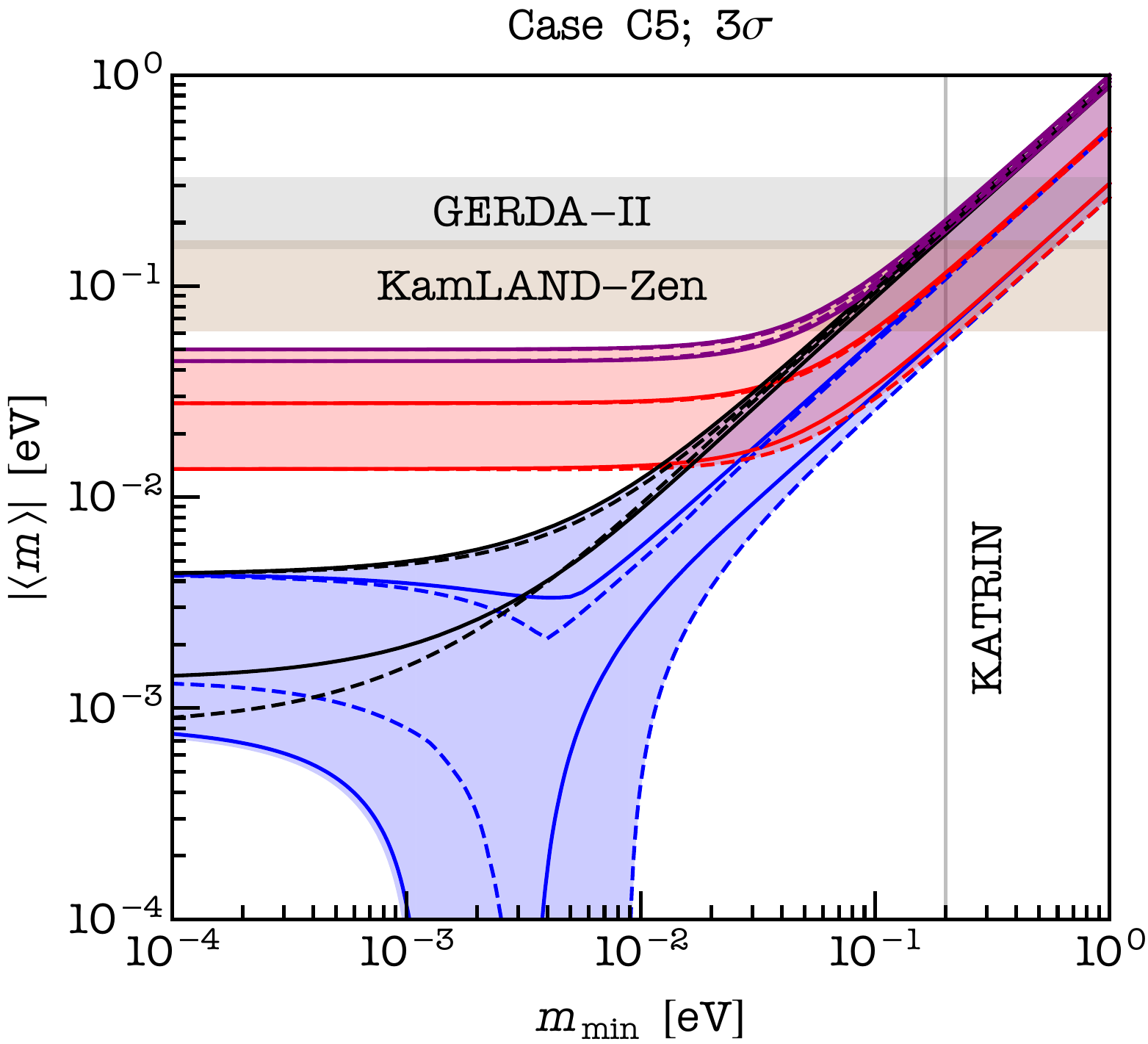}
\includegraphics[width=0.49\textwidth]{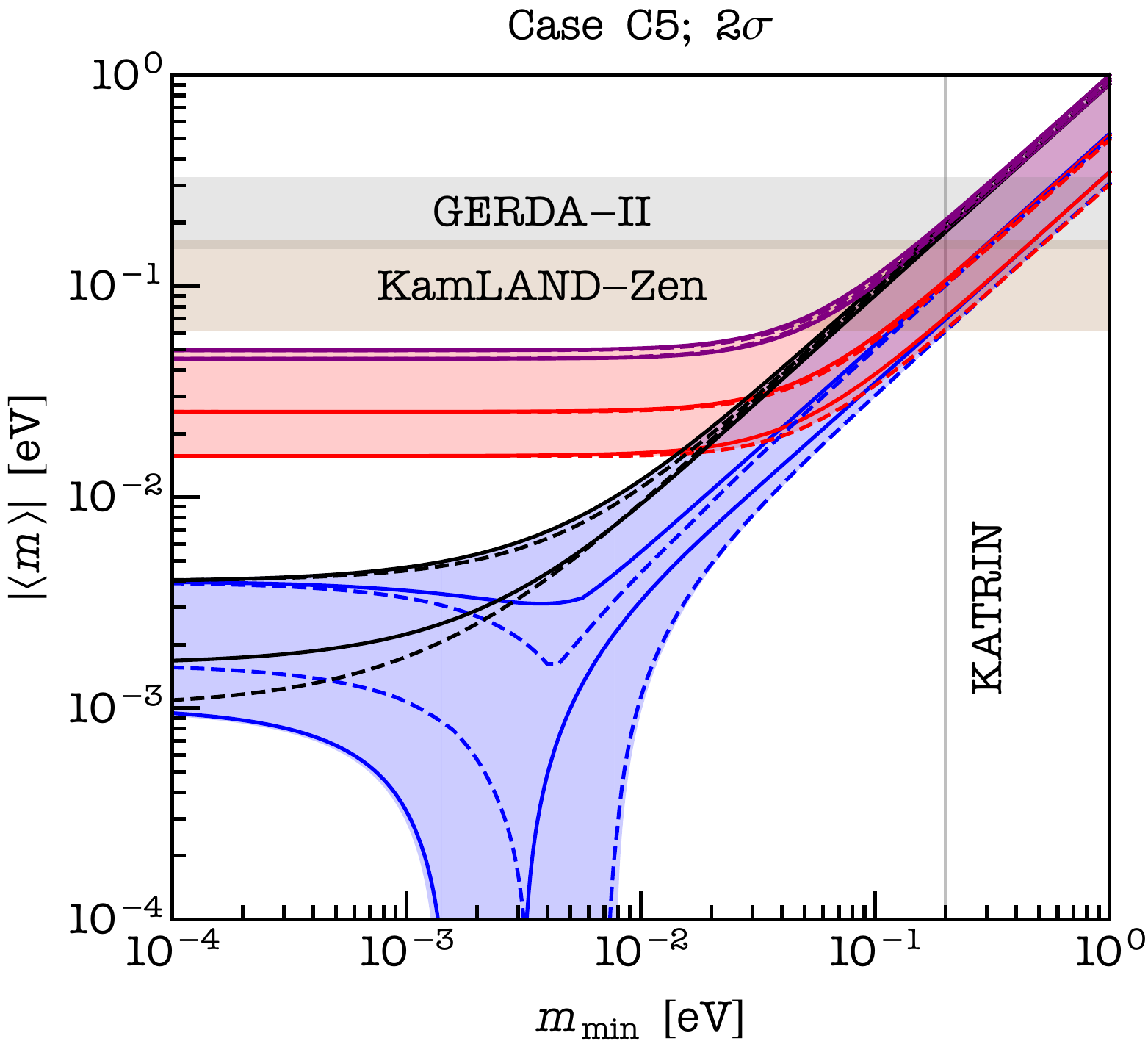}
\caption{The same as in Fig.~\ref{fig:bb0nuB}, but for cases C4 and C5.}
\label{fig:bb0nuC45}
\end{figure}
%
%
%
%

\clearpage
\section{Summary and Conclusions}
\label{sec:conclusions}

 In the present article we have derived predictions for the 
3-neutrino (lepton) mixing and leptonic Dirac and Majorana CP violation 
in a class of models based on  $S_4$ lepton flavour symmetry 
combined with a generalised CP (GCP) symmetry  $H_{\rm CP}$, which are broken to 
residual  $Z_2^{g_e}$ and $Z_2^{g_\nu} \times H^\nu_{\rm CP}$ symmetries in 
the charged lepton and neutrino sectors, respectively, where 
$Z_2^{g_e} = \{1,g_e\}$, $Z_2^{g_\nu} = \{1,g_\nu\}$ and 
$H^\nu_{\rm CP} = \{X_\nu\}$,
$1$ being the unit element of $S_4$. The massive neutrinos 
are assumed to be Majorana particles with their masses generated 
by the neutrino Majorana mass term of the left-handed (LH) flavour 
neutrino fields $\nu_{lL}(x)$, $l=e,\mu,\tau$.
We show that in this class of models the three 
neutrino mixing angles, $\theta_{12}$, $\theta_{23}$ and $\theta_{13}$,
the Dirac and the two Majorana CP violation (CPV)  phases,
$\delta$ and $\alpha_{21}$, $\alpha_{31}$,
are functions of altogether 
three parameters~---~two mixing angles and a phase, 
$\theta^e$, $\theta^\nu$ and $\delta^e$.
 
  The  $S_4$ group has 9 different $Z_2$ subgroups.
Assuming that the LH flavour neutrino and charged lepton 
fields, $\nu_{lL}(x)$ and $l_L(x)$, $l=e,\mu,\tau$, 
transform under a triplet irreducible unitary representation of $S_4$, 
we prove that there are only 3 pairs of subgroups 
$Z_2^{g_e}$ and $Z_2^{g_\nu}$ which can lead to different viable 
(i.e., compatible with the current data)
predictions for the lepton mixing.
For these three pairs, 
$\{g_e,g_\nu\} = \{S,TU\}$, $\{TU,S\}$ and $\{TU,U\}$, 
where $S$, $T$ and $U$ are the generators 
of $S_4$ (see eq. (\ref{eq:S4gener})) taken 
here in the triplet representation of  $S_4$ (eq.~\eqref{eq:generators}).
In what concerns the residual GCP symmetry in the neutrino sector, 
$H^\nu_{\rm CP} = \{X_\nu \}$, we show that 
the constraints on $X_{\nu}$ (following from the 
conditions of consistency between $Z_2^{g_\nu}$ and $H^\nu_{\rm CP}$ 
and of having non-degenerate neutrino mass spectrum, $X_\nu = X^T_\nu$) 
are satisfied in the following cases: 
\begin{enumerate}[i)]
\item for $g_{\nu} = S$, 
if  $H^\nu_{\rm CP} = \{1,S\}$, $\{U,SU\}$ or $\{TST^2U,T^2STU\}$;
\item for $g_{\nu} = U$, 
if $H^\nu_{\rm CP} = \{1,U\}$ or $\{S,SU\}$; 
\item for $g_{\nu} = TU$, 
if $H^\nu_{\rm CP} = \{U,T\}$ or $\{STS,T^2STU\}$.
\end{enumerate}
%
However, $H^\nu_{\rm CP} = \{U,SU\}$ and  $H^\nu_{\rm CP} = \{TST^2U,T^2STU\}$ 
in the case of $g_{\nu} = S$, and $H^\nu_{\rm CP} = \{U,T\}$ and 
$ H^\nu_{\rm CP}=\{STS,T^2STU\}$ in the case of  $g_{\nu} = TU$,
are shown to  lead to the same predictions for the  
PMNS neutrino mixing matrix. Thus, we have found that 
effectively there are 4 distinct groups of cases to be considered. 
We have analysed them case by case and have classified all 
phenomenologically viable mixing patterns they lead to. 
In all four groups of cases the PMNS neutrino mixing matrix is predicted 
to contain one constant element which does not depend 
on the three basic parameters, $\theta^e$, $\theta^\nu$ and $\delta^e$.
The magnitude of this element is equal to $1/\sqrt{2}$ in the ``Group A'' cases 
of  $\{G_e, G_\nu\} = \{Z_2^{TU}, Z_2^{S} \times H^\nu_{\rm CP}\}$ with 
$H^\nu_{\rm CP} = \{1,S\}$, 
and in the ``Group B'' cases of 
$\{G_e, G_\nu\} = \{Z_2^{TU}, Z_2^{S} \times H^\nu_{\rm CP}\}$ with 
$H^\nu_{\rm CP} = \{U,SU\}$; 
and it is equal to $1/2$ in the ``Group C'' cases of 
$\{G_e, G_\nu\} = \{Z_2^{TU}, Z_2^{U}\times H^\nu_{\rm CP}\}$ 
with $H^\nu_{\rm CP} = \{1,U\}$,  
and  in the ``Group D'' cases of
$\{G_e, G_\nu\} = \{Z_2^{TU}, Z_2^{U} \times H^\nu_{\rm CP}\}$ with 
$H^\nu_{\rm CP} = \{S,SU\}$.
In the approach to the neutrino mixing based on $S_4$
flavour and GCP symmetries  
employed by us, the PMNS matrix
is determined up to permutations of columns and rows.
This implies that theoretically any of the elements of the PMNS matrix
can be equal by absolute value
to $1/\sqrt{2}$ in the Group A and Group B cases, 
and to $1/2$ in the Group C and Group D cases. 
However, the data on the neutrino mixing angles
and the Dirac phase $\delta$ imply
that, taking into account the currently
allowed $3\sigma$ ranges of the PMNS matrix elements
(see eqs.~(\ref{eq:PMNS3sigmaNO}) and (\ref{eq:PMNS3sigmaIO})),
only 4 elements, namely, $(U_{\rm PMNS})_{\mu2}$, $(U_{\rm PMNS})_{\mu3}$, 
$(U_{\rm PMNS})_{\tau2}$ or $(U_{\rm PMNS})_{\tau3}$,
can have an absolute value equal to  $1/\sqrt{2} \approx 0.707$,
and only 5 elements, namely,
$(U_{\rm PMNS})_{e2}$, $(U_{\rm PMNS})_{\mu1}$, $(U_{\rm PMNS})_{\tau1}$, 
$(U_{\rm PMNS})_{\mu2}$ or $(U_{\rm PMNS})_{\tau2}$, 
can have an absolute value equal to  $1/2$. 
It should be added that
i) $|(U_{\rm PMNS})_{\tau2}| = 0.707$ lies outside the respective
currently allowed $3\sigma$ range in the case of NO neutrino mass spectrum,
ii) $|(U_{\rm PMNS})_{\mu2}| = 0.707$ is slightly outside the $3\sigma$
allowed range for the IO spectrum, and that
iii) the value of $|(U_{\rm PMNS})_{\mu2}| = 1/2$ is allowed
at $3\sigma$ only for the IO spectrum.

 We have derived predictions for the six parameters
of the PMNS matrix, $\theta_{12}$, $\theta_{23}$ and $\theta_{13}$,
$\delta$, $\alpha_{21}$ and $\alpha_{31}$,
in the potentially viable cases of Groups A\,--\,D.
This was done for both NO and IO neutrino mass
spectra in the cases compatible at $3\sigma$ with the existing data.
We have performed also a statistical analysis of the predictions 
for the neutrino mixing angles and CPV phases for each of these
cases.  We have found that 
in certain cases the predicted values of the neutrino mixing angles 
are ruled out, or are strongly disfavoured, by the existing data
(see subsection~\ref{subsec:results} for details).
These are:
\begin{enumerate}[i)]
\item in Group A, the cases
of 
$|(U_{\rm PMNS})_{\mu3}| = 1/\sqrt{2}$ (strongly disfavoured), and
$|(U_{\rm PMNS})_{\tau3}| = 1/\sqrt{2}$ (strongly disfavoured);
\item in Group D, the cases of
 $|(U_{\rm PMNS})_{e2}| = 1/2$ (ruled out),
$|(U_{\rm PMNS})_{\mu2}| = 1/2$ (strongly disfavoured), and
$|(U_{\rm PMNS})_{\tau2}| = 1/2$ (strongly disfavoured).
\end{enumerate}
%

The results of the statistical analysis in the viable cases
are presented graphically in
Figs.~\ref{fig:caseB1}\,--\,\ref{fig:caseD3}. 
The predicted ranges of the neutrino mixing parameters and 
the their corresponding best fit values are summarised in 
Tables~\ref{tab:rangesAB}\,--\,\ref{tab:bestfitCD}.

 Given the difference in the currently allowed $2\sigma$ 
ranges of $\sin^2\theta_{23}$ (see Table \ref{tab:parameters}),
the prediction for the allowed values of
$\sin^2\theta_{23}$ in certain phenomenologically viable
cases makes the IO (NO) spectrum statistically somewhat
more favourable than the NO (IO) spectrum.
At the same time, we have found that in a large number of
viable cases the results we have obtained for the NO and IO
spectra are very similar. 

 As a consequence of the fact that, in the 
class of models we consider, 
the six PMNS matrix parameters,
$\theta_{12}$, $\theta_{23}$, $\theta_{13}$,
$\delta$, $\alpha_{21}$ and $\alpha_{31}$, are
fitted with the three basic parameters,
$\theta^e$, $\theta^\nu$ and $\delta^e$, 
it is not surprising that we have found that 
there are strong correlations
i) between the values of the Dirac phase $\delta$ and
the values of the two Majorana phases $\alpha_{21}$ and $\alpha_{31}$,
which in turn are correlated between themselves  
(Figs.~\ref{fig:caseB1}, \ref{fig:caseB2}, \ref{fig:caseC2}\,--\,\ref{fig:caseC5}),  
and depending on the case 
ii)~either between the values of $\theta_{12}$ and $\theta_{13}$ 
(Fig.~\ref{fig:caseC1}),
or between the values of  $\theta_{23}$ and $\theta_{13}$  
(Figs.~\ref{fig:caseB3}~and~\ref{fig:caseB4})
or else between the values of  $\theta_{12}$ and $\theta_{23}$
 (Figs.~\ref{fig:caseB1}, \ref{fig:caseB2},
 \ref{fig:caseC2}\,--\,\ref{fig:caseD3}). 
In certain cases our results showed strong correlations 
between the predicted values of 
$\theta_{23}$ and the Dirac phase $\delta$ and/or 
the Majorana phases $\alpha_{21,31}$
(Figs.~\ref{fig:caseC4}\,--\,\ref{fig:caseD3}). 

  In the cases of
i) Group B with $|(U_{\rm PMNS})_{\mu2}| = 1/\sqrt{2}$, or
$|(U_{\rm PMNS})_{\tau 2}| = 1/\sqrt{2}$, 
ii) Group C with 
$|(U_{\rm PMNS})_{\mu1}| = 1/2$, or
$|(U_{\rm PMNS})_{\tau 1}| = 1/2$, or
$|(U_{\rm PMNS})_{\mu2}| = 1/2$, or
$|(U_{\rm PMNS})_{\tau 2}| = 1/2$, and
iii) Group D with $|(U_{\rm PMNS})_{\mu1}| = 1/2$, or
$|(U_{\rm PMNS})_{\tau 1}| = 1/2$,
the cosine of the Dirac phase $\delta$ satisfies a sum rule by which it
is expressed in terms of the three neutrino mixing angles
$\theta_{12}$, $\theta_{23}$ and  $\theta_{13}$.
Taking into account the ranges and correlations of the predicted
values of the three neutrino mixing angles,
$\delta$ is predicted to lie in certain,
in most of the discussed cases rather narrow, intervals 
(subsection~\ref{subsec:results}).
In the remaining viable cases of Groups B and C,  
$\cos\delta$ was shown to satisfy sum rules which depend explicitly, 
in addition to $\theta_{12}$, $\theta_{23}$ and  $\theta_{13}$,
on one of the three basic parameters of the class of models considered,
$\theta^e$ or $\theta^\nu$. In these cases, as we have shown, 
$\cos\delta$ can take any value.

 We have derived also predictions for the Majorana CPV 
phases $\alpha_{21}$ and $\alpha_{31}$ in all viable cases 
of Groups B, C and D (subsection~\ref{subsec:results}). 
With one exception~---~the case of    
$|(U_{\rm PMNS})_{e2}| = 1/2$ of Group C~---~%
the values of $\alpha_{21}$ and $\alpha_{31}$, as we have 
indicated earlier, are 
strongly correlated between themselves. 
In case C1 there is a strong linear correlation
between $\alpha_{31}$ and $\delta$.

 Using the predictions for the Dirac and Majorana CPV phases 
allowed us to derive predictions for the magnitude of the 
neutrinoless double beta decay effective Majorana mass, 
$\meff$, as a function of the lightest neutrino mass 
for all the viable cases belonging to  
Groups B, C and D.
They are presented  graphically in
Figs.~\ref{fig:bb0nuB}\,--\,\ref{fig:bb0nuD}. 
     
 All viable cases in the class of $S_4$
models investigated in the present article 
have distinct predictions for the set of observables 
$\sin^2\theta_{12}$, $\sin^2\theta_{23}$, $\sin^2\theta_{13}$, 
the Dirac phase $\delta$ and the absolute value of 
one element of the PMNS neutrino mixing matrix.
Using future more precise
data on $\sin^2\theta_{12}$, $\sin^2\theta_{23}$, $\sin^2\theta_{13}$ 
and the Dirac phase $\delta$, which will allow 
also to determine the absolute values of the elements of 
the PMNS matrix with a better precision, 
will make it possible to test and discriminate between 
the predictions of all the cases found by us 
to be compatible with the current data 
on the neutrino mixing parameters.

 Future data will show whether Nature 
followed the $S_4 \rtimes H_{\rm CP}$ flavour $+$ GCP symmetry 
``three-parameter path'' for fixing 
the values of the three neutrino mixing angles 
and of the Dirac (and Majorana) CP violation phases 
of the PMNS neutrino mixing matrix. 
We are looking forward to these data.

\section*{Acknowledgements} 
We would like to thank F. Capozzi, E. Lisi, A. Marrone, D. Montanino 
and A. Palazzo for kindly sharing with us the data files for 
one-dimensional $\chi^2$ projections.
This work was supported in part by the INFN
program on Theoretical Astroparticle Physics (TASP), 
by the research grant  2012CPPYP7
under the program  PRIN 2012 funded by the Italian 
Ministry of Education, University and Research (MIUR),
by the European Union Horizon 2020 research and innovation programme
under the  Marie Sk\l{}odowska-Curie grants 674896 and 690575, and by
the World Premier International Research Center Initiative (WPI
Initiative), MEXT, Japan (S.T.P.).

\appendix
\section{Symmetry of $\bo{X_\nu}$}
\label{app:symmetryofX}

 If the neutrino sector respects a residual GCP symmetry $H^\nu_{\rm CP} = \{X_\nu\}$,
the neutrino mass matrix satisfies eq.~\eqref{eq:GCP}, namely,
\be
X_\nu^T\, M_\nu\, X_\nu = M_\nu^*\,.
\label{eq:appXnu}
\ee
The GCP transformation matrices $X_\nu$ must be unitary
due to the GCP invariance of the neutrino kinetic term.
In what follows we show that these matrices are additionally constrained to be symmetric
if the neutrino mass spectrum is non-degenerate, as is known to be the case.

 Expressing $M_\nu$ from eq.~\eqref{eq:mnudiag} and substituting it in 
eq.~\eqref{eq:appXnu} yields
\be
 d_\nu\,\tilde X= \tilde X^*\,  d_\nu\,,
\label{eq:appcomm}
\ee
where $d_\nu \equiv \diag(m_1,m_2,m_3)$ 
and  $\tilde X \equiv U_\nu^\dagger \, X_\nu \, U_\nu^*$ is unitary. 

Being $3\times 3$ unitary, $\tilde X$ can be parametrised  as the product of three complex rotations $U_{ij}$ and a diagonal matrix of phases $\Psi$ as follows:
\be
  \tilde X = \Psi\,U_{23}(\vartheta_{23},\delta_{23})\,U_{13}(\vartheta_{13},\delta_{13})\,
  U_{12}(\vartheta_{12},\delta_{12})\,,
\ee
where $\Psi= \diag(e^{i\psi_1},e^{i\psi_2},e^{i\psi_3})$ and the $U_{ij}(\vartheta_{ij},\delta_{ij})$ are 
complex rotations in the $i$-$j$ plane. 
Explicitly,
\be
  U_{23}(\vartheta_{23},\delta_{23})=\begin{pmatrix}
1 & 0 &0 \\ 
0 & \cos \vartheta_{23} & \sin \vartheta_{23} \,e^{-i\delta_{23}} \\ 
0 & -\sin \vartheta_{23}\,e^{i\delta_{23}} & \cos\vartheta_{23}
\end{pmatrix}\,,
\ee
with a straightforward generalisation to $(ij)=(12),~(13)$.

Imposing eq.~\eqref{eq:appcomm} produces the following relations:
\begin{align}
 e^{i(\psi_1 -\delta_{13})}\, m_1\,\sin\vartheta_{13} &=
 e^{-i(\psi_1 -\delta_{13})}\, m_3\,\sin\vartheta_{13}\,,
\\ 
 e^{i(\psi_2 -\delta_{23})}\, m_2\,\cos\vartheta_{13}\,\sin\vartheta_{23} &=
 e^{-i(\psi_2 -\delta_{23})}\, m_3\,\cos\vartheta_{13}\,\sin\vartheta_{23}\,,
\\ 
 e^{i(\psi_1 -\delta_{12})}\, m_1\,\cos\vartheta_{13}\,\sin\vartheta_{12} &=
 e^{-i(\psi_1 -\delta_{12})}\, m_2\,\cos\vartheta_{13}\,\sin\vartheta_{12}\,.
\end{align}

From the non-degeneracy of the neutrino mass spectrum 
it follows that $\sin\vartheta_{13} =\sin\vartheta_{23} =\sin\vartheta_{12} = 0$.
Thus, $\tilde X$ is constrained to be diagonal
and hence symmetric, $\tilde X^T = \tilde X$.   
This finally implies that also $X_\nu^T = X_\nu$,
i.e., a phenomenologically relevant $X_\nu$ must be symmetric.

\section{Conjugate Pairs of $\bo{S_4}$ Elements}
\label{app:conjugatepairs}

 As detailed in subsection~\ref{subsec:conjsym}, 
residual flavour symmetries  $Z_2^{g_e}$ and $Z_2^{g_\nu}$ 
which are conjugate to each other
lead to the same form of the PMNS matrix.
For $G_f = S_4$, there are nine group elements of order two, given in eqs.~\eqref{eq:C2} and \eqref{eq:C2prime}, which generate $Z_2$ subgroups.
The resulting 81 pairs of elements $\{g_e,g_\nu\}$ can themselves be partitioned,
under the conjugacy relation of eq.~\eqref{eq:similarity}, 
into the following nine equivalence classes:
\begin{itemize}
\item%
$\bo{\{S,S\}}$, $\{TST^2,TST^2\}$, $\{T^2ST,T^2ST\}$;
\item%
$\bo{\{U,U\}}$, $\{SU,SU\}$, $\{T^2U,T^2U\}$, $\{TU,TU\}$, $\{ST^2SU,ST^2SU\}$, $\{STSU,STSU\}$;
\item%
$\bo{\{T^2ST,S\}}$, $\{TST^2,S\}$, $\{T^2ST,TST^2\}$, $\{S,T^2ST\}$, $\{S,TST^2\}$, $\{TST^2,T^2ST\}$;
\item%
$\bo{\{S,U\}}$, $\{S,SU\}$, $\{TST^2,T^2U\}$, $\{T^2ST,TU\}$, $\{TST^2,ST^2SU\}$, $\{T^2ST,STSU\}$;
\item%
$\bo{\{U,S\}}$, $\{SU,S\}$, $\{T^2U,TST^2\}$, $\{TU,T^2ST\}$, $\{ST^2SU,TST^2\}$, $\{STSU,T^2ST\}$;
\item%
$\bo{\{SU,U\}}$, $\{U,SU\}$, $\{ST^2SU,T^2U\}$, $\{STSU,TU\}$, $\{T^2U,ST^2SU\}$, $\{TU,STSU\}$;
\item%
$\bo{\{S,TU\}}$, $\{S,STSU\}$, $\{S,T^2U\}$, $\{TST^2,TU\}$, $\{S,ST^2SU\}$, $\{T^2ST,U\}$, $\{T^2ST,SU\}$, $\{TST^2,U\}$, $\{T^2ST,T^2U\}$, $\{TST^2,SU\}$, $\{T^2ST,ST^2SU\}$, $\{TST^2,STSU\}$;
\item%
$\bo{\{TU,S\}}$, $\{STSU,S\}$, $\{T^2U,S\}$, $\{TU,TST^2\}$, $\{ST^2SU,S\}$, $\{U,T^2ST\}$, $\{SU,T^2ST\}$, $\{U,TST^2\}$, $\{T^2U,T^2ST\}$, $\{SU,TST^2\}$, $\{ST^2SU,T^2ST\}$, $\{STSU,TST^2\}$;
\item%
$\bo{\{TU,U\}}$, $\{STSU,U\}$, $\{STSU,SU\}$, $\{TU,SU\}$, $\{T^2U,U\}$, $\{TU,T^2U\}$, $\{ST^2SU,U\}$, $\{U,TU\}$, $\{TU,ST^2SU\}$, $\{SU,STSU\}$, $\{U,T^2U\}$, $\{T^2U,TU\}$, $\{U,ST^2SU\}$, $\{SU,T^2U\}$, $\{SU,ST^2SU\}$, $\{T^2U,STSU\}$, $\{ST^2SU,STSU\}$, $\{ST^2SU,TU\}$, $\{STSU,ST^2SU\}$, \hphantom{XXXXXXXXXX} $\{STSU,T^2U\}$, $\{SU,TU\}$, $\{ST^2SU,SU\}$, $\{T^2U,SU\}$, $\{U,STSU\}$;
\end{itemize}
where in boldface we have identified a representative pair of elements for each class,
matching the choice made in eqs.~\eqref{eq:nonviable} and \eqref{eq:viable}.

\section{Equivalent Cases}
\label{app:equivalence}
 A necessary condition for two matrices $U_{\rm PMNS}$ and $U'_{\rm PMNS}$ 
to be equivalent is the same magnitude of the fixed element. 
Indeed, in the four cases under consideration the absolute value of one element 
is $1/\sqrt{2}$. For $P_e = P'_e$ and $P_\nu = P'_\nu$, the two matrices
$U_{\rm PMNS}$ and $U'_{\rm PMNS}$ would be equivalent, if the products 
$\O_e^\dagger\,\O_\nu$ and $\O_e^{\prime\dagger}\,\O_\nu^\prime$ could be related 
in the following way:
\be
\O_e^\dagger\,\O_\nu = 
\diag(e^{i\phi_1}, e^{i\phi_2}, e^{i\phi_3})\,
U_{23}(\th^e_\circ,\delta^e_\circ)\,
\O_e^{\prime\dagger}\,\O_\nu^\prime\,
R_{23}(\th^\nu_\circ)\,
\diag(1, i^k, i^k)\,,
\label{eq:equivalence}
\ee
%
with $\phi_i$, $\delta^e_\circ$ and  
$\th^e_\circ$, $\th^\nu_\circ$ being fixed phases and angles, respectively, 
and $k$ is allowed to be $0, 1, 2$ or $3$.
Indeed, if this relation holds, from eq.~\eqref{eq:UPMNS} we have
\begin{align}
U_{\rm PMNS} & =  
P_e\, U_{23}(\th^e,\delta^e)\, 
\diag(e^{i\phi_1}, e^{i\phi_2}, e^{i\phi_3})\,
U_{23}(\th^e_\circ,\delta^e_\circ)\,
\O_e^{\prime\dagger}\,\O_\nu^\prime\,
R_{23}(\th^\nu_\circ)\,
\diag(1, i^k, i^k)\,
R_{23}(\th^\nu)\, P_\nu\, Q_\nu\, \nonumber\\
& = P_e\, \diag(e^{i\phi_1}, e^{i\phi_2}, e^{i\phi_3})\, 
U_{23}(\th^e,\tilde\delta^e)\, 
U_{23}(\th^e_\circ,\delta^e_\circ)\, 
\O_e^{\prime\dagger}\,\O_\nu^\prime\,
R_{23}(\hat\th^\nu)\, 
P_\nu\, \hat Q_\nu\,,
\end{align}
with 
\be
\tilde\delta^e = \delta^e + \phi_2 - \phi_3\,, \quad
\hat\th^\nu = \th^\nu_\circ + \th^\nu\, \quad 
{\rm and} \quad 
\hat Q_\nu = P_\nu^T\, \diag(1, i^k, i^k)\, P_\nu\, Q_\nu\,.
\ee
Now, using
\be
U_{23}(\th^e,\tilde\delta^e)\, U_{23}(\th^e_\circ,\delta^e_\circ) = 
\diag(1, e^{i\alpha}, e^{-i\alpha})\, U_{23}(\hat\th^e, \hat\delta^e)\,,
\ee
where (see Appendix~B in \cite{Girardi:2015rwa})
\begin{align}
\alpha = \arg\left\{\cos\th^e \cos\th^e_\circ - \sin\th^e \sin\th^e_\circ\, 
e^{i(\delta^e_\circ - \tilde\delta^e)}\right\}\,,& \quad
\beta = \arg\left\{\sin\th^e \cos\th^e_\circ\, e^{-i\tilde\delta^e} + 
\cos\th^e \sin\th^e_\circ\, e^{-i\delta^e_\circ}\right\}\,, \nonumber\\
\cos\hat\th^e = \left|\cos\th^e \cos\th^e_\circ - \sin\th^e \sin\th^e_\circ\, 
e^{i(\delta^e_\circ - \tilde\delta^e)}\right|\,,& \quad
\sin\hat\th^e = \left|\sin\th^e \cos\th^e_\circ\, e^{-i\tilde\delta^e} + 
\cos\th^e \sin\th^e_\circ\, e^{-i\delta^e_\circ}\right|
\nonumber \\
{\rm and}  \quad \hat\delta^e &= \alpha - \beta\,, \nonumber
\end{align}
we obtain
\be
U_{\rm PMNS} = 
Q_e\, 
P_e\, U_{23}(\hat\th^e,\hat\delta^e)\, 
\O_e^{\prime\dagger}\,\O_\nu^\prime\,
R_{23}(\hat\th^\nu)\, 
P_\nu\, \hat Q_\nu\,,
\ee
with 
\be
Q_e = P_e\, \diag\left(e^{i\phi_1}, e^{i(\phi_2 + \alpha)}, e^{i(\phi_3 - \alpha)}\right)\, P_e^T 
\ee
being the matrix of unphysical phases.
Thus, up to this matrix, $U_{\rm PMNS}$ and $U'_{\rm PMNS}$ are the same.

 Taking $\{G_e, G_\nu\} = \{Z_2^{TU}, Z_2^{S} \times H^\nu_{\rm CP}\}$ with 
$H^\nu_{\rm CP} = \{U,SU\}$ as a reference case and denoting 
the corresponding diagonalising matrices as $\O_e^\prime$ and $\O_\nu^\prime$, 
we find the values of $\phi_i$, $\delta^e_\circ$, $\th^e_\circ$, $\th^\nu_\circ$ and $k$ 
for which eq.~\eqref{eq:equivalence} holds, if $\O_e$ and $\O_\nu$ are the 
diagonalising matrices 
in one of the three remaining cases under consideration.
We summarise these values in Table~\ref{tab:equivalence}.
\begin{table}
\centering
\renewcommand{\arraystretch}{1.2}
\begin{tabular}{cccc}
\toprule
$g_e~~g_\nu~~H^\nu_{\rm CP}$ &  $S~~TU~~\{U,T\}$ & $S~~TU~~\{STS,T^2STU\}$ & $TU~~S~~\{TST^2U,T^2STU\}$ \\
\midrule
$\phi_1$ & $\pi/6$ & $-\pi/3$ & $-\pi/2$ \\
$\phi_2$ & $-\arctan\sqrt{1+2\sqrt{2}/3}$ & $-\arctan\left(\sqrt{2} + \sqrt{3}\right)$ & $\arccot\left(2\right)$ \\
$\phi_3$ & $\arctan\left(3\sqrt{3} + 2\sqrt{6}\right)$ & $\arccot\left(2\sqrt{2} + \sqrt{3}\right)$ & $\arctan\left(2\right)$ \\
$\delta^e_\circ$ & $\arccot\left(5/\sqrt{3}\right)$ & $\pi/3$ & $\arctan\left(\left(5\sqrt{3} - 6\right)/13\right)$ \\
$\th^e_\circ$ & $\arctan\sqrt{\left(11 - 6\sqrt{2}\right)/7}$ & $\arctan\left(\sqrt{2} + \sqrt{3}\right)$ & $\pi - \arctan\left(2/\sqrt{5}\right)$ \\
$\th^\nu_\circ$ & $\pi - \arctan\left(3 - 2\sqrt{2}\right)$ & $\pi/4$ & $\pi/4$ \\
$k$ & $0$ & $1$ & $3$ \\
\bottomrule
\end{tabular}
\caption{The values of the parameters $\phi_i$, $\delta^e_\circ$, $\th^e_\circ$, $\th^\nu_\circ$ and $k$ for which eq.~\eqref{eq:equivalence}, proving the equivalence 
of the PMNS matrix in a given case to the PMNS matrix in the reference case of 
$\{G_e, G_\nu\} = \{Z_2^{TU}, Z_2^{S} \times H^\nu_{\rm CP}\}$ with 
$H^\nu_{\rm CP} = \{U,SU\}$, holds.}
\label{tab:equivalence}
\end{table}

\section{Correspondence with Earlier Results}
\label{app:sumrulescorrespondence}

 The sum rules for $\cos\delta$ or $\sin^2\th_{23}$ ($\sin^2\th_{12}$ 
in case C1) can formally be obtained from the corresponding sum rules derived 
in \cite{Girardi:2015rwa}. 
In certain cases, this requires an additional input 
which is provided by the residual GCP symmetry  $H^\nu_{\rm CP}$ 
considered in the present article.
Below we provide the correspondence between the phenomenologically viable 
cases of the present study and the cases considered in \cite{Girardi:2015rwa}.
\begin{enumerate}[i)]
\item Cases B1, C4 and D4 of the present study correspond to case C8 in
\cite{Girardi:2015rwa}, since for all these cases $(U_{\rm PMNS})_{\mu2}$ is fixed. 
The sum rule for $\cos\delta$ in case B1, eq.~\eqref{eq:cosdeltaB1}, 
follows from that of case C8 in \cite{Girardi:2015rwa} (see Table~4 therein) for
$\sin^2\th^\circ_{23} = 1/2$, 
while the sum rule in eq.~\eqref{eq:cosdeltaC4}, valid in cases C4 and D4, 
can be obtained from the same sum rule found in \cite{Girardi:2015rwa}, 
but for $\sin^2\th^\circ_{23} = 3/4$. 
As should be, these two values of $\sin^2\th^\circ_{23}$ follow from $G_f = S_4$, when 
it is broken to two different non-equivalent specific pairs of 
residual $\{Z_2^{g_e}, Z_2^{g_\nu}\}$ flavour symmetries 
(see Table~10 in \cite{Girardi:2015rwa}).
\item Cases B2, C5 and D5 correspond to case C1 in \cite{Girardi:2015rwa}, 
since for all of them $(U_{\rm PMNS})_{\tau2}$ is fixed. 
The sum rule for $\cos\delta$ in case B2, eq.~\eqref{eq:cosdeltaB2}, 
follows from that of case C1 in \cite{Girardi:2015rwa} (see Table~4 therein) for
$\sin^2\th^\circ_{23} = 1/2$,
while the sum rule in eq.~\eqref{eq:cosdeltaC5}, valid in cases C5 and D5, 
can be obtained from the same sum rule found in \cite{Girardi:2015rwa}, 
but for $\sin^2\th^\circ_{23} = 1/4$. 
Again, these values of $\sin^2\th^\circ_{23}$ are fixed uniquely by $G_f = S_4$ and 
the specific choice of the residual symmetries 
considered in the present article~%
\footnote{Note that the value of $\sin^2\th^\circ_{23} = 1/2$ is not present in Table~10 
of \cite{Girardi:2015rwa}, since in this reference the best fit values 
of the mixing angles for the NO spectrum quoted in eqs.~(6)\,--\,(8) therein have been used, 
and employing them, one obtains $\cos\delta \approx 2.76$.}.
\item Cases A1 and B3 of the present study correspond to case C2 in \cite{Girardi:2015rwa}, 
since for these cases $(U_{\rm PMNS})_{\mu3}$ is fixed. 
The expression for $\sin^2\th_{23}$ in eq.~\eqref{eq:ss23A1} follows 
from the corresponding expression for case C2 in Table~6 of \cite{Girardi:2015rwa} 
with $\sin^2\th^\circ_{23} = 1/2$. 
This value is in agreement with Table~10 of \cite{Girardi:2015rwa}. 
Moreover, the sum rule for $\cos\delta$ in eq.~\eqref{eq:cosdeltaA1} in case A1 
can be obtained from the sum rule for case C2~%
\footnote{We would like to point out a typo in eq.~(85) in \cite{Girardi:2015rwa}: 
$\cos^2\th^\circ_{23}$ should read $\cos\th^\circ_{23}$. This typo, however, 
does not affect the corresponding sum rule for $\cos\delta$ in eq.~(86) and in Table~4 
of \cite{Girardi:2015rwa}.}
in Table~4 of \cite{Girardi:2015rwa} with $\sin^2\th^\circ_{23} = 1/2$ and 
$\sin^2\hat\th^\nu_{12} = 1/2$. 
The value of $\sin^2\hat\th^\nu_{12}$, 
which was an arbitrary free parameter in \cite{Girardi:2015rwa}, 
is fixed by the GCP symmetry 
employed in the present study. 
Finally, we note that the expression for $\cos\delta$ 
in eq.~\eqref{eq:cosdeltaB3} valid in case B3 
can formally be obtained from the corresponding expression in case C2 of Table~4 
in \cite{Girardi:2015rwa} setting $\hat\th^\nu_{12} = \th^\nu - \pi/4$.
\item Analogously, cases A2 and B4 correspond to case C7 in \cite{Girardi:2015rwa}. 
Equation~\eqref{eq:ss23A2} can be obtained from the corresponding formula 
in Table~6 of \cite{Girardi:2015rwa} for $\sin^2\th^\circ_{23} = 1/2$, 
which agrees with the result in Table~10 therein. 
The sum rule in eq.~\eqref{eq:cosdeltaA2} follows from 
that in case C7 in Table~4 of \cite{Girardi:2015rwa} with $\sin^2\th^\circ_{23} = 1/2$ 
and $\sin^2\hat\th^\nu_{12} = 1/2$, where again the value of $\sin^2\hat\th^\nu_{12}$, 
which in \cite{Girardi:2015rwa} is a free parameter, here is fixed by the GCP symmetry. 
Similarly to the previous clause, eq.~\eqref{eq:cosdeltaB4} can formally be derived 
from the corresponding expression in case C7 of Table~4 in \cite{Girardi:2015rwa} 
setting $\hat\th^\nu_{12} = \th^\nu - \pi/4$.
\item Case C1 corresponds to case C5 in \cite{Girardi:2015rwa}, 
in which all possible residual flavour symmetries $G_e = Z_2$ and $G_\nu = Z_2$ 
have been considered. 
The expression for $\sin^2\th_{12}$ in eq.~\eqref{eq:ss12C1} follows 
from that of case C5 in Table~6 in \cite{Girardi:2015rwa} with $\sin^2\th^\circ_{12} = 1/4$.
This value of $\sin^2\th^\circ_{12}$ is found for $G_f = S_4$ and the specific choice of 
the residual symmetries (see Table~10 in \cite{Girardi:2015rwa}).
Moreover, eq.~\eqref{eq:cosdeltaC1} for $\cos\delta$ can formally be obtained 
from the corresponding formula in case C5 of Table~4 in \cite{Girardi:2015rwa} 
setting $\sin^2\hat\th^e_{23} = \sin^2\th^e$.
\item Cases C2 and D2 correspond to case C4 of \cite{Girardi:2015rwa}. 
The sum rule for $\cos\delta$ in eq.~\eqref{eq:cosdeltaA1}, valid in cases C2 and D2, 
follows from that of case C4 in \cite{Girardi:2015rwa} (see Table~4 therein) for
$\sin^2\th^\circ_{12} = 1/4$, which is in agreement with Table~10 in \cite{Girardi:2015rwa}. 
\item Cases C3 and D3 correspond to case C3 in \cite{Girardi:2015rwa}. 
Equation~\eqref{eq:cosdeltaA2} for $\cos\delta$, which holds in these cases, 
can be obtained from the corresponding sum rule for case C3 from 
Table~4 in \cite{Girardi:2015rwa} with $\sin^2\th^\circ_{13} = 1/4$. 
As it should be, we find this value in Table~10 of \cite{Girardi:2015rwa}.
\end{enumerate}

\end{document}